\documentclass[longauth,oneside]{aa}

\usepackage[varg]{txfonts}
\usepackage{natbib}
\usepackage{graphicx}
\usepackage{subcaption}
\usepackage{rotating}
\usepackage{multirow} 
\usepackage{xcolor} 
\usepackage{amsmath}
\usepackage{ulem} 
\usepackage{blindtext}
\usepackage{color}
\usepackage[title,toc]{appendix}

\usepackage{amssymb}
\newcommand{\msun}{\mbox{M$_{\odot}$\,}}

\newcommand{\kms}{\mbox{$\rm{km}\,s^{-1}$}}

\DeclareMathAlphabet{\mathsc}{OT1}{cmr}{m}{sc}
\def\testbx{bx}%
\DeclareRobustCommand{\ion}[2]{%
\relax\ifmmode
\ifx\testbx\f@series
{\mathbf{#1\,\mathsc{#2}}}\else
{\mathrm{#1\,\mathsc{#2}}}\fi
\else\textup{#1\,{\mdseries\textsc{#2}}}%
\fi}

\newcommand{\Ha} {\mbox{H$\alpha$}\,}

\begin{document}

\title{Intermediate-luminosity red transients: Spectrophotometric properties and connection to electron-capture supernova explosions}

\author{Y.-Z. Cai \inst{\ref{inst1},\ref{inst2},\ref{inst3}\thanks{E-mail: yzcai@mail.tsinghua.edu.cn}}
\and A.~Pastorello \inst{\ref{inst2} \thanks{E-mail: andrea.pastorello@inaf.it}} 
\and M. Fraser \inst{\ref{inst4}  \thanks{E-mail: morgan.fraser@ucd.ie}}
\and M.~T.~Botticella \inst{\ref{inst5} }
\and N.~Elias-Rosa \inst{\ref{inst2},\ref{inst6} }
\and L.-Z. Wang \inst{\ref{inst30},\ref{inst31}} 
\and R. Kotak  \inst{\ref{inst14}}
\and S. Benetti \inst{\ref{inst2}} 
\and E. Cappellaro \inst{\ref{inst2}}
\and M. Turatto \inst{\ref{inst2}}
\and A.~Reguitti \inst{\ref{inst19},\ref{inst20},\ref{inst2}}
\and S. Mattila  \inst{\ref{inst14}}
\and S. J. Smartt   \inst{\ref{inst22}}
\and C. Ashall \inst{\ref{inst111}}
\and S. Benitez \inst{\ref{inst24}}
\and T.-W. Chen \inst{\ref{inst9}}
\and A. Harutyunyan  \inst{\ref{inst37}}
\and E. Kankare \inst{\ref{inst14}}
\and P. Lundqvist   \inst{\ref{inst9}}
\and P. A. Mazzali \inst{\ref{inst41},\ref{inst25}}
\and A. Morales-Garoffolo \inst{\ref{inst17}}
\and P. Ochner  \inst{\ref{inst2},\ref{inst3}}
\and G. Pignata  \inst{\ref{inst19},\ref{inst20}}
\and S.~J. Prentice \inst{\ref{inst15}}
\and T.~M.~Reynolds \inst{\ref{inst14}}
\and X.-W. Shu \inst{\ref{inst21}}
\and M. D. Stritzinger \inst{\ref{inst23}}
\and L. Tartaglia \inst{\ref{inst2}} 
\and G. Terreran \inst{\ref{inst26}} 
\and L. Tomasella \inst{\ref{inst2}}
\and S. Valenti \inst{\ref{inst27}}
\and G. Valerin  \inst{\ref{inst2},\ref{inst3}}
\and G.-J. Wang \inst{\ref{inst35},\ref{inst3535}}
\and X.-F. Wang \inst{\ref{inst1},\ref{inst29}}
\and L. Borsato \inst{\ref{inst2}}
\and E. Callis \inst{\ref{inst4}}
\and G. Cannizzaro \inst{\ref{inst7},\ref{inst8}}
\and S. Chen  \inst{\ref{inst33}}
\and E. Congiu \inst{\ref{inst34}}
\and M. Ergon     \inst{\ref{inst9}}
\and L. Galbany\inst{\ref{inst45}} 
\and A. Gal-Yam \inst{\ref{inst10}}
\and X. Gao   \inst{\ref{inst32}}
\and M. Gromadzki   \inst{\ref{inst11}}
\and S. Holmbo \inst{\ref{inst23}}
\and F. Huang \inst{\ref{inst12}}
\and C. Inserra   \inst{\ref{inst13}}
\and K. Itagaki      \inst{\ref{inst38}}
\and Z. Kostrzewa-Rutkowska \inst{\ref{inst40},\ref{inst8}}
\and K. Maguire  \inst{\ref{inst15}}
\and S. Margheim      \inst{\ref{inst43}}
\and S. Moran \inst{\ref{inst14}}
\and F. Onori \inst{\ref{inst18}}
\and A.~Sagu\'es Carracedo  \inst{\ref{inst39}}
\and K. W. Smith   \inst{\ref{inst22}}
\and J. Sollerman  \inst{\ref{inst9}}
\and A. Somero \inst{\ref{inst14}}
\and B. Wang \inst{\ref{inst44}}
\and D.~R.~Young   \inst{\ref{inst22}}
}

\institute{Physics Department and Tsinghua Center for Astrophysics (THCA), Tsinghua University, Beijing, 100084, China \label{inst1}
\and INAF - Osservatorio Astronomico di Padova, Vicolo dell'Osservatorio 5, 35122 Padova, Italy \label{inst2}
\and Universit\`a degli Studi di Padova, Dipartimento di Fisica e Astronomia, Vicolo dell'Osservatorio 2, 35122 Padova, Italy \label{inst3}
\and School of Physics, O'Brien Centre for Science North, University College Dublin, Belfield, Dublin 4, Ireland   \label{inst4}
\and INAF-Osservatorio Astronomico di Capodimonte, Salita Moiariello 16, 80131, Napoli, Italy  \label{inst5}
\and Institute of Space Sciences (ICE, CSIC), Campus UAB, Carrer de Can Magrans s/n, 08193 Barcelona, Spain \label{inst6}
\and  Chinese Academy of Sciences South America Center for Astronomy, National Astronomical Observatories, CAS, Beijing 100101, People's Republic of China  \label{inst30}
\and CAS Key Laboratory of Optical Astronomy, National Astronomical Observatories, Chinese Academy of Sciences, Beijing 100101, People's Republic of China \label{inst31}
\and Department of Physics and Astronomy, University of Turku, FI-20014 Turku, Finland \label{inst14}
\and Departamento de Ciencias Fisicas, Universidad Andres Bello, Fernandez Concha 700, Las Condes, Santiago, Chile \label{inst19}
\and Millennium Institute of Astrophysics (MAS), Nuncio Monsenor S\`{o}tero Sanz 100, Providencia, Santiago, Chile  \label{inst20}
\and Astrophysics Research Centre, School of Mathematics and Physics, Queen’s University Belfast, Belfast BT7 1NN Northern Ireland, UK  \label{inst22}
\and Institute for Astronomy, University of Hawai'i at Manoa, 2680 Woodlawn Dr. Hawai'i, HI 96822, USA \label{inst111}
\and Institute of Astrophysics of the Canary Islands, C/ V\'ia L\'actea, s/n, Tenerife, Spain\label{inst24}
\and The Oskar Klein Centre, Department of Astronomy, Stockholm University, AlbaNova, SE-10691 Stockholm, Sweden \label{inst9}
\and Telescopio Nazionale Galileo, Fundaci\'on Galileo Galilei - INAF, Rambla Jos\'e Ana Fern\'andez P\'erez, 7, 38712 Bre\~na Baja, TF - Spain \label{inst37}
\and Astrophysics Research Institute, Liverpool John Moores University, IC2, Liverpool Science Park, 146 Brownlow Hill, Liverpool L3 5RF, UK  \label{inst41}
\and Max-Planck-Institut f$\ddot{\mathrm{u}}$r Astrophysik, Karl-Schwarzschild Str. 1, D-85741 Garching, Germany\label{inst25}
\and Department of Applied Physics, University of C\'adiz, Campus of Puerto Real, E-11510 C\'adiz, Spain\label{inst17}
\and School of Physics, Trinity College Dublin, The University of Dublin, Dublin 2, Ireland\label{inst15}
\and Department  of  Physics,  Anhui  Normal  University,  Wuhu,Anhui, 241002, China \label{inst21}
\and Department of Physics and Astronomy, Aarhus University, Ny Munkegade 120, DK-8000 Aarhus C, Denmark  \label{inst23}
\and Center for Interdisciplinary Exploration and Research in Astrophysics (CIERA) and Department of Physics and Astronomy,Northwestern University, Evanston, IL 60208, USA\label{inst26}
\and Department of Physics and Astronomy, University of California, 1 Shields Avenue, Davis, CA 95616-5270, USA \label{inst27}
\and School of Chemistry and Physics, University of KwaZulu-Natal, Westville Campus, Private Bag X54001, Durban, 4000, South Africa \label{inst35}
\and NAOC-UKZN Computational Astrophysics Centre (NUCAC), University of KwaZulu-Natal, Durban, 4000, South Africa\label{inst3535}
\and Beijing Planetarium, Beijing Academy of Science and Technology, Beijing, 100044, China\label{inst29}
\and Department of Astrophysics/IMAPP, Radboud University, P.O. Box 9010, 6500 GL Nijmegen, Netherlands \label{inst7}
\and SRON, Netherlands Institute for Space Research, Sorbonnelaan, 2, NL-3584CA Utrecht, Netherlands \label{inst8}
\and Physics Department, Technion, Haifa 32000, Israel     \label{inst33} 
\and Departamento de Astronom\'{i}a, Universidad de Chile, Camino del Observatorio 1515, Las Condes, Santiago, Chile  \label{inst34} 
\and Institute of Space Sciences (ICE, CSIC), Campus UAB, Carrer de Can Magrans, s/n, E-08193 Barcelona, Spain.
\label{inst45}
\and Department of Particle Physics and Astrophysics, Weizmann Institute of Science, Rehovot 76100, Israel\label{inst10}
\and Xinjiang Astronomical Observatory, 150 Science-1 Street, Urumqi 830011, China \label{inst32} 
\and Astronomical Observatory, University of Warsaw, Al. Ujazdowskie 4, 00-478 Warszawa, Poland \label{inst11}
\and Department of Astronomy, Shanghai Jiao Tong University, Shanghai, 200240, China \label{inst12}
\and School of Physics and Astronomy, Cardiff University, Queens Buildings, The Parade, Cardiff, CF24 3AA, UK \label{inst13}
\and Itagaki Astronomical Observatory, Yamagata, Yamagata 990-2492, Japan \label{inst38}
\and Leiden Observatory, Leiden University, Niels Bohrweg 2, 2333 CA Leiden, The Netherlands   \label{inst40}
\and Gemini Observatory/NSF’s NOIRLab, Casilla 603, La Serena, Chile \label{inst43}
\and INAF-Osservatorio Astronomico d'Abruzzo, via M. Maggini snc, I-64100 Teramo, Italy \label{inst18}
\and The Oskar Klein Centre, Department of Physics, Stockholm University, AlbaNova, SE-10691 Stockholm, Sweden \label{inst39}
\and Xingming Observatory, Mount Nanshan, Xinjiang, China\label{inst44} 
}

\date{Received Month XX, 2021 / Accepted July 29, 2021 }

\abstract{
We present the spectroscopic and photometric study of five intermediate-luminosity red transients (ILRTs), namely AT 2010dn, AT 2012jc, AT 2013la, AT 2013lb, and AT 2018aes. They share common observational properties and belong to a family of objects similar to the prototypical ILRT SN~2008S. These events have a rise time that is less than 15 days and absolute peak magnitudes of between $-11.5$ and $-14.5$ mag. Their pseudo-bolometric light curves
~peak in the range $0.5$ - $9.0 \times10^{40}~\mathrm{erg~s}^{-1}$ and their total radiated energies are on the order of (0.3 - 3) $\times$~10$^{47}$~erg. After maximum brightness, the light curves show a monotonic decline or a plateau, resembling those of faint supernovae IIL or IIP, respectively. At late phases, the light curves flatten,  roughly following the slope of the $^{56}$Co decay. If the late-time power source is indeed radioactive decay, these transients produce $^{56}$Ni masses on the order of $10^{-4}$ to $10^{-3}$~\msun.  The spectral energy distribution of our ILRT sample, extending from the optical to the mid-infrared (MIR) domain, reveals a clear IR excess soon after explosion and non-negligible MIR emission at very late phases. The spectra show prominent H lines in emission with a typical velocity of a few hundred km~s$^{-1}$, along with Ca~II features. In particular, the [Ca~{\sc ii}] $\lambda$7291,7324 doublet is visible at all times, which is a characteristic feature for this family of transients. The identified progenitor of SN~2008S, which is luminous in archival {\it Spitzer} MIR images, suggests an intermediate-mass precursor star embedded in a dusty cocoon. We propose the explosion of a super-asymptotic giant branch star forming an electron-capture supernova as a plausible explanation for these events. 
}
 
\authorrunning{Y-Z. Cai et al.} 
\titlerunning{Intermediate-Luminosity Red Transients}
\maketitle

\section{Introduction}\label{sec:intro} 

Modern surveys are discovering new types of stellar transients. Some events show intermediate luminosities lying between classical novae and core-collapse (CC) supernovae (SNe), and are collectively named
`gap transients' ($-$10 mag $<$M$_{V}$$<$ $-$15 mag; e.g. \citealt{Kulkarni2009aaxo.conf..312K, Kasliwal2012PASA...29..482K,Pastorello2019NatAs...3..676P}) or even intermediate-luminosity optical transients (e.g. \citealt{Berger2009noao.prop..164B, Soker2012ApJ...746..100S,  Soker2020ApJ...893...20S, Soker2021RAA....21...90S}).     

A fraction of these transients are called  `SN impostors', as they mimic some observational properties of H-rich interacting SNe, but deep late-time imaging reveals that the progenitor stars survived (see, e.g. \citealt{VanDyk2000PASP..112.1532V}). Their typical peak absolute magnitudes are between $-13$ and $-15$ mag, and their spectra show narrow emission lines similar to Type IIn SNe~\citep{Schlegel1990MNRAS.244..269S, Filippenko1997ARA&A..35..309F}. Giant eruptions of luminous blue variables (LBVs) are considered a plausible interpretation for classical SN impostors. 
During giant eruptions, their luminosity increases by over 3 mag with respect to the quiescent phase, accompanied by severe mass loss, sometimes exceeding 1 \msun~\citep{Humphreys1994PASP..106.1025H, Humphreys1999PASP..111.1124H}. Well-studied examples are the historical giant eruptions of two Galactic LBVs: $\eta$ Carinae \citep[e.g. ][]{Humphreys1994PASP..106.1025H,  Soker2001A&A...377..672S, Soker2001MNRAS.325..584S, Soker2003ApJ...597..513S} and P Cygni \citep[e.g. ][]{deGroot1969BAN....20..225D, deGroot1969CoKon..65..203D}, whose physical parameters (including ejected mass, spatial geometry, and radiated energy)  have been well constrained. 

Another class of gap transients has double-peaked light curves and fast-evolving spectra, showing SN IIn-like spectra  at early phases (e.g. blue continuum and narrow H emission lines) and spectra similar to M-type stars at late phases (i.e. a much redder continuum and molecular bands). They are conventionally designated as luminous red novae \citep[LRNe; e.g. see  ][]{Kochanek2014MNRAS.443.1319K,Pejcha2016MNRAS.461.2527P,Smith2016MNRAS.458..950S, Blagorodnova2017ApJ...834..107B, Lipunov2017MNRAS.470.2339L, MacLeod2017ApJ...835..282M, Pejcha2017ApJ...850...59P,Cai2019A&A...632L...6C,Pastorello2019A&A...625L...8P, Pastorello2019A&A...630A..75P, Pastorello2021A&A...646A.119P, Pastorello2021A&A...647A..93P}. The LRN phenomenon is usually interpreted as a post-common-envelope-ejection phase in a contact binary system, and may be followed by stellar coalescence \citep[e.g. see  ][]{Munari2002A&A...389L..51M, Soker2003ApJ...582L.105S, Soker2006MNRAS.373..733S, Tylenda2011A&A...528A.114T, Soker2016MNRAS.462..217S, Smith2016MNRAS.458..950S,Lipunov2017MNRAS.470.2339L, Mauerhan2018MNRAS.473.3765M,Pastorello2019A&A...630A..75P,Soker2021RAA....21..112S}.  

The third group of gap transients are named intermediate-luminosity red transients \citep[ILRTs; ][ and references therein]{Pastorello2019NatAs...3..676P}. These show a slow rise time ($\approx$ 2 weeks) to maximum ($-$11.5 mag $<$M$_{V}$$<$ $-$14.5 mag), followed by a linear decline or a plateau lasting about 2-4 months. When late-time light curves are observed, they decline following the expectation of $^{56}$Co decay. Their light curves resemble those of very faint SNe IIP/L. The spectra are similar to Type IIn SNe with narrow Balmer lines in emission, but are characterised by prominent Ca~{\sc ii} lines (e.g. Ca H\&K, [Ca {\sc ii}] doublet, and Ca {\sc ii} near-infrared triplet); in particular, the [Ca~{\sc ii}] $\lambda\lambda$ 7291, 7324 doublet is always visible at all phases. ILRTs are linked to 8-15 \msun~progenitors and are strong mid-infrared (MIR) emitters, suggesting their progenitors are embedded in dusty cocoons \citep[see ][]{Prieto2008ApJ...681L...9P, Bond2009ApJ...695L.154B, Botticella2009MNRAS.398.1041B, Berger2009ApJ...699.1850B, Thompson2009ApJ...705.1364T}. However, their origin remains mysterious and different scenarios are contemplated. These transients have been proposed to be LBV-like outbursts of moderate-mass stars \citep[e.g. ][]{Smith2009ApJ...697L..49S, Andrews2020arXiv200913541A}~or stellar mergers \citep[e.g. ][]{Kasliwal2011ApJ...730..134K}. A third suggested scenario is that we are actually observing faint CC SN explosions triggered by electron captures in the core of super-asymptotic giant branch (S-AGB) stars, and these are therefore labelled electron-capture (EC) SNe \citep[see, e.g.  ][]{Botticella2009MNRAS.398.1041B, Pumo2009ApJ...705L.138P}. In recent years, a number of ILRTs have been studied, including SN 2008S \citep[see, e.g. ][]{Prieto2008ApJ...681L...9P, Botticella2009MNRAS.398.1041B, Smith2009ApJ...697L..49S, Kochanek2011ApJ...741...37K, Szczygie2012ApJ...750...77S, Adams2016MNRAS.460.1645A}, NGC 300-2008OT1 (NGC 300 OT hereafter) \citep[e.g. ][]{Berger2009ApJ...699.1850B, Bond2009ApJ...695L.154B, Prieto2009ApJ...705.1425P, Kashi2010ApJ...709L..11K, Humphreys2011ApJ...743..118H}, M85 OT2006-1 \footnote{For M85 OT2006-1 (hereafter M85 OT), the classification as ILRT  or LRN is still controversial \citep[][]{Kulkarni2007Natur.447..458K, Pastorello2007Natur.449E...1P, Rau2007ApJ...659.1536R}.} \citep[][]{Kulkarni2007Natur.447..458K, Pastorello2007Natur.449E...1P, Rau2007ApJ...659.1536R},  PTF10fqs \citep{Kasliwal2011ApJ...730..134K}, AT~2017be \citep{Stephens2017TNSTR..33....1S, Adams2018PASP..130c4202A, Cai2018MNRAS.480.3424C}, and AT~2019abn \citep[][Valerin et al. 2021, in preparation]{Jencson2019ApJ...880L..20J,Williams2020A&A...637A..20W}. 

In this paper, we analyse a sample of five ILRTs \footnote{International Astronomical Union (IAU) names for the sampled five ILRTs are used throughout this paper.}, including AT 2010dn, AT 2012jc, AT 2013la, AT 2013lb, and AT 2018aes. In general, they were initially classified as LBV-like outbursts or, more generically, SN impostors. Subsequent follow-up campaigns allowed us to confirm their classifications as ILRTs. In this work, we also compared ILRTs with other claimed EC SN candidates, such as SN 2018zd \citep[][ Callis et al. 2021, in preparation]{2018ATel11379....1Z, Arcavi2018TNSCR2082....1A, Zhang2020MNRAS.498...84Z, Hiramatsu2021NatAs.tmp..107H,Kozyreva2021MNRAS.503..797K}, SN~2018hwm \citep[][]{Reguitti2021MNRAS.501.1059R}, and SN~2015bf \citep[][]{Lin2021MNRAS.505.4890L}, as well as underluminous SNe IIP \citep[i.e. SN~1999br, SN~2005cs; ][]{Pastorello2004MNRAS.347...74P, Pastorello2006MNRAS.370.1752P, Pastorello2009MNRAS.394.2266P}.

The present paper is organised as follows: general information on the sampled ILRTs is given in Sect. \ref{sec:info}. Photometric and spectroscopic analysis is reported in Sects. \ref{sec:photometry} and  \ref{sec:spectra}, respectively.  The rate of ILRTs is estimated in Sect. \ref{ILRTrates}.  Finally, a discussion on the observational and theoretical properties of ILRTs is given in Section \ref{sec:discussion}. \\

\section{Basic sample information  }
\label{sec:info}

\subsection{Host galaxies: distance, reddening and metallicity}
\label{sec:hostgalaxy}

AT~2010dn was discovered by the amateur astronomer K. Itagaki in May 2010 at 31.523 UT (hereafter UT will be used throughout this paper) at an unfiltered magnitude 17.5 mag (CBAT\footnote{\url{http://www.cbat.eps.harvard.edu/iau/cbet/002200/CBET002299.txt}}). The object was detected at RA=$10^{h}18^{m}19\fs89$, Dec=$+41\degr26\arcmin28\arcsec.80$~[J2000], $61\arcsec$ north and $33\arcsec$ east from the centre of NGC~3184. We adopt $d=14.40 \pm 0.33$ Mpc ($\mu = 30.79 \pm 0.05$~$\mathrm{mag}$) as the  distance to the host galaxy, as obtained  from Cepheids \citep{Ferrarese2000ApJ...529..745F}. The Galactic reddening  $E(B-V)_{\mathrm{Gal}}=0.017$~$\mathrm{mag}$ is from \citet{Schlafly2011ApJ...737..103S}. The presence of narrow interstellar Na I D ($\lambda\lambda$ 5890, 5896) absorption in the transient spectra at the host galaxy redshift is usually considered as evidence for additional internal extinction \citep[e.g. ][]{Munari1997A&A...318..269M, Turatto2003fthp.conf..200T, Pozzo2006MNRAS.368.1169P, Sahu2006MNRAS.372.1315S, Poznanski2011MNRAS.415L..81P}. In this case, the equivalent width (EW) of Na I D varies with time, suggesting that this dust is likely due to circumstellar material (CSM). For this reason, we neglect the possible host component of extinction and adopt a total reddening E($B-V$)$_{\mathrm{Total}}$ = E($B-V$)$_{\mathrm{Gal}}$ = 0.017~$\mathrm{mag}$, in agreement with \citet{Smith2011MNRAS.415..773S}.   

AT 2012jc (also known as PSN J14535395+0334049, NGC~5775-2012OT1, SNhunt120 and LSQ12brd) was found by S. Howerton and the Catalina Real-Time Transient Survey (CRTS\footnote{\url{http://crts.caltech.edu/index.html}}) in March 2012 at 27.460 UT \citep{Berger2012ATel.4009....1B, Howerton2012ATel.4004....1H}. Its coordinates are: RA=$14^{h}53^{m}53\fs95$, Dec=$+03\degr34\arcmin04\arcsec.90$~[J2000], and it is located at $85\arcsec$ north and $55\arcsec$ west from the centre of NGC 5775. Based on the weighted average of several estimates using the Tully-Fisher method \citep[e.g. ][]{Tully2013AJ....146...86T, Sorce2014MNRAS.444..527S, Tully2016AJ....152...50T} from the NASA/IPAC Extragalactic database (NED\footnote{NED; \url{http://nedwww.ipac.caltech.edu/}}), a distance of $d = 18.62 \pm 1.12$ Mpc (hence $\mu = 31.35 \pm 0.13$~$\mathrm{mag}$) can be estimated for NGC 5775. 
A Galactic reddening E($B-V$)$_{\mathrm{Gal}}$ = 0.037~$\mathrm{mag}$ \citep{Schlafly2011ApJ...737..103S} is adopted. We measure a constant  $EW\simeq0.9$~\AA~for the Na~{\sc i} D absorption at the redshift of NGC 5775 in the early spectra of the transient. Following \citet{Turatto2003fthp.conf..200T} and assuming R$_\mathrm{V}$ = 3.1 \citep{Cardelli1989ApJ...345..245C}, we obtain a host galaxy reddening of $E(B-V)_{\mathrm{Host}}=0.144$~$\mathrm{mag}$. Hence, the total reddening towards AT 2012jc is $E(B-V)_{\mathrm{Total}}$~=~0.181~$\mathrm{mag}$.  

The discovery of AT 2013lb (PSN J15213475-0722183; NGC 5917-2013OT1) was announced by the CHilean Automatic Supernovas sEarch in January 2013 at 27.340 UT \citep[CHASE\footnote{\url{http://www.das.uchile.cl/proyectoCHASE}},][]{Pignata2009AIPC.1111..551P, Margheim2013ATel.4798....1M}. The coordinates are RA=$15^{h}21^{m}34\fs75$, Dec=$-07\degr22\arcmin18\arcsec.30$~[J2000], and $19\arcsec$ north and $31\arcsec$ east from the centre of NGC~5917. For this galaxy, we adopt a kinematic distance, assuming a standard cosmological model with $H_{0}=73$ \kms $\rm{Mpc}^{-1}$, $\Omega_{\mathrm{M}}=0.27$, $\Omega_{\mathrm{\Lambda}}=0.73$ \citep[][]{Spergel2007ApJS..170..377S}, which is used throughout this paper. From the radial velocity corrected for Local Group infall onto the Virgo Cluster $V_{\mathrm{Vir}}$ = 2024 $\pm$ 7 \kms~ \citep[see][reported by {\textit HyperLeda} \footnote{\url{http://leda.univ-lyon1.fr/}}]{Sandage1990ApJ...365....1S, Theureau1998A&A...340...21T, Terry2002A&A...393...57T}, 
we obtain a luminosity distance $d = 27.73 \pm 1.90$ Mpc ($\mu = 32.21 \pm 0.15$~$\mathrm{mag}$). A Galactic reddening $E(B-V)_{\mathrm{Gal}}=0.085$~$\mathrm{mag}$ \citep{Schlafly2011ApJ...737..103S} is reported for this object. Additionally, spectroscopy indicates  a negligible host galaxy extinction based on non-detection of the Na~{\sc i} D absorption feature at the redshift of the host galaxy.  

 AT 2013la (also named PSN~J13100734+3410514; PS1-14ln; UGC~8246-2013OT1) was discovered by B. Wang and X. Gao\footnote{\url{http://www.cbat.eps.harvard.edu/unconf/followups/J13100734+3410514.html}} in December 2013 at  20.932 UT, at RA=$13^{h}10^{m}07\fs34$, Dec=$+34\degr10\arcmin51\arcsec40$~[J2000] \citep[e.g. ][]{Tartaglia2014ATel.5737....1T}. The object is $0.8\arcsec$ south and $32.7\arcsec$ east from the centre of UGC~8246. Averaging several recent Tully-Fisher estimates \citep[e.g. ][]{Tully2013AJ....146...86T, Sorce2014MNRAS.444..527S, Tully2016AJ....152...50T}, we obtain $d = 15.21 \pm 0.11$ Mpc (hence, $\mu = 30.91 \pm 0.02$~$\mathrm{mag}$) for UGC 8246. 
We adopt the same line-of-sight reddening as \citet{Barsukova2014arXiv1412.7090B}, that is,   $E(B-V)_{\mathrm{Total}}=0.009$~$\mathrm{mag}$. We note that, in analogy to AT~2010dn, a variable $EW$ of Na I D suggests the presence of additional circumstellar dust, which is not accounted for in this reddening estimate. We hence adopt $E(B-V)_{\mathrm{Total}}=E(B-V)_{\mathrm{Gal}}$.  

AT~2018aes (Kait-18M) was discovered by the Lick Observatory Supernova Search (LOSS\footnote{\url{http://w.astro.berkeley.edu/bait/kait.html}}) in March 2018 at 11.535 \citep{Yunus2018TNSTR.338....1Y}. It was detected at RA=$13^{h}48^{m}17\fs76$, Dec=$+03\degr56\arcmin44\arcsec.20$~[J2000], $18.9\arcsec$ south and $25.7\arcsec$ east of the nucleus of NGC~5300 \citep{Andrews2018ATel11441....1A}. We average a few recent Tully-Fisher distances estimates from NED \citep[e.g.][]{Tully2013AJ....146...86T, Sorce2014MNRAS.444..527S, Tully2016AJ....152...50T}, obtaining a weighted average distance $d = 18.02 \pm 0.61$ Mpc ($\mu = 31.28 \pm 0.07$~$\mathrm{mag}$). 
As there is some persistent spectroscopic signatures of interstellar Na~I~D at the host galaxy redshift, following \citet{Turatto2003fthp.conf..200T},  we estimate an internal reddening $E(B-V)_{\mathrm{Host}}=0.160$~$\mathrm{mag}$,  while we adopt a Milky Way contribution $E(B-V)_{\mathrm{Gal}}=0.020$~$\mathrm{mag}$ \citep{Schlafly2011ApJ...737..103S} for AT 2018aes. Hence, we obtain a total colour excess $E(B-V)_{\mathrm{Total}}=0.180$~$\mathrm{mag}$. 

Although we did not perform precise estimates of metallicity at the locations of these ILRTs, an estimate of the characteristic oxygen abundance 
at $0.4R_{25}$ \footnote{$R_{25}$, also known as the de Vaucouleurs radius, is defined as the radius along the semi-major axis where the surface brightness in B-band is 25 magnitudes per square arcsecond ($\mu_{\rm{B}}=25$~mag/arcsec$^{2}$) \citep[e.g. ][]{deVaucouleurs1991rc3..book.....D,Corwin1994AJ....108.2128C}.} can be done using a statistical approach and the following relation from \citet{Pilyugin2004A&A...425..849P}:

\begin{equation}\label{metallicity}
12 + \mathrm{log}\left(\mathrm{O/H}\right)~$=$~6.93~\left(\pm~0.37\right)~-~0.079~\left(\pm~0.018\right)~\times~M_\mathrm{B},
\end{equation}

\noindent which links the characteristic oxygen abundance  to the absolute $B$-band magnitude ($M_\mathrm{B}$) of the galaxy. The oxygen abundances of the ILRT host galaxies span a very narrow range, from 8.2 to 8.6~$\mathrm{dex}$, which is nearly solar or marginally subsolar \cite[adopting a solar metallicity of 12 + log(O/H) = 8.69 dex; see e.g. ][]{Asplund2009ARA&A..47..481A, vonSteiger2016ApJ...816...13V,Vagnozzi2019Atoms...7...41V}. We note that the hosts are all spiral galaxies and the positions of the ILRTs are towards their outer edges.

\begin{figure*}[htp]
\centering
\includegraphics[width=.3\textwidth]{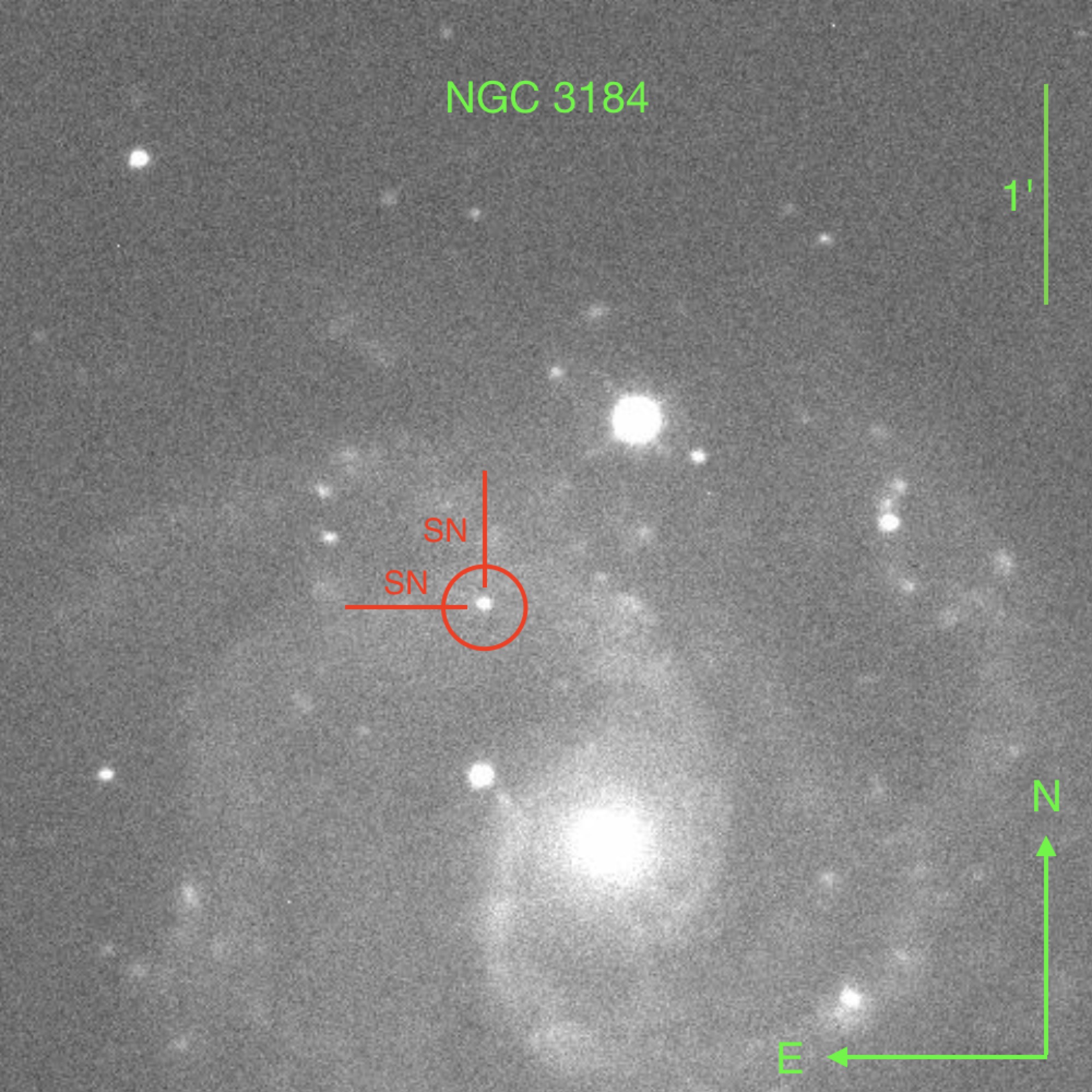}\quad
\includegraphics[width=.3\textwidth]{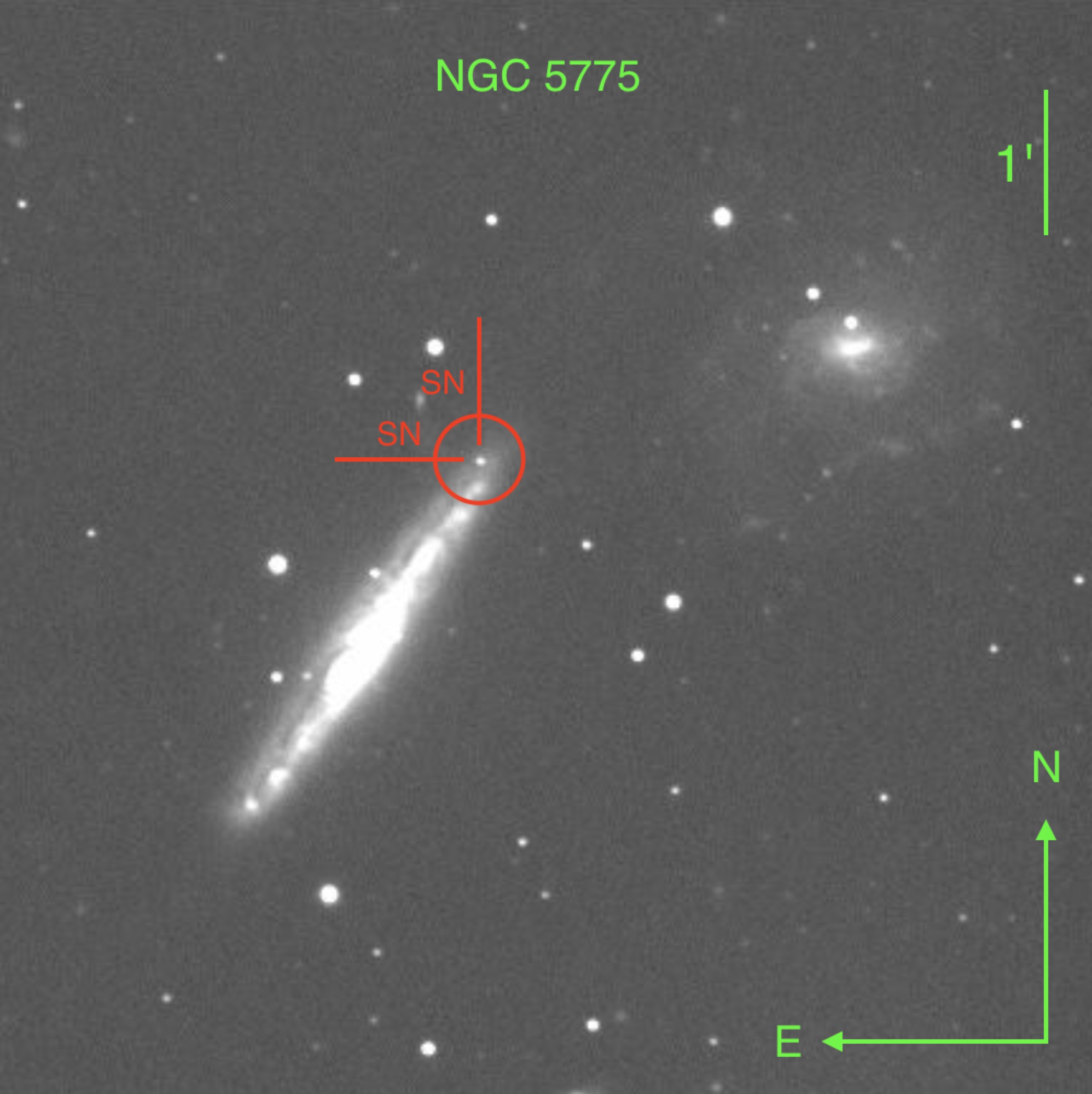}\quad
\includegraphics[width=.3\textwidth]{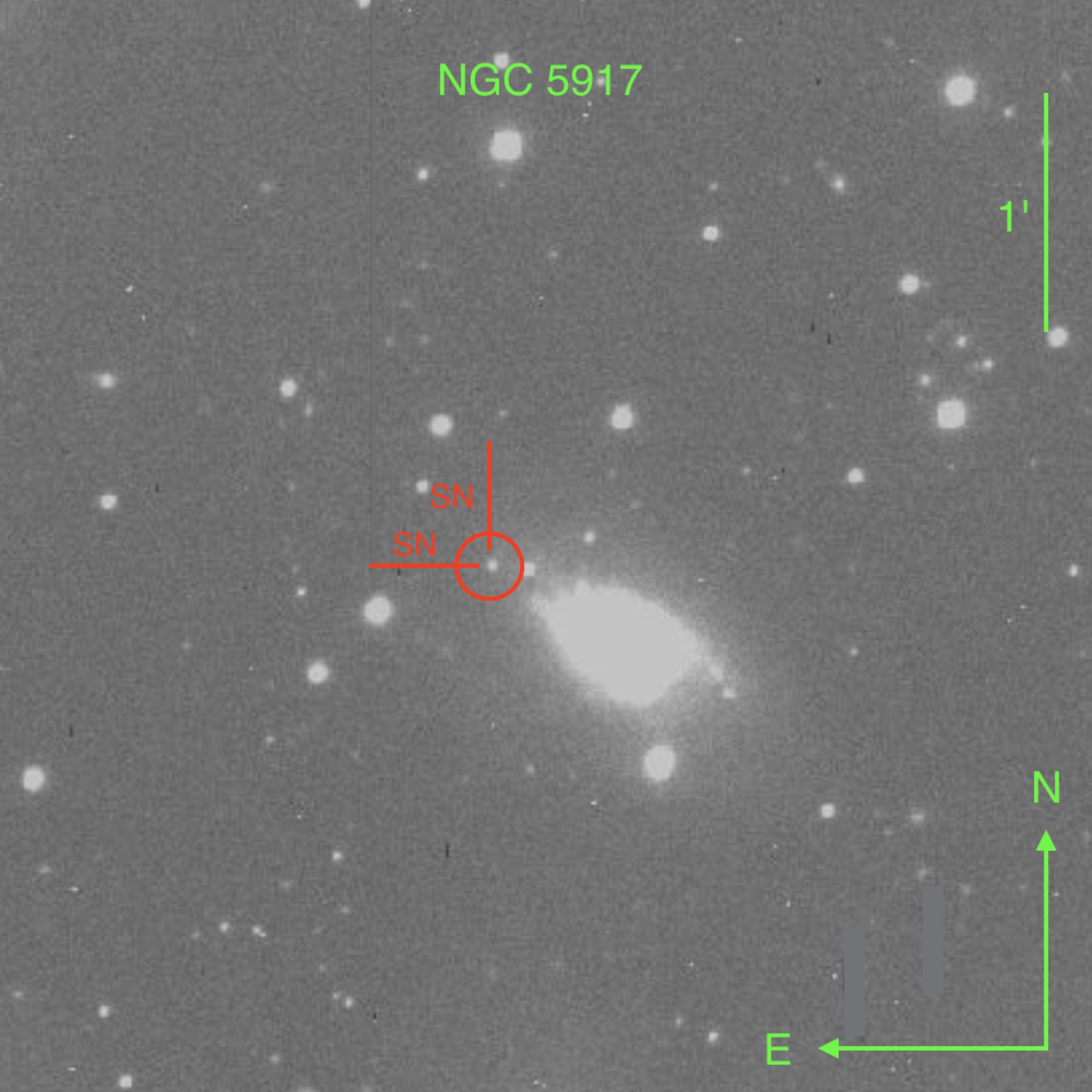}\quad

\medskip

\includegraphics[width=.3\textwidth]{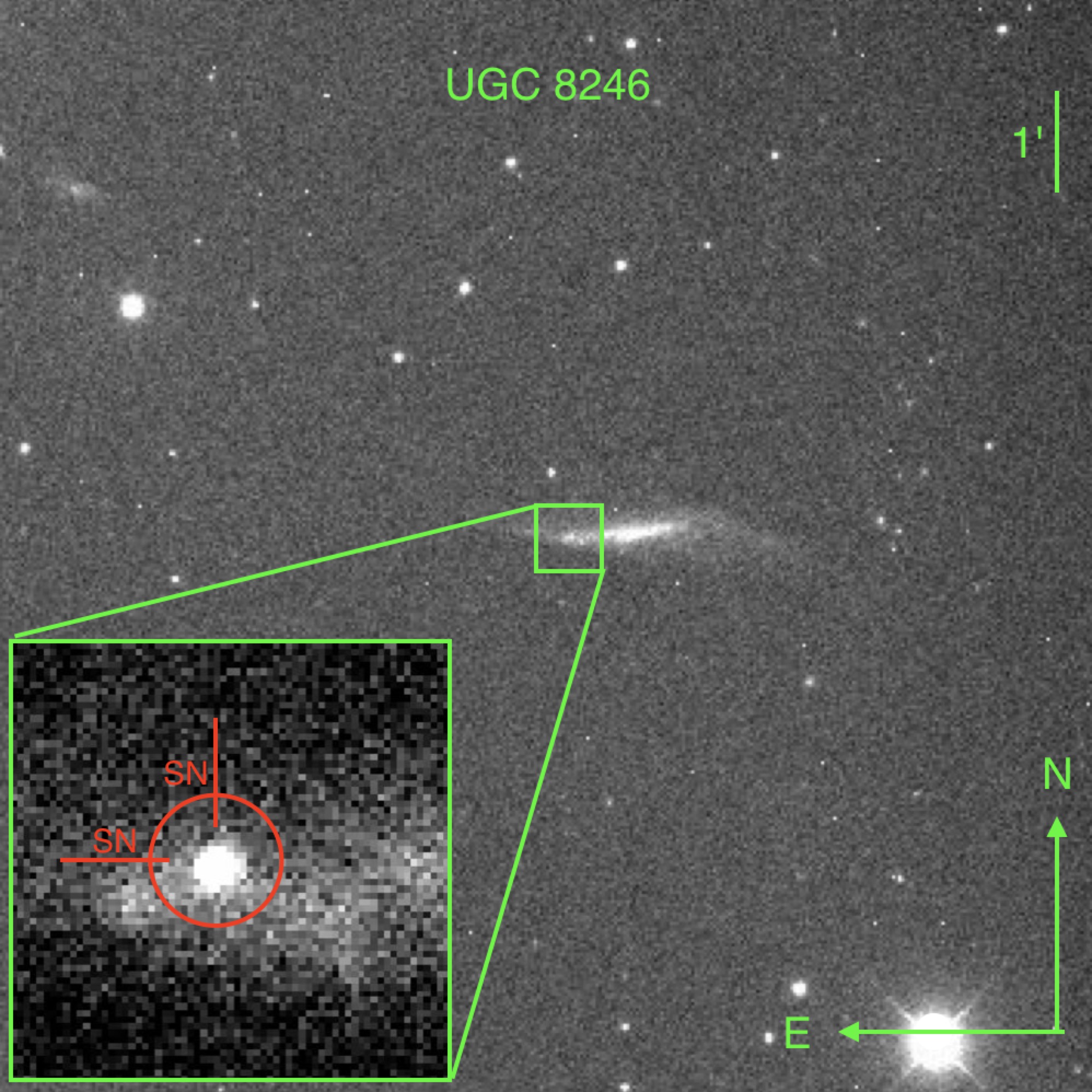}\quad
\includegraphics[width=.3\textwidth]{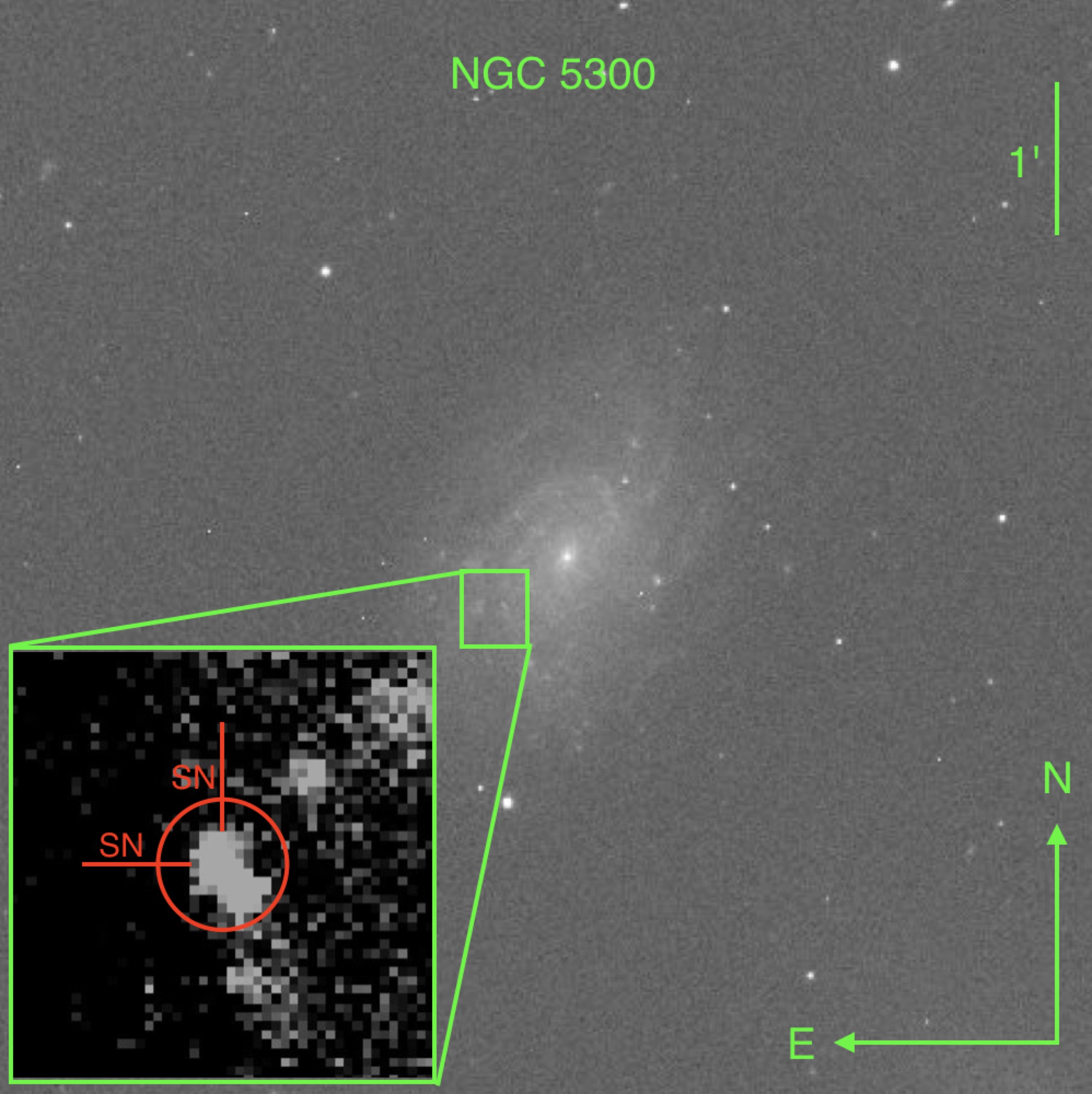}\quad

\caption{Top-left: AT~2010dn in one of the spiral arms of the host galaxy. Sloan $r$-band image taken in June 2010 at 19 with the 2m robotic Faulkes Telescope North (FTN) operated by Las Cumbres Observatory \citep[LCO; ][]{Brown2013PASP..125.1031B}. Top-middle: AT 2012jc at the north edge of its host galaxy. Johnson-Cousins $R$-band image obtained in April 2012 at 06 with the 0.41m  PROMPT5 telescope. Top-right: AT 2013lb in the outskirts of its host galaxy. Johnson-Cousins $R$-band image obtained in February 2013 at 20 with the 2m fully robotic Liverpool Telescope (LT) with RATCam. Bottom-left: AT 2013la at the east edge of its galaxy. A zoom-in of the transient position is shown in the lower-left corner. Sloan $r$-band image obtained in February 2014 at 12 with the LT equipped with IO:O. Bottom-right: AT~2018aes and its host galaxy. LT/IO:O Sloan $r$-band image obtained in May 2018 at 17 with a blow-up of the transient site  in the lower-left inset.}
\label{pic1:fields}
\end{figure*}

A summary of the ILRTs and their host galaxy parameters is listed in Table~\ref{table_parameters}, while in Figure~\ref{pic1:fields} we show the location of the transients in their host galaxies. \\

\begin{table*}
 \centering
  \caption{Basic information for the ILRT host galaxies. }
  \label{table_parameters}
  {
  \scalebox{0.8}{
  \begin{tabular}{@{}lccccccccc@{}}
     \hline
     
     \hline
     Object & Host Galaxy$^a$ &$m_\mathrm{B}$$^b$ &Redshift& Distance  & Distance Modulus & Radial Distance$^c$& Metallicity&$E(B-V)_{\mathrm{Gal}}$ &$E(B-V)_{\mathrm{Host}}$  \\
                 &                &  ($\mathrm{mag}$)            & & ($\mathrm{Mpc}$) & ($\mathrm{mag}$)  & ($\mathrm{kpc}$)&($\mathrm{dex}$)      &($\mathrm{mag}$)   &($\mathrm{mag}$)\\
     \hline
     
     \hline
     AT 2010dn      &  NGC 3184 [SABc] &10.3& 0.00198  &14.40 (0.33)  & 30.79 (0.05)  &4.8 (0.1)   & 8.55 (0.74) &0.017&0\\ 
     AT 2012jc  &  NGC 5775 [SBc]    &12.2& 0.00561 & 18.62 (1.12)  & 31.35 (0.13)  & 9.1 (0.5)  & 8.44 (0.71)& 0.037& 0.144 (0.080)\\
     AT 2013lb  &  NGC 5917 [Sb]      &13.7& 0.00635 &27.73 (1.90)   & 32.21 (0.15)  & 2.7 (0.1)  & 8.39 (0.70)& 0.085& 0\\
     AT 2013la  & UGC 8246 [SBc]     &14.6& 0.00271 & 15.21 (0.11)  & 30.91 (0.02)  & 4.0 (0.3)  & 8.22 (0.66) & 0.009& 0\\
     AT 2018aes     & NGC 5300 [SABc]  &12.1& 0.00391 &18.02 (0.61)   & 31.28 (0.07)    & 2.8 (0.1) & 8.45 (0.72) & 0.020& 0.160 (0.085)\\
     \hline \hline
     SN 2008S        & NGC 6946 [SABc]  &8.2& 0.00013 & 5.70 (0.21)& 28.78 (0.08)     & 5.6 (0.2)   & 8.56 (0.74) &  0.360 &0.320 (0.031)\\
     NGC300 OT     & NGC 0300 [Scd]    &8.8& 0.00048 &1.88 (0.12)& 26.37 (0.14)      & 2.5  (0.2)   & 8.32 (0.69) &  0.011& 0.250 (0.148)\\
     PTF10fqs         & NGC 4254 [Sc]      &10.2& 0.00803 &14.26 (3.41)& 30.77 (0.52)     & 6.9 (1.7)    & 8.56 (0.74)& 0.040 & 0\\
     AT 2017be        & NGC 2537 [SBm]  &13.1& 0.00144 & 7.82 (0.54)& 29.47 (0.15)  &  -   & 8.22 (0.66)  & 0.048 &0.040 (0.02)\\
     M85 OT            & NGC 4382 [S0-a]   &10.1& 0.00243 &15.85 (0.88)& 31.00 (0.12)    &2.5 ( 0.1)   & 8.58 (0.75) & 0.027 & 0.113 (0.108)\\
     \hline
     
     \hline
  \end{tabular}
}  }
  \begin{flushleft} 
  $Notes:$ Our sample is reported in the top half of the table and other published objects are in the lower half. The uncertainties on distance, distance modulus, radial distance, and metallicity are obtained from error propagations, while host galaxy extinction is computed from the standard deviation of measurements. \\
  $^a$ Galaxy type from {\textit HyperLeda}. \\
  $^b$ $B$-band apparent magnitude of the host galaxy,  from  NED.\\
  $^c$ Projected distance of the ILRT location from the host galaxy nucleus.\\
  \end{flushleft}
\end{table*}

\subsection{Data reduction}
\label{sec:datareduction}
Routine follow-up campaigns were triggered soon after the announcement of the ILRT discoveries, using the instruments available to our collaboration.  Information on the instrumental configurations is reported in Table~\ref{table_setup} (Appendix~\ref{sec:facilities}). For some transients, additional early-time unfiltered data were collected from amateur astronomers. We also collected historical data available in public archives.  

The raw images were first pre-reduced, applying overscan, bias and flat-field corrections through standard \textsc{iraf}\footnote{\url{http://iraf.noao.edu/}} tasks \citep{Tody1986SPIE..627..733T, Tody1993ASPC...52..173T}. When necessary, multiple-exposure frames were median-combined to increase the {\it signal-to-noise} ratio (S/N). Photometric measurements were performed through a dedicated pipeline, {\sl ecsnoopy}\footnote{{\sl ecsnoopy} is a package for SN photometry using PSF fitting and/or template subtraction developed by E. Cappellaro. A package description can be found at \url{http://sngroup.oapd.inaf.it/snoopy.html.}}. {\sl ecsnoopy} is a \textsc{python}-based script, which makes use of a series of packages for photometry and template subtraction \citep[e.g. {\sc sextractor}\footnote{\url{www.astromatic.net/software/sextractor/}}, {\sc daophot}\footnote{\url{http://www.star.bris.ac.uk/~mbt/daophot/}}, {\sc hotpants}\footnote{\url{http://www.astro.washington.edu/users/becker/v2.0/hotpants.html/}}; ][]{Bertin1996A&AS..117..393B, Stetson1987PASP...99..191S, Becker2015ascl.soft04004B}. Individual instrumental magnitudes were measured with the point-spread function (PSF) fitting technique. We first subtracted the background contaminating the object using a low-order polynomial fit. A standard PSF template was constructed by fitting the profiles of several isolated stars (usually from 5 to 10 stars) in the SN frames. The PSF model was then fitted to the target source, and the goodness of the fit was evaluated by inspecting the residuals at the SN location. When the object was faint or had a complex background, we used the template-subtraction technique to remove the background contamination from our measurements. Specifically, we applied it to measurements of AT~2010dn, AT 2013la,  and AT~2018aes. The errors were estimated via artificial star tests, in which several fake stars were placed close to the SN location. The fake stars were fitted with the PSF method, and the standard deviation of these measurements provided us the instrumental errors. These were combined in quadrature with the PSF-fit error, finally providing the total photometry error.          

We applied the instrumental zero points (ZPs) and colour terms (CTs) to the instrumental magnitudes obtained through observations of standard fields obtained in photometric nights. Specifically, the Johnson-Cousins magnitudes were determined with reference to the \citet{Landolt1992AJ....104..340L} catalogue, and  Sloan-filter data were calibrated via the SDSS DR 13 catalogue \citep{Albareti2017ApJS..233...25A}. In order to obtain an accurate calibration, a sequence of reference stars in the field of each ILRT was used to correct the ZPs in non-photometric nights. 
When available, the Sloan magnitudes of the reference stars in the ILRT field were directly taken from SDSS, while the Johnson-Cousins magnitudes were obtained from the SDSS data using the conversion relations of \citet{Chonis2008AJ....135..264C}.             
     
Near-infrared (NIR) data reduction includes flat fielding, distortion correction and sky subtraction. We performed pre-reduction on NOT/NOTCam and NTT/SOFI raw images using dedicated pipelines for NOTCam (version 2.5) and SOFI \citep[PESSTO pipeline, version 2.2.10; ][]{Smartt2015A&A...579A..40S} respectively. Standard {\sc iraf}~tasks were used to reduce LT/SupIRCam and TNG/NICS images. Instrumental magnitudes were measured via {\sl ecsnoopy}, and then calibrated using the Two Micron All Sky Survey \citep[2MASS\footnote{\url{http://irsa.ipac.caltech.edu/Missions/2mass.html/}}, ][]{Skrutskie2006AJ....131.1163S} catalogue (assuming negligible colour corrections). 

The Spitzer Space Telescope was equipped with the InfraRed Array Camera (IRAC; two channels: CH1=3.6 $\mu$m and CH2=4.5 $\mu$m). We used the Level 2 post-BCD (Basic Calibrated Data) images, which were reduced with the {\it Spitzer} pipeline\footnote{The IRAC Instrument Handbook is available on the website:  \url{https://irsa.ipac.caltech.edu/data/SPITZER/docs/irac/iracinstrumenthandbook/}}. These pBCD images are composed of a mosaic image, after rejection of most instrumental artefacts, an uncertainty image, and an associated coverage image which is the map of how many frames per position are available to generate the mosaic image. We note that AT 2010dn was targeted at 14 epochs by the Spitzer Space Telescope during 2010 -- 2015 and the data are available from {\it Spitzer} Heritage Archive (SHA)\footnote{\url{http://irsa.ipac.caltech.edu/applications/Spitzer/SHA/}}. We took the {\it Spitzer} images from 2019 September 14 as templates\footnote{Note that there were some images taken before AT~2010dn exploded, but as these were taken during the cryogenic phase of the {\it Spitzer} mission they would be less suitable for our data.}, which were matched and transformed to each epoch of the observed images and then subtracted using {\sc hotpants} \citep[][]{Becker2015ascl.soft04004B}. We performed aperture photometry on the subtracted images using a small (4$\times$$0.6\arcsec$ pBCD pixel) aperture, and then applied an aperture correction and a conversion from MJy/sr to mJy/pix to get the fluxes. With this approach, we obtained ten detections for the CH1 and CH2 channels over a time-span of five years (from 2010 to 2015). For a few other epochs we have non-detections down to 3$\sigma$. The resulting magnitudes are reported in Table~\ref{table:SpitzerMag} (Appendix~\ref{sec:spectphot}).      

\section{Photometry}
\label{sec:photometry}
Our follow-up campaigns for each of the five transients started soon after their discovery, and lasted several months. The photometric measurements of the transients are reported in Tables \ref{2010dn_opt_LC} - \ref{AT2018aes_nir_LC} (Appendix \ref{sec:spectphot}), while individual light curves are shown in  Figure~\ref{pic1:lcs}. As a reference epoch, we selected the  $R/r$-band maximum, obtained using a third-order polynomial fit to the observed light curves between about $-$20 and $+$20 days from maximum (see below).
 
\subsection{Apparent light curves }
\label{sec:AppMag}


\begin{figure*}[htp]
	\centering
	\includegraphics[width=.44\textwidth]{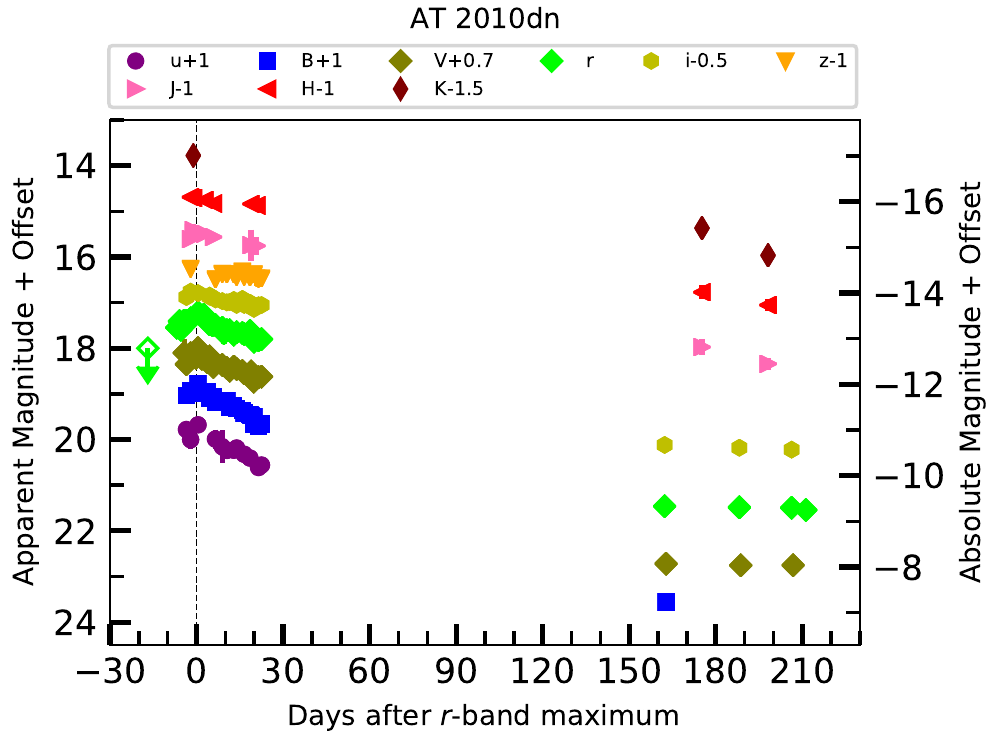}\quad
	\includegraphics[width=.44\textwidth]{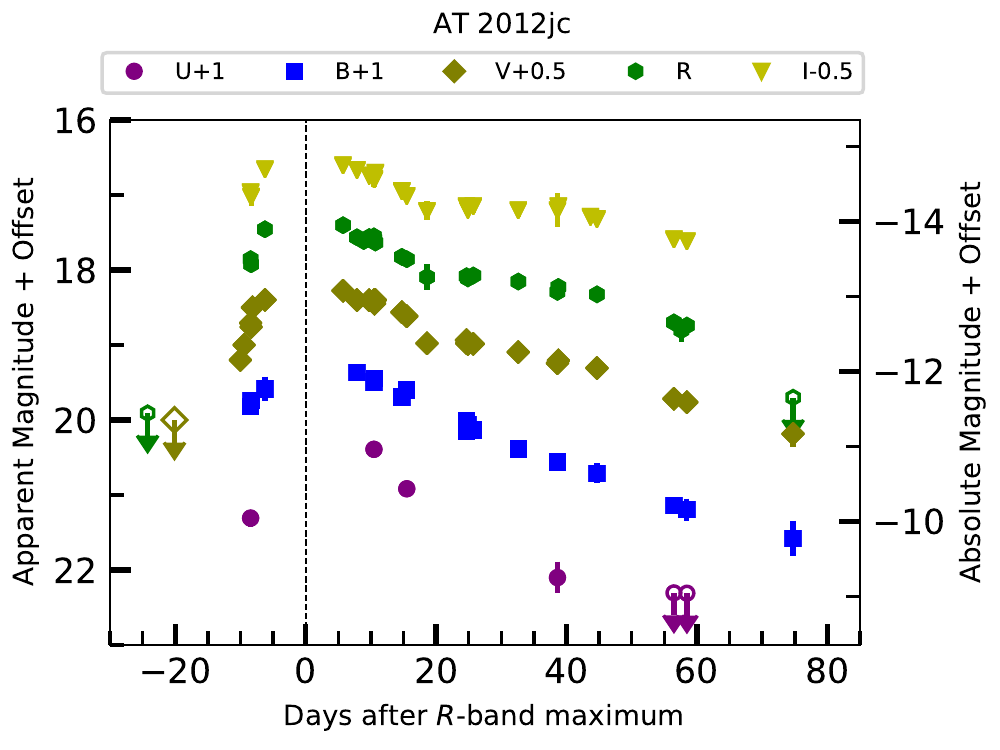}\quad
	\medskip	
	\includegraphics[width=.44\textwidth]{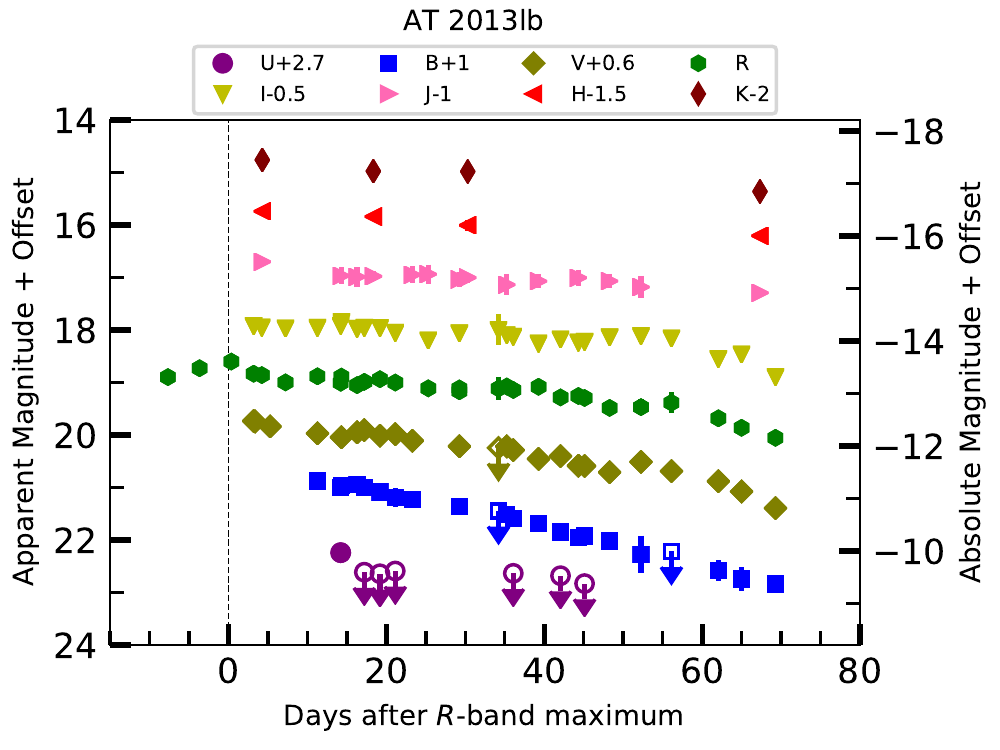}\quad
	\includegraphics[width=.44\textwidth]{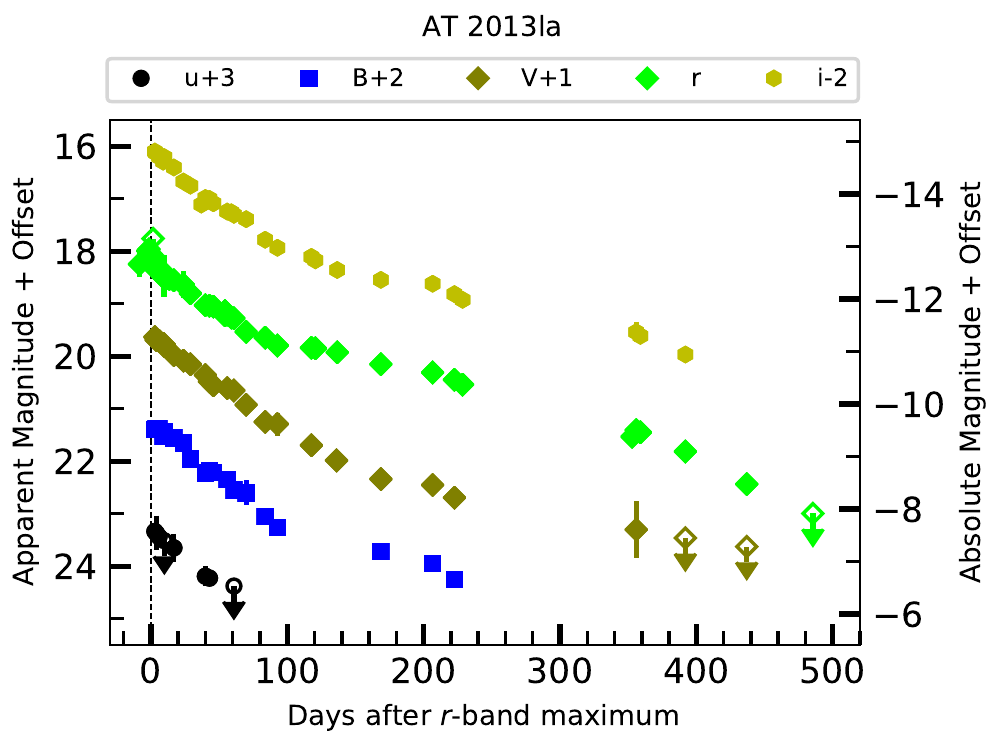}\quad
	\medskip	
	\includegraphics[width=.44\textwidth]{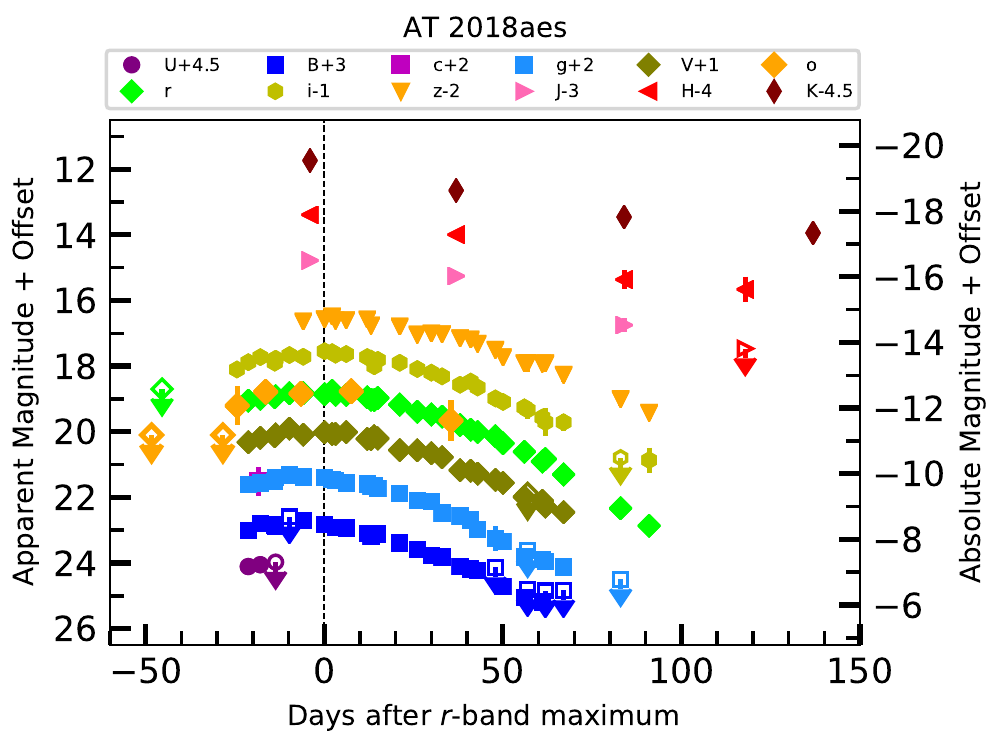}\quad
	\caption{Multi-band light curves of AT 2010dn (top-left), AT 2012jc (top-right), AT 2013lb (middle-left), AT 2013la (middle-right), and AT 2018aes (bottom). The dashed vertical lines indicate the $R/r$-band maximum time.  Upper limits are marked by empty symbols with arrows. The light curves are shifted with some constants for clarity and the shift amount is reported at the top. The errors of most magnitudes are smaller than the plotted symbol sizes.}
	\label{pic1:lcs}
\end{figure*}

\begin{table}
	\center
	\caption{Decline rates of the light curves of individual ILRTs (mag/100d), along with their uncertainties.  }
	\label{lc_para}
	\scalebox{0.7}{
		\begin{tabular}{@{}cccc@{}}
			\hline
			
			\hline
			Filter & Phase {\sc i} ($\gamma$$_1$) & Phase {\sc ii} ($\gamma$$_2$)  &Phase {\sc iii} ($\gamma$$_3$)      \\  
			\hline
			
			\hline
			\multicolumn{4}{c}{AT 2010dn}\\ \hline
			$u$  & 6.32 (0.54) & 4.25 (0.56)  &  --   \\
			$B$  & 6.23 (0.92)  & 4.02 (0.38) &  --   \\
			$V$  & 6.36 (1.24) & 2.14 (0.36)  & 0.08 (0.05)     \\
			$r$  & 5.17 (0.96)  & 1.50 (0.38) &  0.08 (0.05)    \\
			$i$ &  6.16 (0.43) & 1.03 (0.51) & 0.29 (0.15)      \\
			$z$ & 0.89 (0.06) & 0.89 (0.06)  &  --   \\
			$J$ & 1.45 (0.11) & 1.45 (0.11)  & 1.41 (0.00)      \\
			$H$ & 0.72 (0.05) & 0.72 (0.05)   & 0.80 (0.00)      \\
			$K$ & -- & -- & 1.75 (0.00)     \\
			
			\hline \hline
			
			\multicolumn{4}{c}{AT 2012jc}\\ \hline
			$U$  & 5.78 (0.79) &  5.78 (0.79)  & -- \\
			$B$  & 4.16 (0.29) & 2.77 (0.19)   & 2.77 (0.19)  \\
			$V$  & 3.17 (0.37) & 2.20 (0.10)  &  2.20 (0.10)  \\
			$R$  & 4.98 (0.44)  & -0.11 (0.35)  & 2.68 (0.27)  \\
			$I$  & 4.14 (0.44)  & 0.09 (0.20)  & 2.24 (0.08) \\
			
			\hline \hline
			\multicolumn{4}{c}{AT 2013lb}\\ \hline
			$B$  & --           & 3.41 (0.09)  & 3.41 (0.09) \\
			$V$  & 1.90 (0.12)  & 1.90 (0.12)  & 4.94 (0.52) \\
			$R$  & 1.10 (0.11)  & 1.10 (0.11)  & 5.13 (0.18) \\
			$I$  & 0.59 (0.10)  & 0.59 (0.10)  &  5.12 (1.35) \\
			$J$  & 0.56 (0.10)  & 0.56 (0.10)  & 0.56 (0.10) \\
			$H$  & 0.74 (0.10)  & 0.74 (0.10)  & 0.74 (0.10)  \\
			$K$  & 0.91 (0.12)  & 0.91 (0.12)  & 0.91 (0.12)  \\
			
			\hline \hline
			\multicolumn{4}{c}{AT 2013la}\\ \hline
			$u$  & 2.22 (0.04) & 2.22 (0.04)  & --\\
			$B$  & 2.08 (0.07) & 0.71 (0.11)  & --   \\
			$V$  & 1.79 (0.03) & 0.59 (0.27)  & --    \\
			$r$  & 1.70 (0.05) & 0.62 (0.04)  & 1.17 (0.12)   \\
			$i$  & 1.72 (0.07) & 0.37 (0.11)  & 1.14 (0.09) \\
			
			\hline \hline
			\multicolumn{4}{c}{AT 2018aes}\\ \hline
			$B$ & 2.21 (0.21) & 3.78 (0.13)    & 5.02 (1.99)   \\
			$g$ & 1.85 (0.22) & 4.02 (0.23)   & 4.97 (0.90) \\
			$V$ & 2.25 (0.25) & 4.33 (0.27)    &  5.01 (2.05)  \\
			$r$ & 2.55 (0.34) & 3.75 (0.13)   & 6.51 (0.04) \\
			$i$ & 2.17 (0.68) & 4.03 (0.21)    & 4.48 (0.39)   \\
			$z$ & 2.52 (0.89) & 3.72 (0.30)   & 5.02 (0.14) \\
			$J$ & 2.32 (0.28) & 2.32 (0.28)    & 2.32 (0.28)   \\
			$H$ & 2.01 (0.26) & 2.01 (0.26)   & 2.01 (0.26) \\
			$K$ & 1.57 (0.21) & 1.57 (0.21)    & 1.57 (0.21)    \\
			
			\hline
			
			\hline
			
		\end{tabular}
	}
	\medskip
	
\end{table}

In Figure~\ref{pic1:lcs} we show the apparent light curves of the five events, which are similar to those of other ILRTs in the literature. Most of the objects have a rise time of about two weeks, but AT~2018aes has a longer rise of about 24.1 d. After maximum, the light curve evolution of the sample resembles that of SNe IIP/IIL. We divide the light curve evolution into three different phases: Phase {\sc i} ($\gamma$$_1$), Phase {\sc ii} ($\gamma$$_2$), and Phase {\sc iii} ($\gamma$$_3$), each with a different decline rate. The values obtained through linear fits are reported in Table \ref{lc_para}. In general, ILRTs decline quite rapidly in all bands during Phase {\sc i}, with the blue bands usually fading faster than the red bands. After Phase {\sc i}, a sort of plateau (see AT 2012jc and AT 2013la) or a linear decline (see AT 2013lb and AT 2018aes) is observed in ILRTs. When late-time observations are available (i.e. AT 2013la, SN~2008S, and NGC~300~OT), a slow evolution is observed, consistent with that expected from the $^{56}$Co decay. We discuss the implications of this in Sect. \ref{bolo_sect}.  \\

\subsection{Colour evolution}
\label{sec:color}

\begin{figure*}[htp]
	\flushleft
	\includegraphics[width=.93\textwidth]{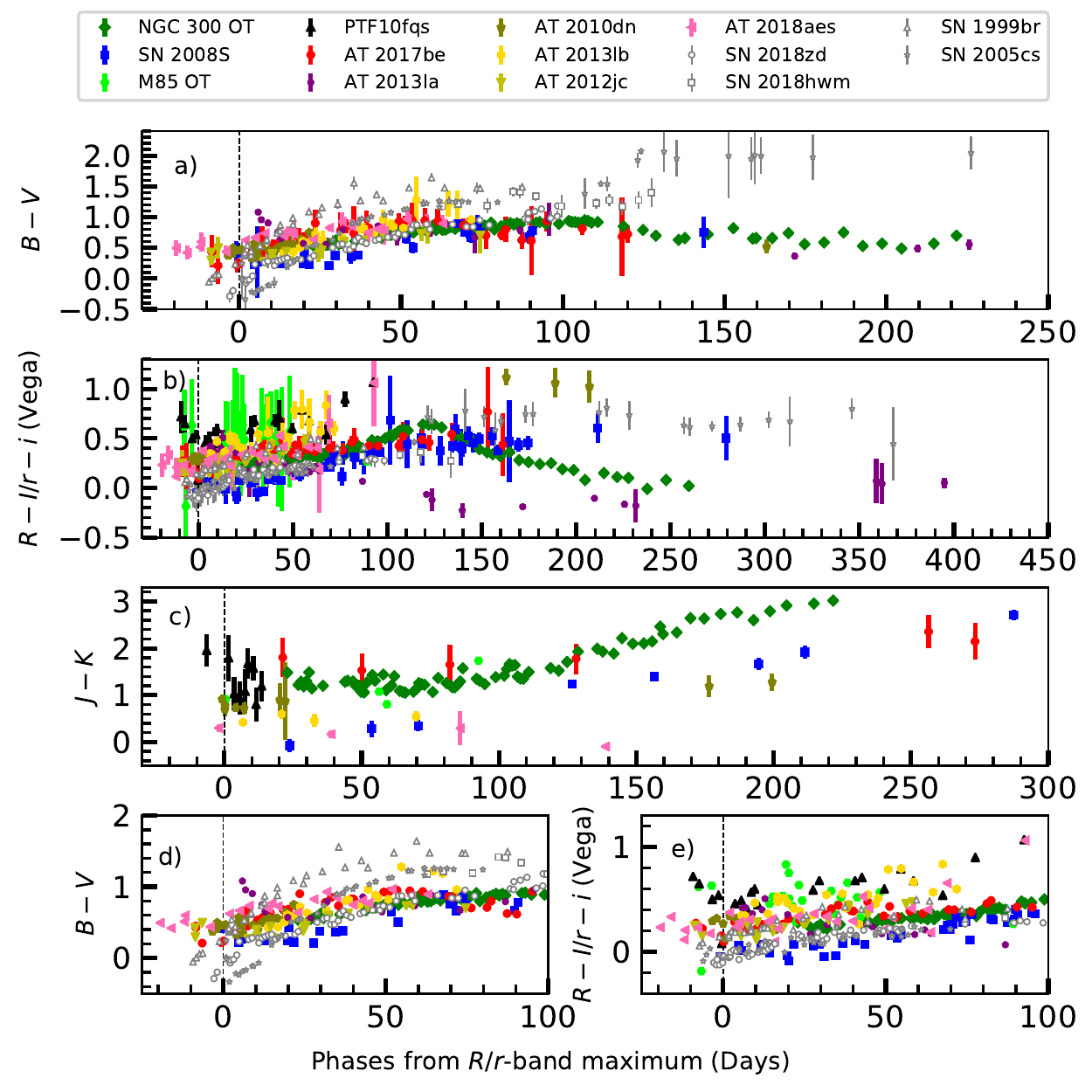}\quad
	\caption{Colour evolution of ILRTs, along with those of the comparison objects: SN
			2018zd \citep{Zhang2020MNRAS.498...84Z}, SN 2018hwm \citep{Reguitti2021MNRAS.501.1059R}, SN 1999br \citep{Pastorello2004MNRAS.347...74P}, and SN 2005cs \citep{Pastorello2006MNRAS.370.1752P, Pastorello2009MNRAS.394.2266P}.  (a) $B-V$ colour curves of ILRTs; (b) $R-I/r-i$ colour curves; (c) $J-K$ colour curves; (d) and (e) show the same colours as panels (a) and (b) until +100 days without error bars for clarity.  All phases are with respect to the $R/r$-band maximum.}
	\label{pic:colour}
\end{figure*}

The colour evolution of the five transients is shown in Figure~\ref{pic:colour}, along with those of other ILRTs and comparison objects from the literature. For the ILRTs, the $B-V$ colour evolves steadily from $\sim$ 0.2 - 0.4 mag at early phases to $\sim0.7$ - 1.0 mag at around 100 days past maximum, suggesting that the temperature decreases with time (see panel {\sl a} in Figure~\ref{pic:colour}). At the later phases ($>$100 days), $B-V$ becomes bluer again, from 0.8 to 0.5~mag. At similar epochs, the $R-I$/$r-i$ colours increase from $\sim$ 0.1 - 0.3 mag to 0.5 - 1.0 mag (Figure~\ref{pic:colour}, panel b). Very late-time colours show a large dispersion. AT 2013la is somewhat discrepant, with the $r-i$ colour becoming bluer (reaching $\sim-0.2$ mag) than at early epochs. The $J-K$ colour usually shows a minimum at 30-70 days. The best-sampled NIR dataset is that of NGC~300~OT, which allows us to estimate a minimum value of $\sim$ 1.1 mag. Later, $J-K$ rises to 3.0 mag at $\sim$ 230 days (Figure~\ref{pic:colour}, panel c).   
As shown in Figure~\ref{pic:colour}, the comparison objects are bluer (e.g. SN~2018zd and SN~2005cs: $B-V$ $\approx$ $-$0.2 mag) at early phase and become redder (e.g. SN~2005cs: $B-V$ $\approx$ 2.0 mag at tail phase) over time.  
 \\

\begin{figure*}[htp]
\centering
\includegraphics[width=.93\textwidth]{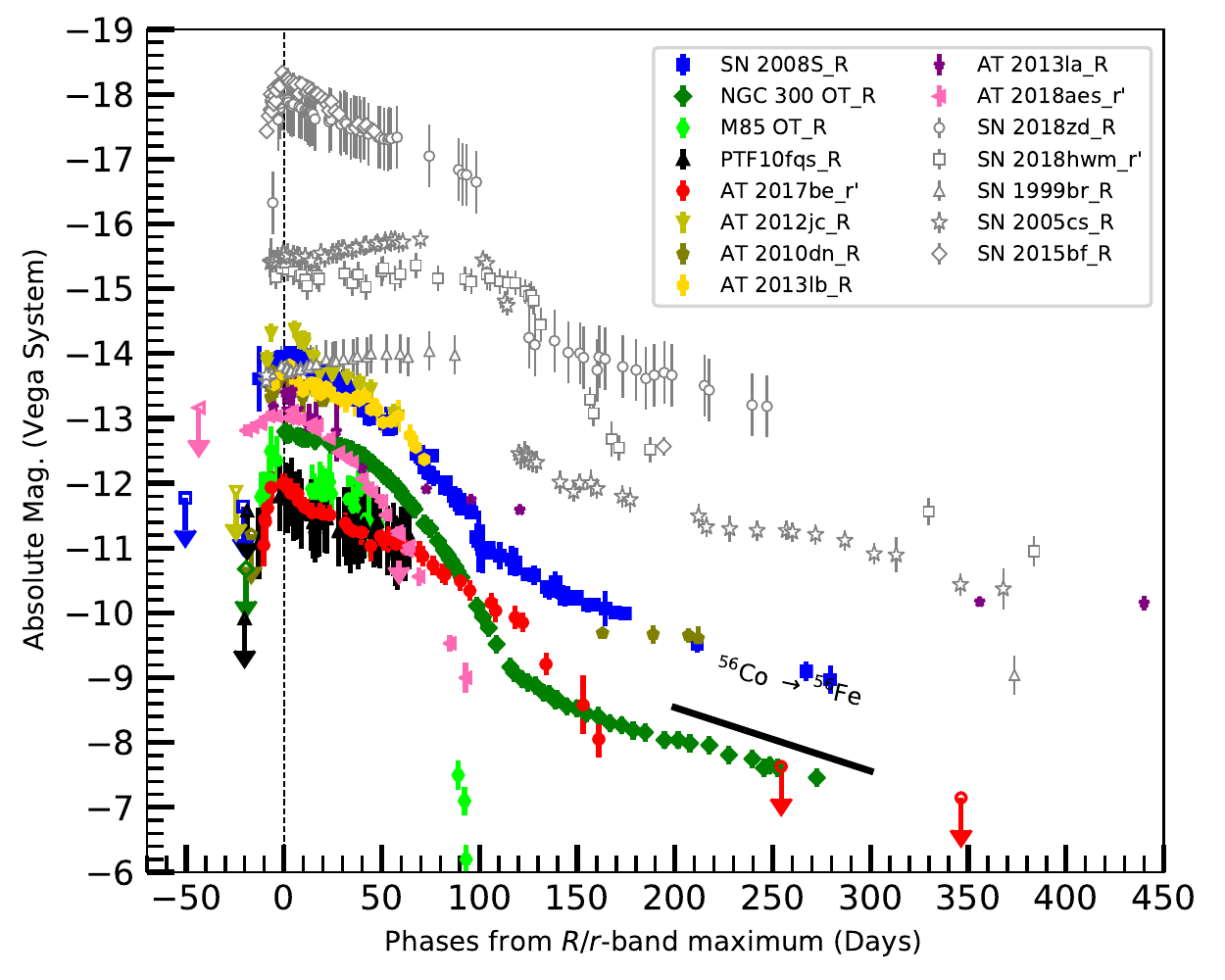}\quad
\caption{Comparison of the $R/r$-band absolute light curves of ILRTs and  SN 2018zd, SN 2018hwm, SN 1999br, and SN 2005cs.  Reddenings and distance moduli of the comparison objects are taken from their respective papers (also see Table~\ref{Dist_Reden_II} in Appendix~\ref{appendix:Dist_Reden}). The dashed vertical line indicates the time of $r$-band maximum light. Upper limits are marked by empty symbols with down arrows.}
\label{pic:absLC}
\end{figure*}


\subsection{Absolute light curves}
\label{sec:AbsMag}
The $R/r$-band absolute light curves of our ILRT sample, calibrated in the Vega system, are shown in Figure~\ref{pic:absLC}. We used third-order polynomial fits to estimate the $R/r$-band peak absolute magnitudes\footnote{These fits were performed on the absolute light curves about $-$20 to $+$20 days from maximum.}, and the resulting values are listed in Table~\ref{table:peak}. All transients have rise times to their maximum $\lesssim$ 2 weeks, apart from AT 2018aes, which reaches the light curve peak in $\sim$ 3 weeks. The peak absolute magnitudes span a range from $\sim -11~(\pm~0.5)$ ~to $-15~(\pm~0.5)$~mag,  still within the  range expected for gap transients \citep[$-10 < M < -15$ mag; ][]{Pastorello2019NatAs...3..676P}. If we consider all objects in this study, we obtain an averaged absolute peak magnitude of $M_{\rm R} = -13.04 \pm 0.91$ mag. As shown in Figure~\ref{pic:absLC}, we compared the $R/r$-band absolute light curves of ILRTs with three proposed EC SN candidates: SN~2018zd, SN~2018hwm, and SN 2015bf; and two subluminous Type IIP SNe: SN~1999br and SN~2005cs. They show a wide range of peak magnitudes, with the faint SN~1999br ($\sim$ $-$ 13.8 mag), the relatively luminous SN~2018zd and SN 2015bf  ($\sim$ $-$ 18.0 mag), and intermediate-luminosity SN~2018hwm and SN~2005cs ($\sim$ $-$ 15.5 mag). SN~1999br, SN~2018hwm, and SN~2005cs are characterised by long-lasting plateaus of above 100 days. At the end of the plateau, these comparison objects show a sudden drop in luminosity of 2.5-4.0 mag. Finally, they all settle onto the radioactive tails.

\begin{table*}
  \caption{Light-curve parameters for ILRTs.}
  \label{table:peak}
  
  {
  \scalebox{0.9}{
  \begin{tabular}{@{}lccccccccc@{}}
     \hline
     
     \hline
      Object &MJD$_{\rm{non.d.}}$  & MJD$_{\rm{exp.}}$   & MJD$_{\rm{first.d.}}$ &MJD$_{\rm{R/r, peak}}$ &$M_{\rm{R/r, peak}}$   &$L_{\rm{peak}}$   & $E_{\rm{rad.}}$  & $^{56}$Ni Mass\\
                 &                             &                       &                            &                                   &  ($\mathrm{mag}$)  &  ($\rm{10^{40}~erg~s^{-1}}$) & ($\rm{10^{47}~erg}$) & (\msun)\\
    
     \hline
     
     \hline
     AT 2010dn   & 55338.00 &   55343 (5) &  55348.04  &55354.9 (3.1) & $-$13.51 (0.27) & 3.65 (0.09) & - &$2.5 \times 10^{-3}$ - 2.7 $\times 10^{-3}$ &\\    
     AT 2012jc  & 56003.39 &   56008 (5) &  56013.49  &56024.5 (2.0)& $-$14.50 (0.13)  & 9.97 (1.16)& 2.94 (0.38) &- & \\
     AT 2013lb  &56112.03   &      --    &56319.34  &56324.5 (3.0) & $-$13.73 (0.11)   & 4.12 (0.52)& 1.61 (0.24) & - &\\
     AT 2013la  &56463.22   &    --      &56646.93  &56652.0 (2.1) & $-$12.97 (0.39)  & 1.95 (0.05) & 1.10 (0.04) &  $> 4.8 \times 10^{-3}$ & \\  
     AT 2018aes   &58184.61   & 58187 (2)  &58188.54  &58211.1 (7.6)& $-$12.88 (0.04)   & 2.16 (0.07) & 1.02 (0.07) &- &\\
     \hline 
     SN 2008S   &54481.50    & 54486 (4)    &54489.50    &54502.5 (2.0) & $-$14.24 (0.03) & 5.36 (0.22) &2.93 (0.47) & $3.2 \times 10^{-3}$ - 3.5$\times 10^{-3}$  &\\
     NGC 300 OT  &54504.00   &     --     &54580.15     &54600.0 (2.0) & $-$12.77 (0.06)  & 1.76 (0.53) & 0.69 (0.36)&$> 1.4 \times 10^{-3}$ &\\
     PTF10fqs   &55295.20   &  55299 (4)    &55302.39    &55315.3 (4.0) & $-$11.55 (0.35) & 0.59 (0.30) &0.30 (0.15) & -&\\
     AT 2017be  &57751.48   &   57755 (4)   &57759.51    &57767.8 (2.0) & $-$12.01 (0.14) & 0.93 (0.08) &0.43 (0.06) & $6.7 \times 10^{-4}$ - 7.2$\times 10^{-4}$  & \\
     M85 OT    &52672.00   &     --   &53742.00  &53753.6 (2.0)& $-$12.21 (0.90)  & 0.70 (0.13) & 0.30 (0.13) &$< 1.0 \times 10^{-3}$ &\\
     \hline
     
     \hline
  \end{tabular}

  }
  }
\medskip

$Notes:$ Object name (column 1), last non-detection MJD (column 2), explosion MJD (column 3), first detection MJD (column 4), maximum MJD (column 5), peak magnitude (column 6), peak bolometric ($B$ to $I$) luminosity (column 7), radiated energy (column 8), $^{56}$Ni mass (column 9;  We note that for some objects we do not report the $^{56}$Ni mass because of a lack of observed light curve tails). Uncertainties are reported in parentheses.\\
  \\
\end{table*}

\subsection{Pseudo-bolometric light curves} \label{bolo_sect}

\begin{figure*}[htp]
	\centering
	\includegraphics[width=.93\textwidth]{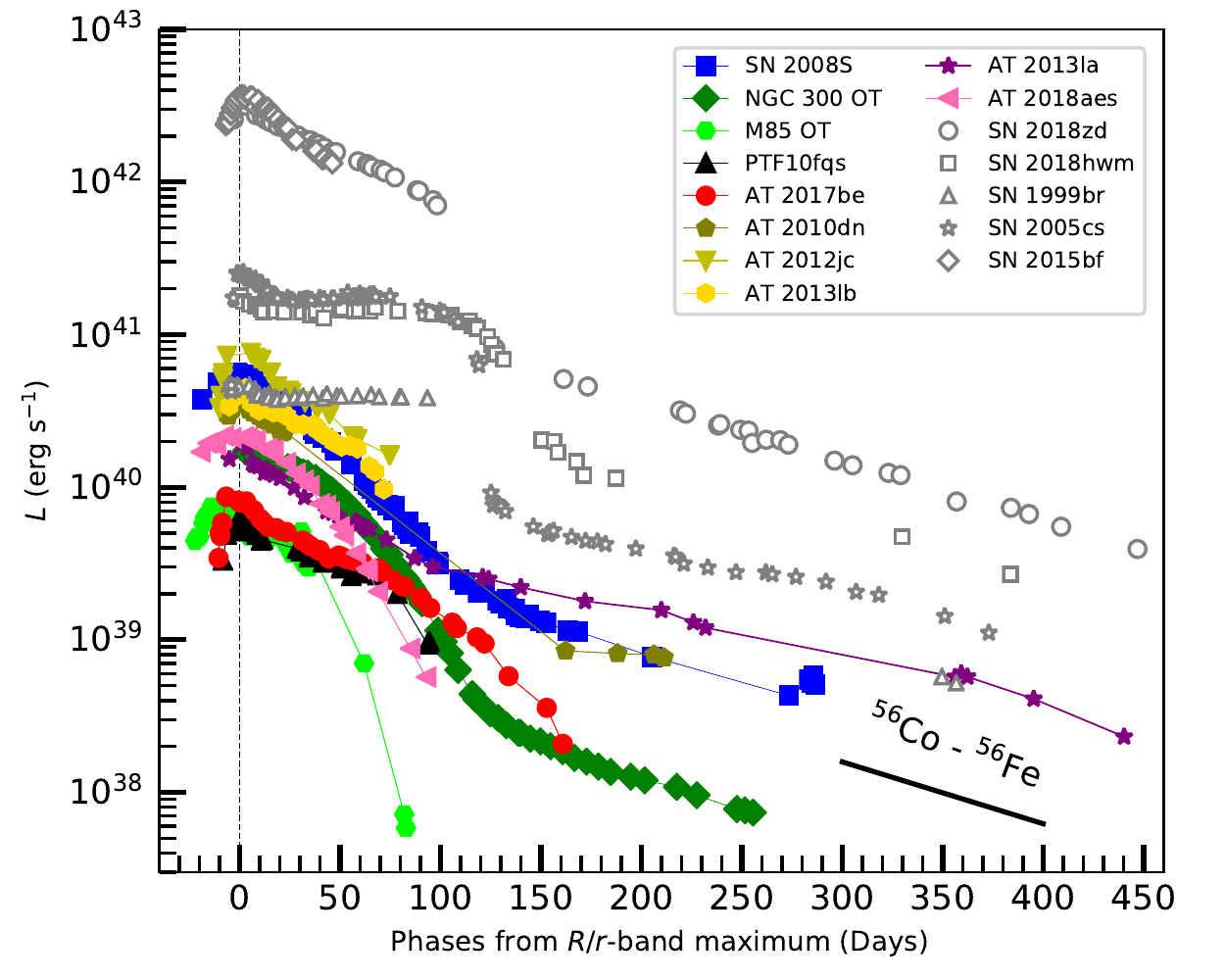}\quad
	\caption{Pseudo-bolometric $B$ to $I/i$ light curves of ILRTs and comparison objects. The dashed vertical line indicates the time of $R/r$-band maximum light.}
	\label{fig:PseudoBolom}
\end{figure*}

The bolometric light curve is computed by integrating the spectral energy distribution (SED) over the whole electromagnetic spectrum. However, in most cases,  observations at wavelengths shorter  than the $u$ band, and longer than $I/i$-band are not available. For this reason, to achieve meaningful comparisons among  ILRTs, we computed pseudo-bolometric light curves including only the contribution from the $B$ to the $I/i$ band. We first converted the extinction-corrected magnitudes to flux densities, and finally integrated the SEDs at their effective wavelengths, assuming a negligible flux contribution outside of the integration region. The resulting pseudo-bolometric light curves are presented in Figure~\ref{fig:PseudoBolom}, while peak luminosities are reported in Table~\ref{table:peak}. They show faint peak luminosities ranging from $0.5 \times 10^{40}$~erg~s$^{-1}$ to $9.0 \times10^{40}$~erg~s$^{-1}$. For comparison, the claimed EC SNe SN~2018zd and SN~2015bf have peak luminosities of $\sim$ $3.7 \times 10^{42}$~erg~s$^{-1}$ and the faint Type IIP SN~1999br is about $4.7 \times 10^{40}$~erg~s$^{-1}$ (see Figure~\ref{fig:PseudoBolom}). The pseudo-bolometric light curve shapes of ILRTs are broadly similar to Type IIP (in particular, PTF10fqs and AT 2017be) and/or Type IIL (SN~2008S and AT 2012jc) SNe, as demonstrated in Figure~\ref{fig:PseudoBolom}. In analogy to Type IIP SNe, the light curves for ILRTs are influenced by a number of factors, such as the presence of a CSM, the H recombination timescale, and $^{56}$Ni distribution. Finally, we perform a non-parametric fit using the {\it ReFANN}\footnote{ReFANN is a nonlinear interpolating tool based on Artificial Neural Networks that can reconstruct functions from data (see https://github.com/Guo-Jian-Wang/refann).} code \citep[][]{Wang2020ApJS..246...13W, Wang2021MNRAS.501.5714W} to reconstruct the pseudo-bolometric light curve, and integrate it over the covered photometric evolution. The resulting radiated energies are in the range of (0.30 - 2.94) $\times$~10$^{47}$~erg, with values reported in Table~\ref{table:peak}. These values should be regarded as lower limits due to our incomplete wavelength coverage and limited temporal coverage. These observed ILRTs radiated energies are fractions of about 10$^{-4}$ - 10$^{-3}$ of the theoretical predictions for the explosion energy of EC SNe \citep[about 10$^{50}$~erg, ][]{Kitaura2006A&A...450..345K, Wanajo2009ApJ...695..208W, Tominaga2013ApJ...771L..12T}. \citet{Stritzinger2020A&A...639A.103S} pointed out that theoretical EC SN simulations may overpredict the explosion energy. Therefore, this discrepancy should be considered in future observations and modelling efforts.

{In order to provide reliable estimates for the ejected $^{56}$Ni masses, we need to compute the bolometric light curves. Unfortunately, most objects do not have MIR observations, and optical and/or NIR light-curve information  is sometimes incomplete at late times. SN~2008S and NGC~300~OT are exceptions, as they were followed until late phases, and also have MIR coverage. Observations of these two objects revealed that the SEDs of ILRTs shift from the optical to the MIR domain, with the MIR being dominant at late phases \citep[e.g.][]{Botticella2009MNRAS.398.1041B, Kochanek2011ApJ...741...37K}. 
To mitigate the limited observational information available for other objects, we adopted SN~2008S as a template for other ILRTs, and assumed that all others share the same SED evolution. Hence, we used SN~2008S to determine the  bolometric corrections for all ILRTs that had incomplete wavelength coverage.
We estimated the optical and NIR luminosity contribution of SN 2008S, using the bolometric light curve model ($L_{\rm{bol}} = L_{0}\times \exp (-t/t_0)$ + $L_1$, with $L_0 \simeq 10^{7.3} L_\odot$, $L_1 \simeq 10^{5.8}L_\odot$ and $t_0\simeq 48$~days) from \citet{Kochanek2011ApJ...741...37K}. We first computed the $\frac{L_{\rm{opt}}}{L_{\rm{bol}}}$, $\frac{L_{\rm{NIR}}}{L_{\rm{bol}}}$, and $\frac{L_{\rm{opt}} + L_{\rm{NIR}}}{L_{\rm{bol}}}$ ratios at late phases for SN~2008S. The $L_{\rm{opt}}$, $L_{\rm{NIR}}$, and $L_{\rm{opt}} + L_{\rm{NIR}}$ contributions for other ILRTs at the same phases were then calculated, assuming the same ratios as for SN 2008S. Specifically, 
$L_{\rm{bol}}$ was computed starting from the measured $L_{\rm{opt}} + L_{\rm{NIR}}$ for AT~2010dn, from $L_{\rm{opt}}$  for AT 2013la, and $L_{\rm{NIR}}$ for AT~2017be, and then by applying the adopted bolometric corrections.

Assuming that the late-time evolution of ILRTs are powered by the radioactive decay chain $^{56}$Ni $\to$ $^{56}$Co $\to$ $^{56}$Fe, the late-time luminosity of ILRTs can be used to constrain the synthesised $^{56}$Ni mass. We use the well-observed Type II SN 1987A as a reference \citep{Catchpole1988MNRAS.231P..75C, Whitelock1988MNRAS.234P...5W, Catchpole1989MNRAS.237P..55C}~and estimate the ejected $^{56}$Ni mass of ILRTs using Equation~\ref{Nimass}:

\begin{equation}\label{Nimass}
M\left(^{56}\mathrm{Ni}\right)_\mathrm{ILRT}~$=$ M\left(^{56}\mathrm{Ni}\right)_\mathrm{SN~1987A} \times \left(L_\mathrm{ILRT}\left(t\right)\over L_\mathrm{SN~1987A}\left(t\right)\right),
\end{equation}

\noindent where $M(^{56}\mathrm{Ni})_\mathrm{SN~1987A}\sim0.073~\mathrm{M_{\odot}}$\footnote{This value is computed through a weighted mean of values reported in \citet[][]{Arnett1989ApJ...340..396A} and \citet[][]{Bouchet1991AJ....102.1135B}.} is the $^{56}$Ni mass synthesised by SN 1987A, and $L_\mathrm{ILRT}$ and $L_\mathrm{SN~1987A}$ are late-time luminosities of an individual ILRT and SN~1987A, respectively. Hereafter, we  only consider five ILRTs that have observations in the nebular phase. Because of the poor constraints on the explosion epoch of ILRTs, we  use the last non-detection 
and the first detection to fix the earliest and the latest possible extremes for the explosion epochs. Hence, we obtain  upper and lower limits of $^{56}$Ni masses for the ILRTs, reported in Table~\ref{table:peak}. 
SN 2008S ejected the largest amount of $^{56}$Ni  ($3.2$ - $3.5\times 10^{-3}$~\msun), while AT~2017be has the lowest $^{56}$Ni mass ($6.7$ - $7.2 \times 10^{-4}$~\msun). 
All inferred $^{56}$Ni masses are of the order of $10^{-4}$ to $10^{-3}$~\msun, which, as discussed in Sect. \ref{sec:discussion}, is in agreement with the predictions for EC SNe.  \\      
 
\subsection{Evolution of the SED of ILRTs: the test case of AT~2010dn} 
  
It is well known that SN~2008S, NGC~300~OT, and AT 2012jc already showed an IR excess soon after their discovery \citep{Botticella2009MNRAS.398.1041B, Humphreys2011ApJ...743..118H, Stritzinger2020A&A...639A.103S}. If we assume their progenitors to be embedded in complex and extended dusty environments, we should expect such IR excesses to be frequently observed for ILRTs.
This can be verified in the case of AT~2010dn, for which MIR observations are available.  Using the light curves presented in Table~\ref{2010dn_opt_LC}, \ref{2010dn_nir_LC}, and \ref{table:SpitzerMag}, we constructed the SEDs of AT~2010dn at different epochs using photometry from the optical to the MIR. The first optical to MIR SED is obtained at $t\sim$30.6 days after maximum, which clearly reveals an IR excess over a single black body (BB) model (see the top panel of Figure \ref{pic:sedFIT}). Hence, the SED was fitted with two-component (hot+warm components) BB functions instead of a single one. The hot component has a temperature $T_{\mathrm{hot}}$=5390 $\pm$ 70~K and a radius $R_{\mathrm{hot}} \approx 2.87 \times 10^{14}$ cm, while the warm component has  $T_{\mathrm{warm}}$ $\approx$ 970 K and  $R_{\mathrm{warm}} \approx 4.63 \times 10^{15}$ cm. 
In comparison, SN 2008S had $T_{\mathrm{hot}}$= 8076 $\pm$ 150 K, $R_{\mathrm{hot}} = (2.1 \pm 0.1) \times 10^{14}$ cm, $T_{\mathrm{warm}}$ $\approx$ 585 K, $R_{\mathrm{warm}} \approx  9.9 \times 10^{15}$ cm at maximum \citep{Botticella2009MNRAS.398.1041B}, while~AT 2012jc had a BB temperature of $T_{\mathrm{hot}}$ $\approx$ 6430 K and $T_{\mathrm{warm}}$ $\approx$  800 K at + 26.8 days \citep{Stritzinger2020A&A...639A.103S}. The hot-component estimates are consistent with emission from the photosphere, while the warm component peaking in the MIR domain is likely due to circumstellar dust.  For AT~2010dn, additional epochs of SEDs along with the corresponding best-fit BB functions are shown in Figure \ref{figSED2010dn} (Appendix \ref{sec:spectphot}), while the first BB parameters are reported in Table~\ref{AT2010dnSed}. The $T_{\mathrm{hot}}$ increases to $\sim$ 6490 K and, when the luminosity fades, $T_{\mathrm{hot}}$ also declines to $\sim$ 3540 K at $+$199.2 d. The $R_{\mathrm{hot}}$~shows a slow evolution (from 2.4 to 2.9 $\times 10^{14}$ cm) until $+$21.5 d, and declines to 1.6 $\times 10^{14}$ cm at  $+$199.2 d. Three epochs with K band data reveal a possible second warm component with an almost constant temperature of 1100-1200 K (see Figure \ref{figSED2010dn}).   

We also measured the residual flux at the location of AT~2010dn in {\it Spitzer} MIR images taken more than 3 years (+1333 days) after the outburst, and constructed a very late-time SED. This SED can be reproduced by a single BB function with a characteristic temperature of 440~K (see the bottom panel of Figure \ref{pic:sedFIT}).  We caution that this temperature should be regarded as an indicative value because of the limited coverage of MIR data.  Similarly, AT 2012jc has very late-phase (+1155 days) {\it Spitzer} MIR photometric fluxes suggesting a BB temperature of about $600\,\rm{K}$. Based on the similarity of these three cases, we suggest that this SED evolution seems to be typical of ILRTs. A plausible explanation is that a dusty environment has formed through mass-loss events prior to the ILRT outbursts, in analogy to what is frequently observed in a few CC SNe \citep[e.g. SN 1995N, SN 1998S, SN 2010jl, SN 2014ab, SN 2015da; see ][]{Gerardy2002ApJ...575.1007G, Andrews2011AJ....142...45A, Fransson2014ApJ...797..118F, Tartaglia2020A&A...635A..39T, Moriya2020A&A...641A.148M}. The slowly declining IR flux lasting around 4 years suggests that the outer dust shell was not destroyed by the radiation field emitted by the luminous outburst \citep[regardless of the physical mechanism that produced the ILRTs; ][]{Botticella2009MNRAS.398.1041B}.

\begin{figure}[htp]
	\includegraphics[width=.48\textwidth]{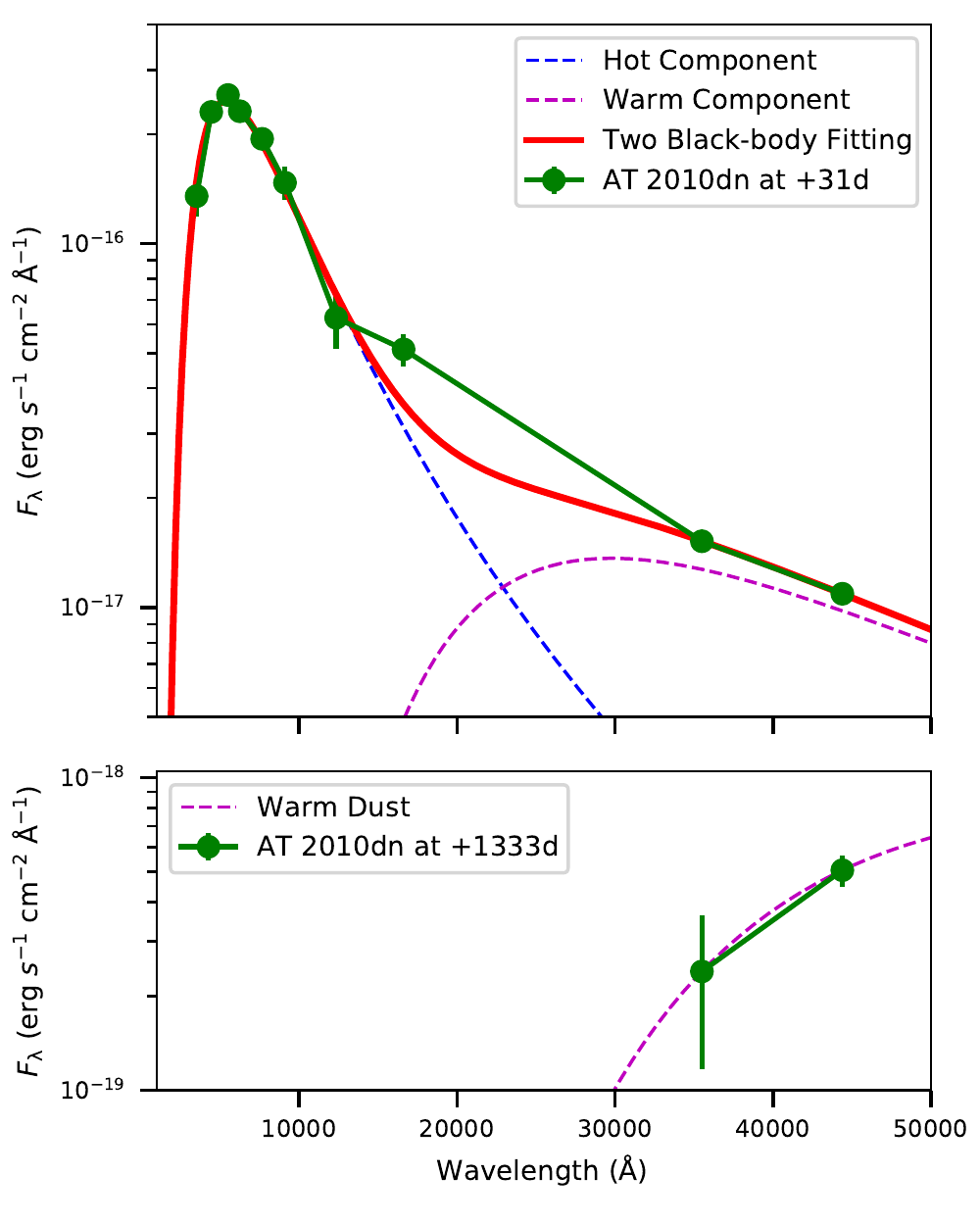}\quad
	\caption{Top panel: Two-component black body fits to the SED of AT~2010dn obtained from optical to MIR at the epoch of +30.6 days. Bottom panel: Single black body fit for AT~2010dn in the MIR domain at very late phase of +1332.5 days. Although this fit could reproduce the late MIR data, the fitted temperature should be considered as an indicative value. The MIR measurements were obtained from the Spitzer Space Telescope + IRAC images at 3.6 $\mu$m and 4.5 $\mu$m.   }
	\label{pic:sedFIT}
\end{figure}

\begin{table*}
\caption{Log of spectroscopic observations of ILRTs.}
\label{speclog}
\begin{tabular}{@{}cccccccc@{}}
\hline

\hline
Date & MJD  & Phase$^a$  & Telescope+Instrument & Grism  & Spectral range & Resolution & Exp. time   \\ 
       &          & (d) &                               &                            & ($\AA$)                  & ($\AA$)           & (s)                \\ 
\hline

\hline
\multicolumn{8}{c}{AT 2010dn}\\  \hline  
20100602 & 55349.9 & -5.0 &    TNG+LRS                      & LR-R                         & 5090 - 8340  &  10.5  & 2100 \\
20100603 & 55350.9 & -4.0 &    TNG+LRS                      & LR-B                         & 3300 - 7990  &  13.5  &  3000 \\
20100604 & 55351.9 & -3.0 &    TNG+LRS                     & VHR-R                       & 6170 - 7780  & 3.2   &  2700\\
20100605 & 55352.9 & -2.0 &    TNG+LRS                     & VHR-I                         & 7290 - 8870  & 3.5   &  2700\\
20100607 & 55354.9 & +0.0 &    WHT+ISIS                    & R300B/R158R            & 3160 - 10300  &  4.1/6.8  &  1800/1800 \\
20100609 & 55356.9 & +2.0 &    NOT+ALFOSC               & gm4                           & 3630 -  8960 &  18  &  3600\\
20100611 & 55358.9 & +4.0 &    TNG+LRS                      & LR-B                          & 3520 -  7990 &   11  &  2700 \\
20100616 & 55363.9 & +9.0 &    TNG+LRS                      & VHR-R                       & 6190 - 7820  &  3.2  &  3000\\
20100617 & 55364.9 & +10.0 &    TNG+LRS                      & VHR-R                       & 6190 - 7820  & 3.2   &  3000\\
20100707 & 55384.9 & +30.0 &    WHT+ISIS                      & R300B/R316R           & 3160 - 8860  &  4.1/3.1  &  1200/1200 \\
\hline
\hline
\multicolumn{8}{c}{AT 2012jc}\\  \hline
20120329  & 56015.1 & $-8.4$ &  Ekar1.82 + AFOSC   & gm4   &  3500 - 8060  &   24   & 3600\\
20120331  & 56017.3 & $-6.2$ &  Ir$\rm{\acute{e}}$n$\rm{\acute{e}}$e du Pont + WFCCD   & blue (400/mm)   & 3600 -  9120  &  5.3  &1500\\
20120414  & 56031.2 & $+7.7$ &  NTT + EFOSC2   &   Gr$\#$11/Gr$\#$16  &  3340 - 9920   &  21.3/21  &1800\\
20120501  & 56048.3 & $+24.8$ &  NTT + EFOSC2   &   Gr$\#$11/Gr$\#$16  & 3340 - 9920 &   14/13   &1800\\
20120623  & 56101.2 & $+77.7$ &  Magellan II + LDSS & VPH-All               &  3940 - 10300  &  6.5   &900\\
\hline 
\hline
\multicolumn{8}{c}{AT 2013lb}\\  \hline 
20130207  & 56330.1   & $+5.6$ &   Ekar1.82 + AFOSC  & gm4         & 3900 - 8200   &   14     & 2700\\
20130207  & 56330.4   & $+5.9$ &   NTT + EFOSC2       & Gr$\#$13  & 3650 - 9090  &   17     &1213\\
20130207  & 56330.4   & $+5.9$ & Gemini-S + GMOS-S & R400        & 4210 - 8440  &   10    & 900\\
20130209  & 56332.3   & $+7.8$ &   NTT + EFOSC2       & Gr$\#$11   & 3340 - 7460  &   22    &1394\\
20130220  & 56343.3   & $+18.8$ &   NTT + EFOSC2       & Gr$\#$11  & 3340 - 7460  &   22    &2700\\
20130305  & 56356.3   & $+31.8$ &   NTT + EFOSC2       & Gr$\#$11/Gr$\#$16  & 3340 - 9990 &14/13   &2700\\
20130422  & 56404.3 & $+79.8$  &   Gemini-S + GMOS-S  & R400 & 4210 - 8440 &    10    &450\\
\hline 
\hline
\multicolumn{8}{c}{AT 2013la}\\  \hline
20140108  & 56665.2 & $+13.2$ &   Ekar1.82 + AFOSC  & gm4       &  3400 - 8200 & 14   &1800 \\
20140115  & 56672.2 & $+20.2$ &   Pennar1.22 + B\&C  &300tr       &  3350 - 7930 & 10    &1800\\
20140203  & 56692.2 & $+40.2$ &  TNG+LRS               & LR-B        &  3430 - 8050 & 10.5  &3600 \\
20140205  & 56694.2 & $+42.2$ &   GTC + OSIRIS       & R500R     &  5150 - 9990 & 15.5  &2700\\
20140228  & 56717.2 & $+65.2$ &   GTC + OSIRIS       & R1000B   & 3650 - 7850  &  7      &900\\
20140322  & 56739.1 & $+87.1$ &  TNG+LRS               & LR-B        &  3430 - 8050 & 10.5  &2700\\
20140716  & 56854.9 & $+202.9$ &   GTC + OSIRIS       & R1000B   &  3650 - 7850 &  7     &1500\\
20140814  & 56883.9 & $+231.9$ &   GTC + OSIRIS       & R1000B   &  3650 - 7850  & 7     &1900 \\
20141220  & 57012.3 & $+360.3$ &   GTC + OSIRIS       & R1000B   &  3650 - 7850  &  7    &1800 \\
\hline
\hline
\multicolumn{8}{c}{AT 2018aes}\\  \hline
20180331  & 58208.1 & $-3.0$ &  GTC + OSIRIS & R1000B/R  &  3630 - 10080  &   7.5   &600/600\\
20180426  & 58234.1 & $+23.0$ &  GTC + OSIRIS & R1000B/R  &  3630 -10340   &   7.5   &600/600\\
\hline

\hline
\end{tabular}

\medskip
$^a$ Phases are relative to the $R/r$-band peak.\\ 
\end{table*}

 \section{Spectroscopy}
\label{sec:spectra}
Our spectral sequences for AT~2010dn, AT 2012jc, AT 2013lb, AT 2013la and AT~2018aes were obtained using multiple instrumental configurations, which are listed in Appendix~\ref{sec:facilities} (Table~\ref{table_setup}). Basic parameters for the spectra are reported in Table~\ref{speclog}.  

The spectra were processed with the standard procedures in {\sc iraf}. The preliminary reduction steps included bias, overscan and flat-field corrections. We then extracted 1D spectra from the 2D frames using the task {\sc apall}.  Wavelength calibration of the 1D spectra was performed with arc lamp spectra obtained with the same instrumental configuration as the science ones. The spectra of spectro-photometric standard stars were used to estimate a sensitivity function curve necessary to flux-calibrate the spectra of the transients. The accuracy of the flux calibration was then checked against coeval broadband photometry, and correction factors were applied in case of discrepancy. The spectra of the standard stars were also used to correct for the telluric absorption bands (e.g. O$_2$ and H$_2$O) from the ILRTs spectra. The resulting ILRT spectra are shown in Figure \ref{pic1:SpectraAll}.  \\

 \begin{figure*}[htp]
 	\centering
 	\includegraphics[width=.44\textwidth]{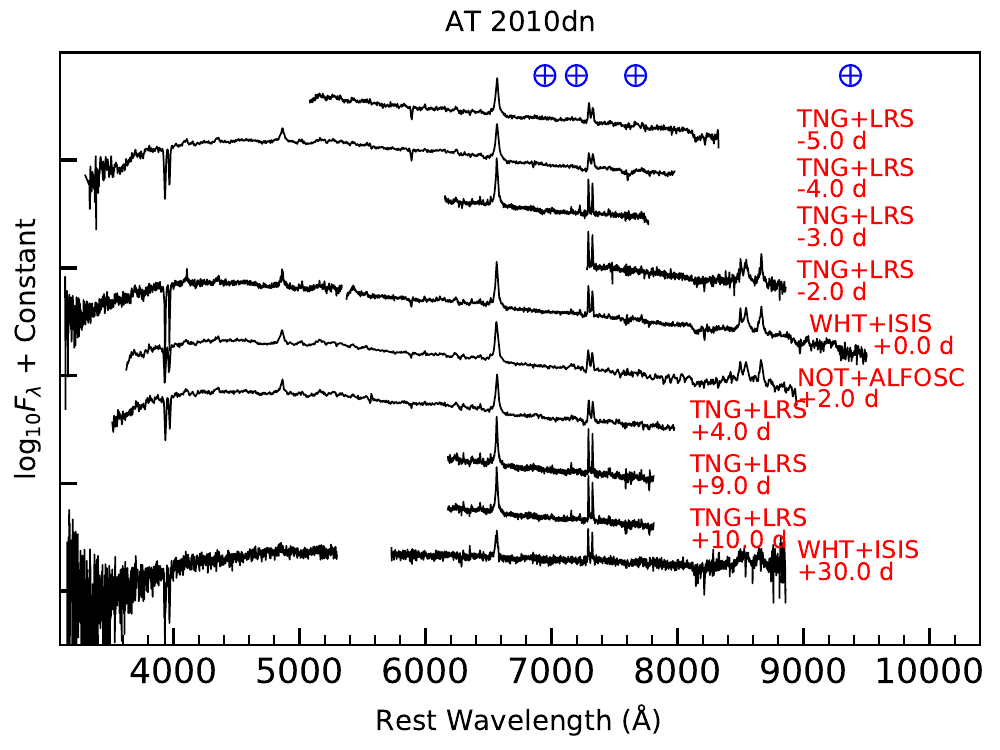}\quad
 	\includegraphics[width=.44\textwidth]{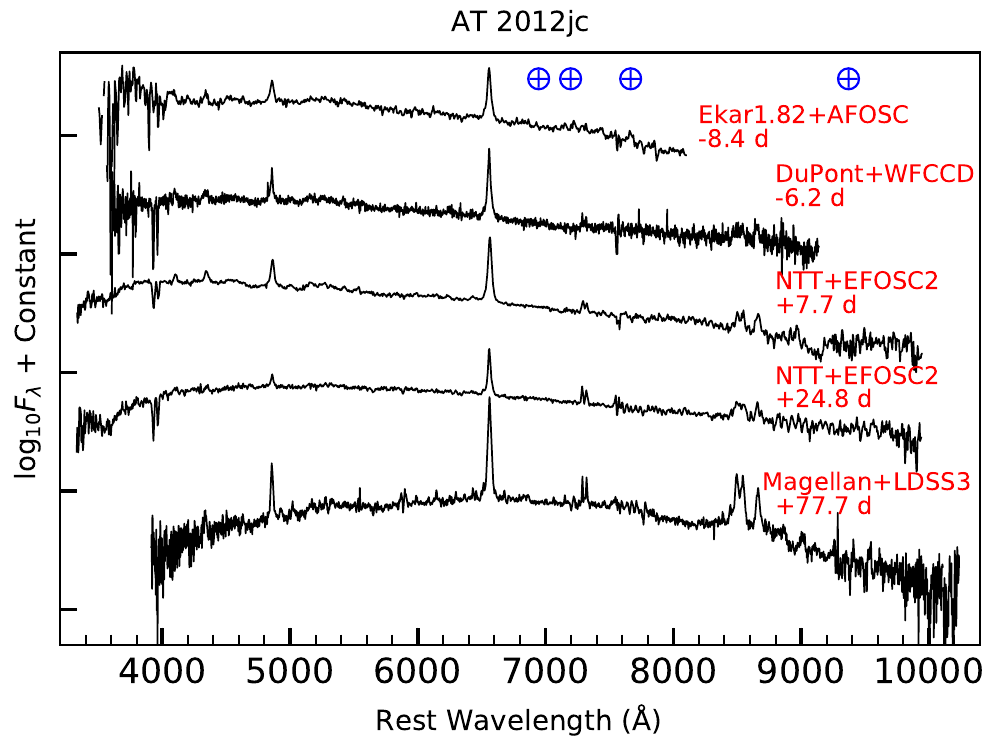}\quad
 	\medskip	
 	\includegraphics[width=.44\textwidth]{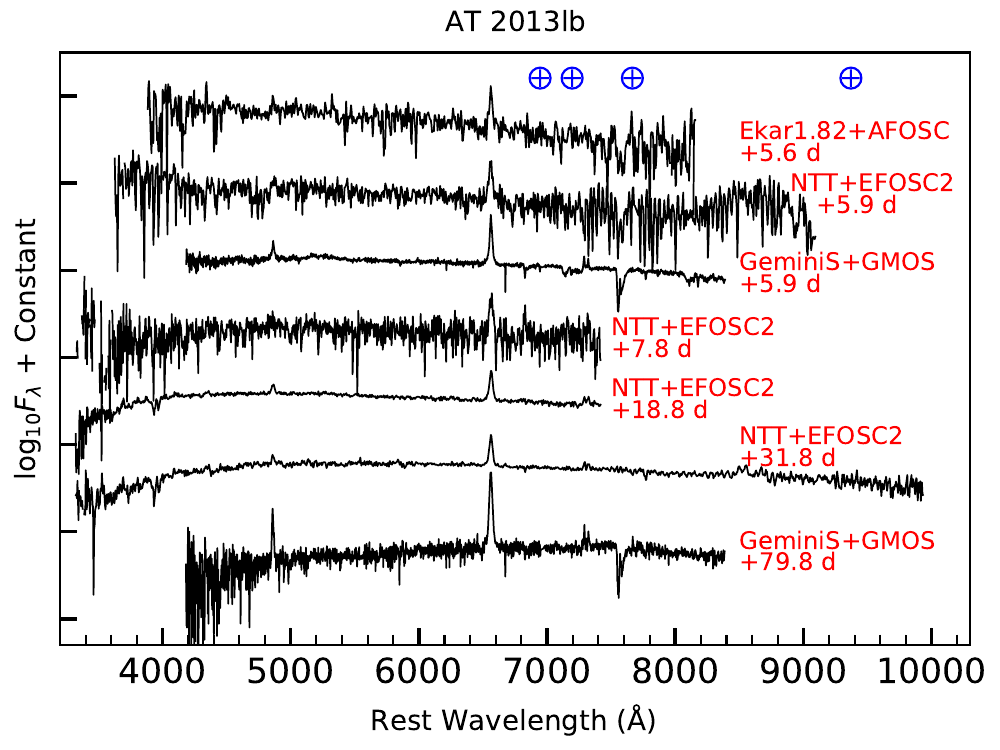}\quad
 	\includegraphics[width=.44\textwidth]{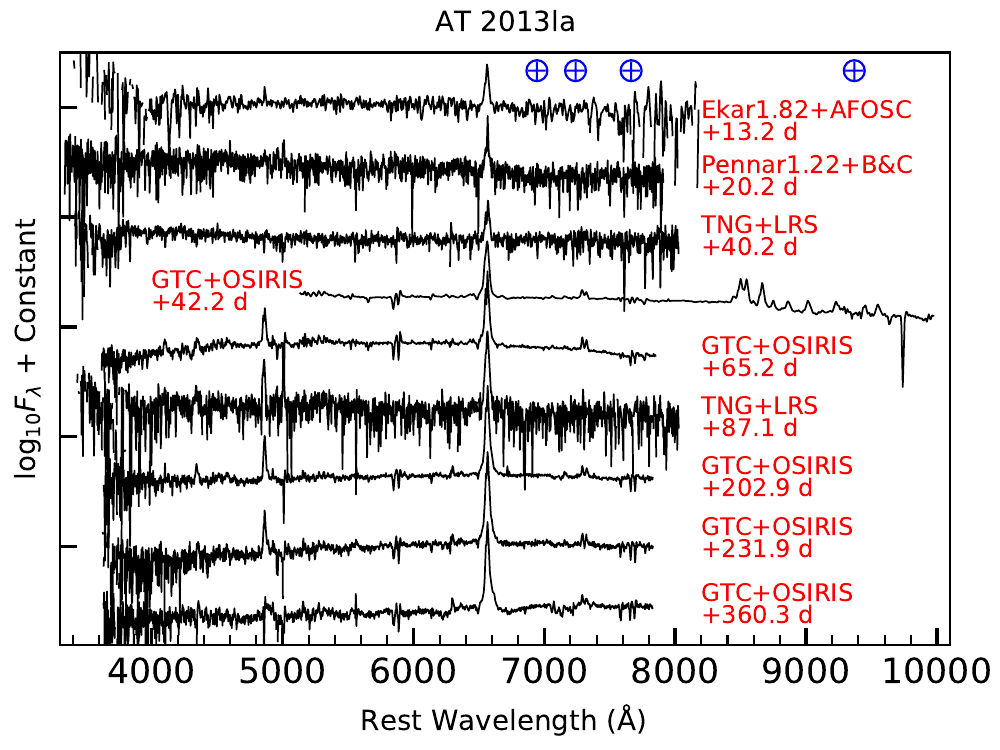}\quad
 	\medskip	
 	\includegraphics[width=.44\textwidth]{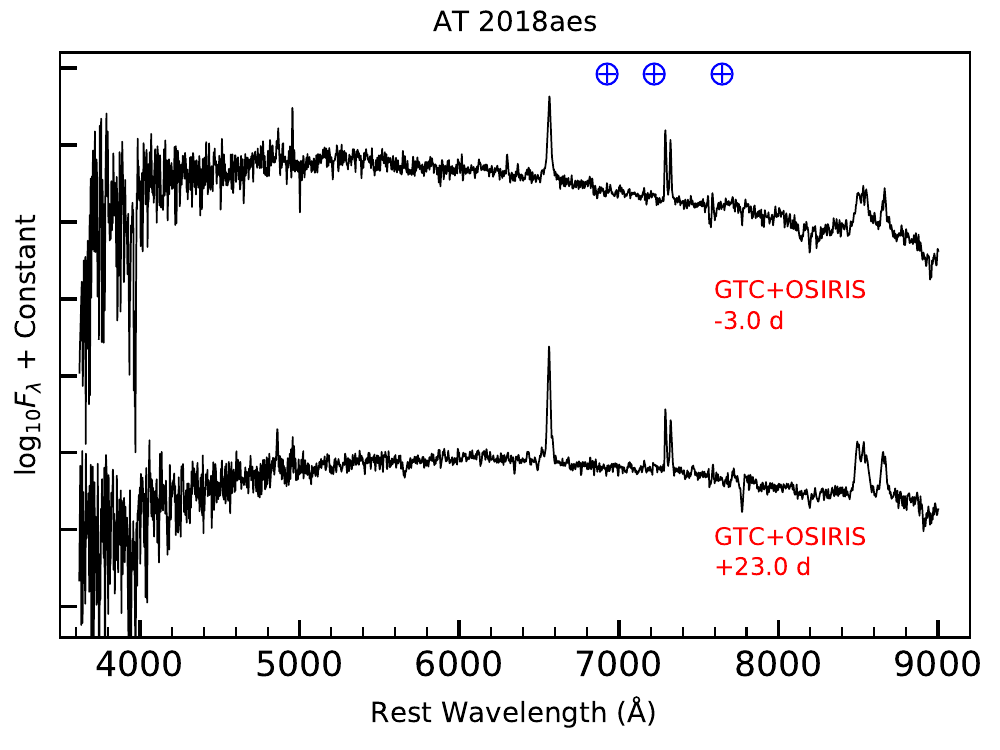}\quad
 	\caption{Spectral evolution of AT 2010dn (top-left), AT 2012jc (top-right), AT 2013lb (middle-left), AT 2013la (middle-right), and AT 2018aes (bottom).  The epochs marked to the right of each spectrum are relative to the $R/r$-band maximum. The $\bigoplus$ symbols mark the position of the strongest telluric absorption bands. The spectra are only corrected for redshift.}
 	\label{pic1:SpectraAll}
 \end{figure*}

\begin{table}
\caption{Physical parameters inferred from the optical spectra of the ILRTs sample.}
\label{spectra_para}
\scalebox{0.7}{
\begin{tabular}{@{}ccccccc@{}}
\hline

\hline
Date & MJD & Phase$^a$  &   $ v$ (H$\alpha$) $^b$ &    $ L$ (H$\alpha$) $^c$    & $L_{\rm{H_{\alpha}}}$/$L_{\rm{[\ion{Ca}~II]}}$ $^c$  &$T_{\mathrm{BB}}$    \\ 
 &  & (d) & (km~s$^{-1} $)  & (10$^{38}$erg~s$^{-1}$)& &  (K) \\ 
\hline

\hline
\multicolumn{6}{c}{AT 2010dn}\\ \hline
20100602  & 55349.9 & $-5.0$  &   695    &    2.12   & 2.9 &7095 $\pm$ 355     \\
20100603  & 55350.9  & $-4.0$ &   671    &   1.85    & 2.6 &6890 $\pm$ 240   \\
20100604  & 55351.9 & $-3.0$  &   588    &    3.47   & 2.5 &6540 $\pm$ 330    \\
20100605 & 55352.9 & $-2.0$   &   --       &    --   & --  &6435 $\pm$ 320    \\
20100607 &  55354.9 & $+0.0$ &  577     &   2.29    &  2.6 &6760 $\pm$ 340      \\
20100609 & 55356.9 & $+2.0$  & $<$823   &  2.63   &    2.5 &6830 $\pm$ 340    \\
20100611 & 55358.9 & $+4.0$  &  563    &   1.86      & 2.3 &6690 $\pm$ 195     \\
20100616 & 55363.9& $+9.0$   &  442    &   2.12   & 2.2 &6820 $\pm$ 320     \\
20100617 & 55364.9 & $+10.0$&  438    &   1.81  & 2.1 &6710 $\pm$ 280     \\
20100707 & 55384.9 & $+30.0$&  $<$142    &   0.58  & 1.8 &5020 $\pm$ 250    \\

\hline \hline

\multicolumn{6}{c}{AT 2012jc}\\ \hline
20120329  & 56015.1 &  $-8.4$  &  $<$1097   &  5.30   & --  &8380 $\pm$ 420    \\
20120331  & 56017.3 & $-6.2$   &  614       &  5.30   & 15.4   &7385 $\pm$ 740    \\
20120414  & 56031.2 & $+7.7$  &  $<$960     &  5.27   & 10.2    &7995 $\pm$ 205   \\
20120501  & 56048.3 & $+24.8$ & $<$617        &  2.97    & 5.5 &6780 $\pm$ 340    \\
20120623  & 56101.2 & $+77.7$ &  842          &  2.36    & --  &5445 $\pm$ 525   \\

\hline \hline
\multicolumn{6}{c}{AT 2013lb}\\ \hline
20130207  & 56330.1   & $+5.6$   & $<$640        &     3.33  &  --  &6865 $\pm$ 685    \\
20130207  & 56330.4  & $+5.9$    & $<$777        &    3.36    & 13.4 &6450 $\pm$ 645   \\
20130207  & 56330.4  & $+5.9$    & $<$460       &    3.29 & --   & 6250 $\pm$ 315    \\
20130209  & 56332.3   & $+7.8$   &  $<$1006     &    --        & -- & 5850 $\pm$ 585    \\
20130219  & 56343.3  & $+18.8$    &  $<$1006     &    2.29     &8.2&  6390 $\pm$ 640    \\
20130304  &  56356.3  & $+31.8$ & $<$617        &    1.87     &4.8 & 6450 $\pm$ 325    \\
20130422  & 56404.3  & $+79.8$  & 680   &    2.33   & -- &  4375 $\pm$ 220    \\

\hline \hline
\multicolumn{6}{c}{AT 2013la $^d$}\\ \hline
20140108  & 56665.2 & $+13.2$  &  $<$640   &   6.31   & --  &5880 $\pm$ 295    \\
20140115  & 56672.2 & $+20.2$  &  $<$457   &   5.92    & -- &7110 $\pm$ 1420    \\
20140203  & 56692.2 & $+40.2$  &    778      &     --   &  --   &6050 $\pm$ 1210    \\
20140205  & 56694.2 & $+42.2$  &  $<$709   &    5.33    & -- &5345 $\pm$ 270    \\
20140228  & 56717.2 & $+65.2$  &   400        &    4.99  & --  &5290 $\pm$ 265    \\
20140322  & 56739.1 & $+87.1$  &  $<$480   &    5.04    &-- &6005 $\pm$ 1200    \\
20140716  & 56854.9 & $+202.9$ &   446       &    3.98   & -- &5055 $\pm$ 255    \\
20140814  & 56883.9 & $+231.9$ &   503      &    3.87    & -- &4285 $\pm$ 225    \\
20141220  & 57012.3 & $+360.3$ &   499       &    2.68   & -- &4485 $\pm$ 180    \\

\hline \hline
\multicolumn{6}{c}{AT 2018aes}\\ \hline
20180331  & 58208.1 & $-3.0$    &  708  &     0.85    & 2.7 &6620 $\pm$ 330   \\
20180426  & 58234.1 & $+23.0$ & 609   &     0.77     &  2.2  &5675 $\pm$ 285    \\

\hline

\hline

\end{tabular}
}
\medskip

$^a$ Phases are relative to the $R/r$-band light curve peaks.\\ 
$b$ Most of our measurements are limited by the instrumental resolution constraints, hence the FWHM velocities may have large error bars, and in many cases they should be regarded as upper limits. \\
$c$ We adopt conservative errors of $\sim$20\% for the H$\alpha$ luminosity and the $L_{\rm{H_{\alpha}}}$/$L_{\rm{[\ion{Ca}~II]}}$ ratio because of the uncertainty in the flux calibration of the spectra. \\
$d$ H$\alpha$ has a complex profile in AT 2013la, and was therefore fitted with multiple components. The narrowest Gaussian emission and absorption components are usually below the instrumental resolution limit. For this reason, only the FWHM velocities of the broader Lorentzian components are reported here.    
\end{table}

\subsection{Spectroscopic evolution and line identification}
The five objects of our sample have spectra showing little evolution over the period of their spectral monitoring (from $\sim-8$~days to 1 year  after maximum). At all epochs, the spectra have an almost featureless continuum, which is  relatively blue at early phases, becoming redder with time. Narrow emission lines of the Balmer series are superposed on the continuum, with H$\alpha$ and H$\beta$ being the most prominent spectral features. Weak Fe {\sc ii} lines are also observed, along with Na I D and Ca H\&K in absorption. Apart from the H lines, the most prominent spectral emissions are the [Ca {\sc ii}] doublet ($\lambda\lambda$ 7291, 7324) and the Ca NIR triplet ($\lambda\lambda$ 8498, 8542, 8662). In addition, O {\sc i}, Fe {\sc ii}, Sc {\sc ii} and Ba {\sc ii} lines are also tentatively identified in some spectra with decent S/N and resolution (e.g. see a GTC/OSIRIS spectrum at + 41.9 d of AT 2013la). Detailed line identification performed on the best-quality spectra of the five transients is given in Figure~\ref{pic:spcttra_identification}. We note that the lines of the [Ca {\sc ii}] doublet are visible in all spectra with good S/N shown in this paper and are therefore considered a characteristic feature of ILRTs. In Figure \ref{pic:spcttra_comparison}, all ILRTs share almost the same spectral features, supporting a remarkable overall homogeneity in their observables. However, as shown in Figure \ref{pic:spcttra_comparison}, the spectra of the comparison objects show much bluer continuum than ILRTs. In addition, H$\alpha$ has a classical broad P-Cygni profile and there is no obvious [Ca {\sc ii}] doublet feature.

During the follow-up campaign of AT 2013la, we collected spectra covering all the evolutionary phases. Hence, we take AT 2013la as a representative object to describe the spectral evolution of ILRTs. The spectral continuum evolves quite slowly, from initially blue (e.g. at phase $\sim$ +13.2 d) to much redder at late phases (+360.0 d). AT 2013la resembles other ILRT spectra in its continuum evolution and characteristic lines. However, we see additional strong He I (5876 $\AA$) absorption feature in all GTC/OSIRIS spectra, most clearly  at +42.2, +65.2, +202.9, +231.9, and +360.3 days (see~middle right panel in Figure~\ref{pic1:SpectraAll}). We note that narrow P-Cygni absorptions imposed on H$\alpha$ lines are also detected in the highest resolution spectra (see e.g. the GTC/OSIRIS spectra of AT 2013la in Figures~\ref{pic1:SpectraAll}~and~\ref {pic:spcttra_phases}). Such spectra allow us to infer the wind velocity via measurement of the position of the deep minimum of the P-Cygni profiles. This blueshifted absorption component on top of the H$\alpha$ emission (see Figure~\ref{pic:spcttra_phases}) has velocities of $\sim$ 360 - 410 \kms,  and may originate from a dense, slow-moving wind. This is discussed more widely in Sect. \ref{profiles}.

We assume that the SEDs of ILRTs in the optical domain covered by our spectra are approximated by black bodies. Therefore, we estimate the temperature through a black body fit to the spectral continuum. The inferred continuum temperatures are reported in Table~\ref{spectra_para}, while the temperature evolution is shown in the top panel of Figure~\ref{pic:spectraparameter}. For all objects in our sample, the temperature rapidly declines from $\sim$ 7000 - 8500 K near maximum, to nearly 5000 K at $\sim$ 100 days. At late phases, the temperature decreases more slowly to $\sim$ 4200 - 4500 K at about 1 year after maximum. Overall, all our ILRTs show a similar temperature evolution. We note that the comparison objects show much higher temperatures than ILRTs at early phases, and this is consistent with their early-time bluer colours.

\begin{figure*}[htp]
\centering
\includegraphics[width=.9\textwidth]{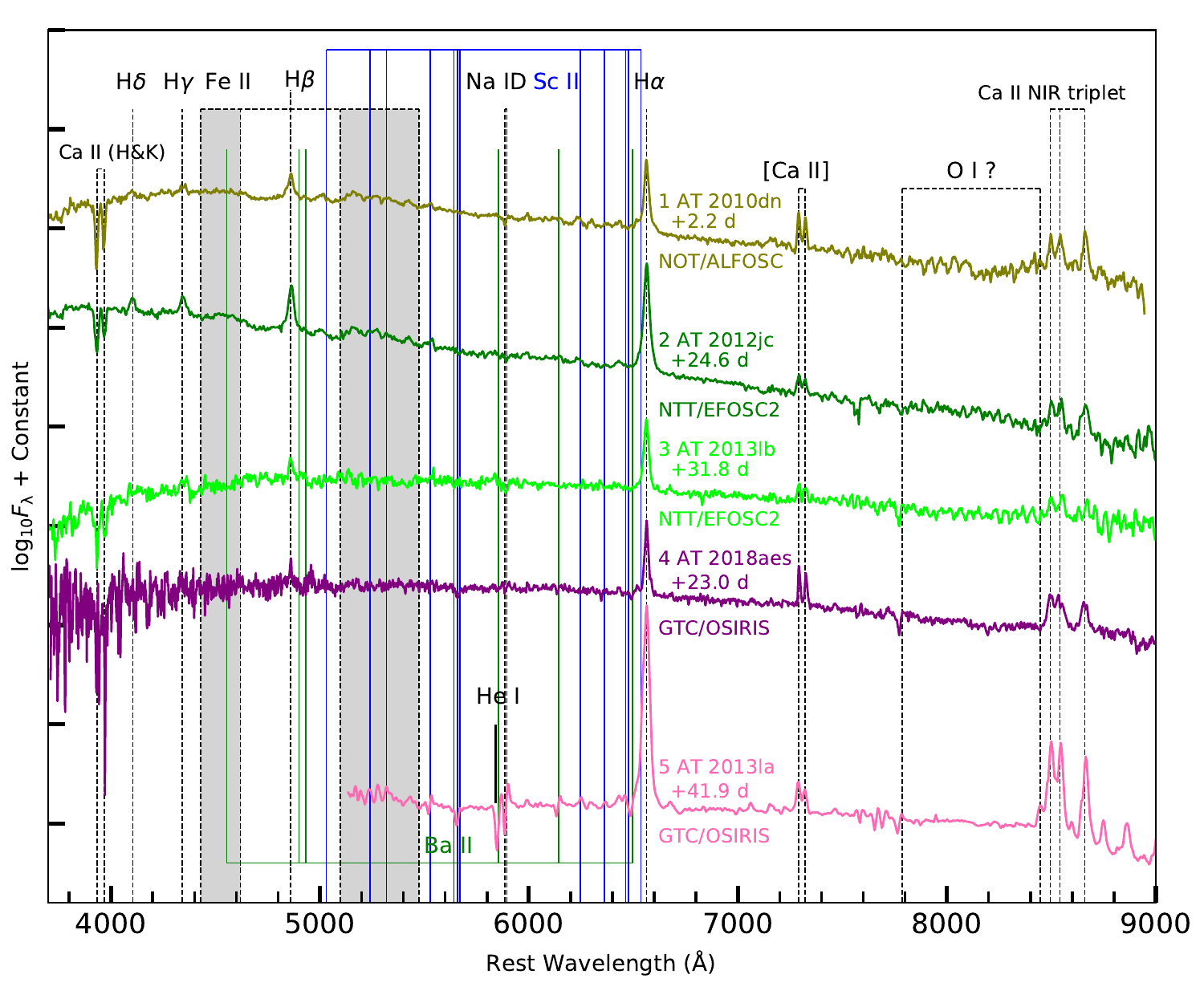}\quad
\caption{Line identification in the spectra of the five transients presented in this paper. The spectra have been corrected for  redshift and reddening. Phases are relative to their $R/r$-band maximum.}
\label{pic:spcttra_identification}
\end{figure*}

\begin{figure*}[htp]
	\centering
	\includegraphics[width=.95\textwidth]{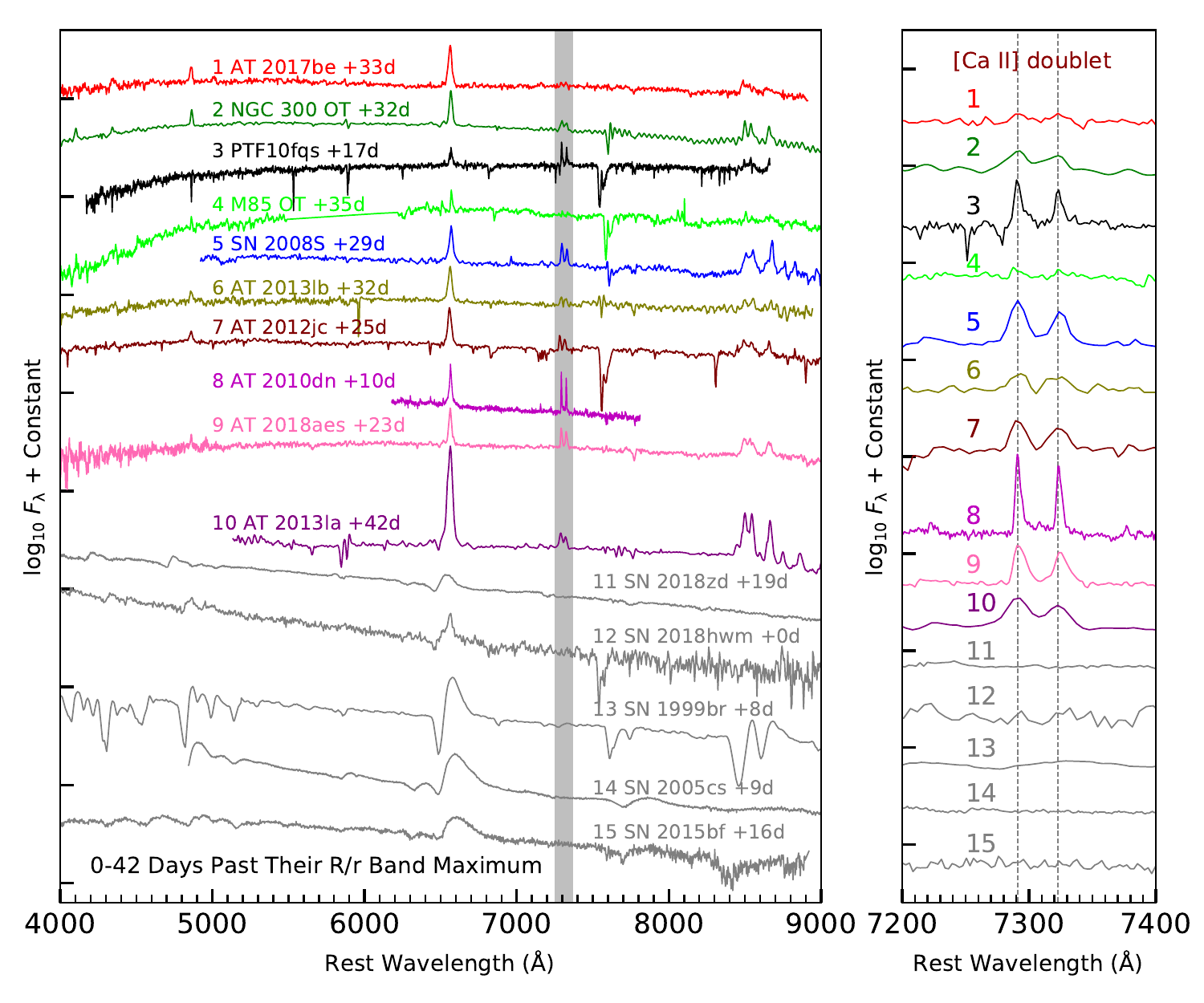}\quad
	\caption{Spectral comparison of a sample of ILRTs and SN~2018zd, SN~2018hwm, SN~2015bf,  SN~1999br, and SN~2005cs. The spectra of M85 OT, SN 2008S, PTF10fqs, AT~2017be, SN~2018zd, SN~2018hwm, SN~2015bf, SN~1999br, and SN~2005cs are taken from \citet{Kulkarni2007Natur.447..458K}, \citet{Botticella2009MNRAS.398.1041B}, \citet{Kasliwal2011ApJ...730..134K},  \citet{Cai2018MNRAS.480.3424C}, \citet{Zhang2020MNRAS.498...84Z}, \citet{Reguitti2021MNRAS.501.1059R}, \citet{Lin2021MNRAS.505.4890L}, \citet{Pastorello2004MNRAS.347...74P}, and   \citet{Pastorello2006MNRAS.370.1752P,Pastorello2009MNRAS.394.2266P}. The right panel shows the region of [Ca {\sc ii}], marked with the grey-shaded region in the left panel. All spectra were obtained at similar epochs from the $R/r$-band maxima, and were corrected for redshift and reddening. }
	\label{pic:spcttra_comparison}
\end{figure*}

\begin{figure}[htp]
\centering
\includegraphics[width=.45\textwidth]{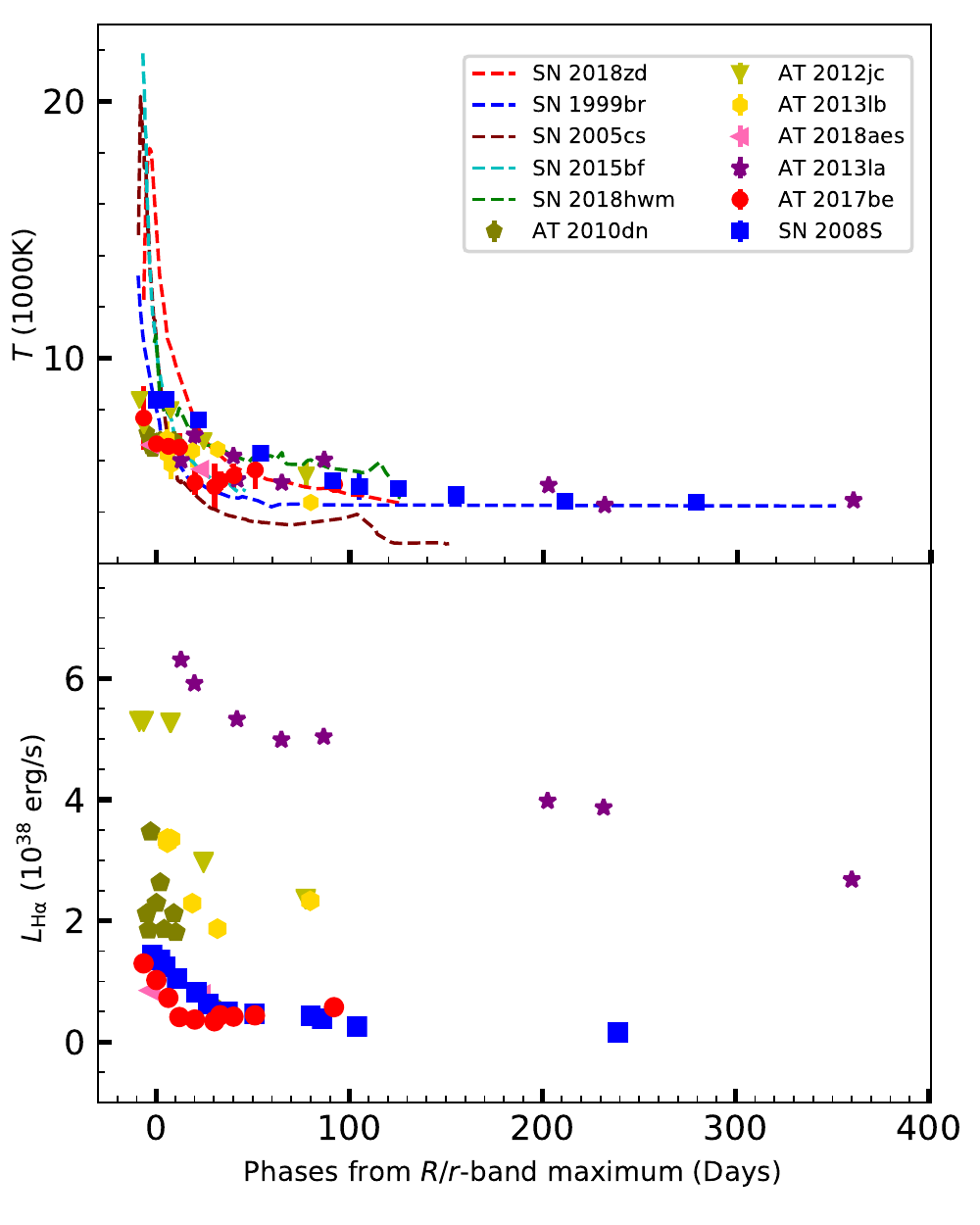}\quad
\caption{Evolution of the black body temperature (top) and the H$\alpha$ luminosity (bottom) for a sample of ILRTs, and SN~2018zd \citep{Zhang2020MNRAS.498...84Z}, SN~2018hwm \citep{Reguitti2021MNRAS.501.1059R}, SN~2015bf \citep{Lin2021MNRAS.505.4890L}, SN~1999br \citep{Pastorello2004MNRAS.347...74P}, and SN~2005cs \citep{Pastorello2006MNRAS.370.1752P, Pastorello2009MNRAS.394.2266P}.    }
\label{pic:spectraparameter}
\end{figure}

\subsection{H$\alpha$ and Ca {\sc ii}  line evolution} \label{profiles} 
 
In order to understand the evolution of individual spectral features, we performed a detailed analysis of the spectra of our ILRT sample at three critical phases\footnote{These critical phases are fixed on the basis of relatively large changes in the colour evolution.}: around 0, +30, and +70 days from maximum (see  Figure~\ref{pic:spcttra_phases}). We measure the full width at half maximum (FWHM) of H$\alpha$~through a single Lorentzian function fit, as in general this type of fit accurately describes the line profiles of narrow-lined SNe \citep[e.g. SNe IIn; see ][]{Taddia2013A&A...555A..10T, Smith2017hsn..book..403S, Nyholm2017A&A...605A...6N}. The H$\alpha$ profile is dominated by a narrow component in the spectra of most objects (e.g. AT~2010dn, AT 2012jc, AT 2013lb, and AT 2018aes). However, AT 2013la shows H$\alpha$ with a more complex profile, with narrow, blueshifted P-Cygni absorption lines. For AT 2013la, we therefore fitted the H$\alpha$ emission lines in our higher resolution spectra  (i.e. at phases +65.2d, +202.9, +231.9 and +360.3d) via multiple components: an intermediate-width Lorentzian profile, a narrow Gaussian emission and a Gaussian absorption. We also fitted the H$\alpha$ line at phase +42.2d with only a broader Lorentzian component plus a narrow Gaussian. The narrow component of H$\alpha$ arises from the unshocked photoionised CSM, while the intermediate-width component likely originates from the shocked CSM. Finally, the broad component is produced by the faster ejecta. The FWHM velocities ($v_{\mathrm{FWHM}}$) of H$\alpha$, after correcting for instrumental resolution ($v_{\mathrm{FWHM}}$ = $\sqrt{v_{\mathrm{observation}}^2 - v_{\mathrm{instrument}}^2}$), are reported in Table~\ref{spectra_para}  and plotted in Figure~\ref{pic:Ha_t}. We note that most spectra of our sample have modest resolution. Hence, in most instances our measurements have to be considered upper limits for the expansion velocities.  We also measured the H$\alpha$ velocity for the comparison objects in Figure~\ref{pic:Ha_t}, which were inferred from the minima of the broad P-Cygni absorptions. These show a remarkably large velocity evolution, ranging from $\sim$ 8000 \kms~to $\sim$ 1000 \kms. In contrast, ILRTs show a slow evolution in H$\alpha$ velocity ($<$ 1000 \kms) over the entire monitoring campaigns. In Figure~\ref{pic:Ha_profile}, we show the temporal evolution of H$\alpha$ profiles in the velocity space for AT~2013la, SN~2018zd, and SN~2005cs. AT~2013la shows a modest velocity evolution in H$\alpha$ profile, while SN~2018zd and SN~2005cs experience significant velocity evolution along with the gradual emergence of broad P-Cygni features. 

\begin{figure}[htp]
	\centering
	\includegraphics[width=.45\textwidth]{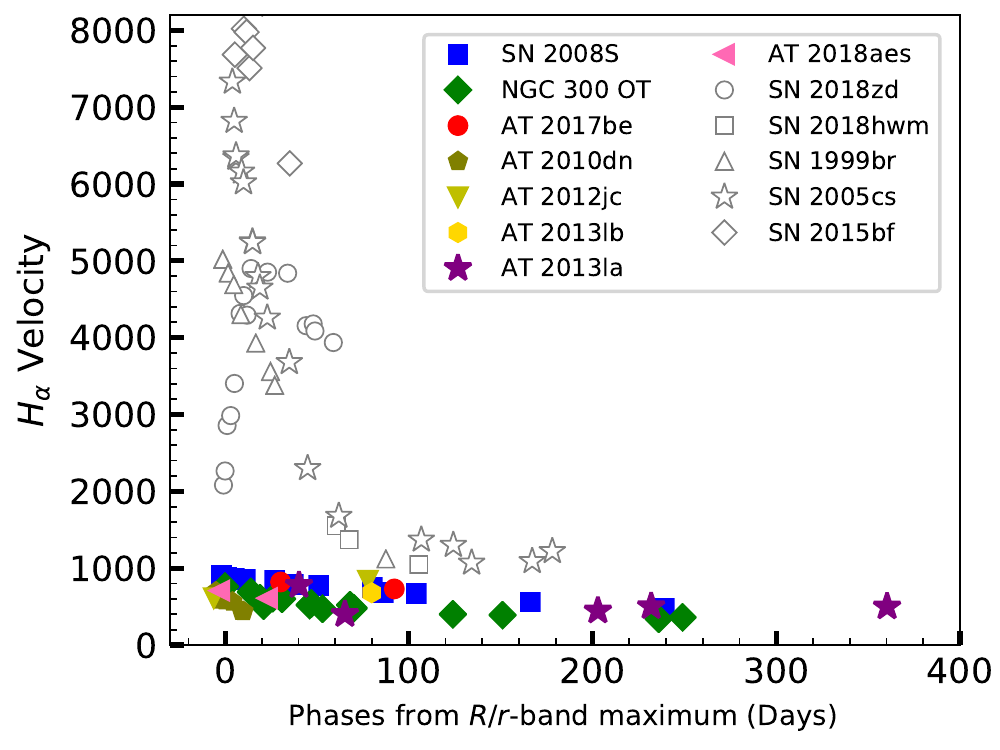}\quad
	\caption{ H$\alpha$ velocity evolution of a sample of ILRTs, SN~2018zd \citep{Zhang2020MNRAS.498...84Z}, SN~2018hwm \citep{Reguitti2021MNRAS.501.1059R}, SN~2015bf \citep{Lin2021MNRAS.505.4890L}, SN~1999br \citep{Pastorello2004MNRAS.347...74P}, and SN~2005cs \citep{Pastorello2006MNRAS.370.1752P, Pastorello2009MNRAS.394.2266P}. The H$\alpha$ velocities were estimated by measuring the FWHM for ILRTs, while those for SNe II were determined from the position of the	minimum of the absorption component.   }
	\label{pic:Ha_t}
\end{figure}

\begin{figure}[htp]
	\centering
	\includegraphics[width=.45\textwidth]{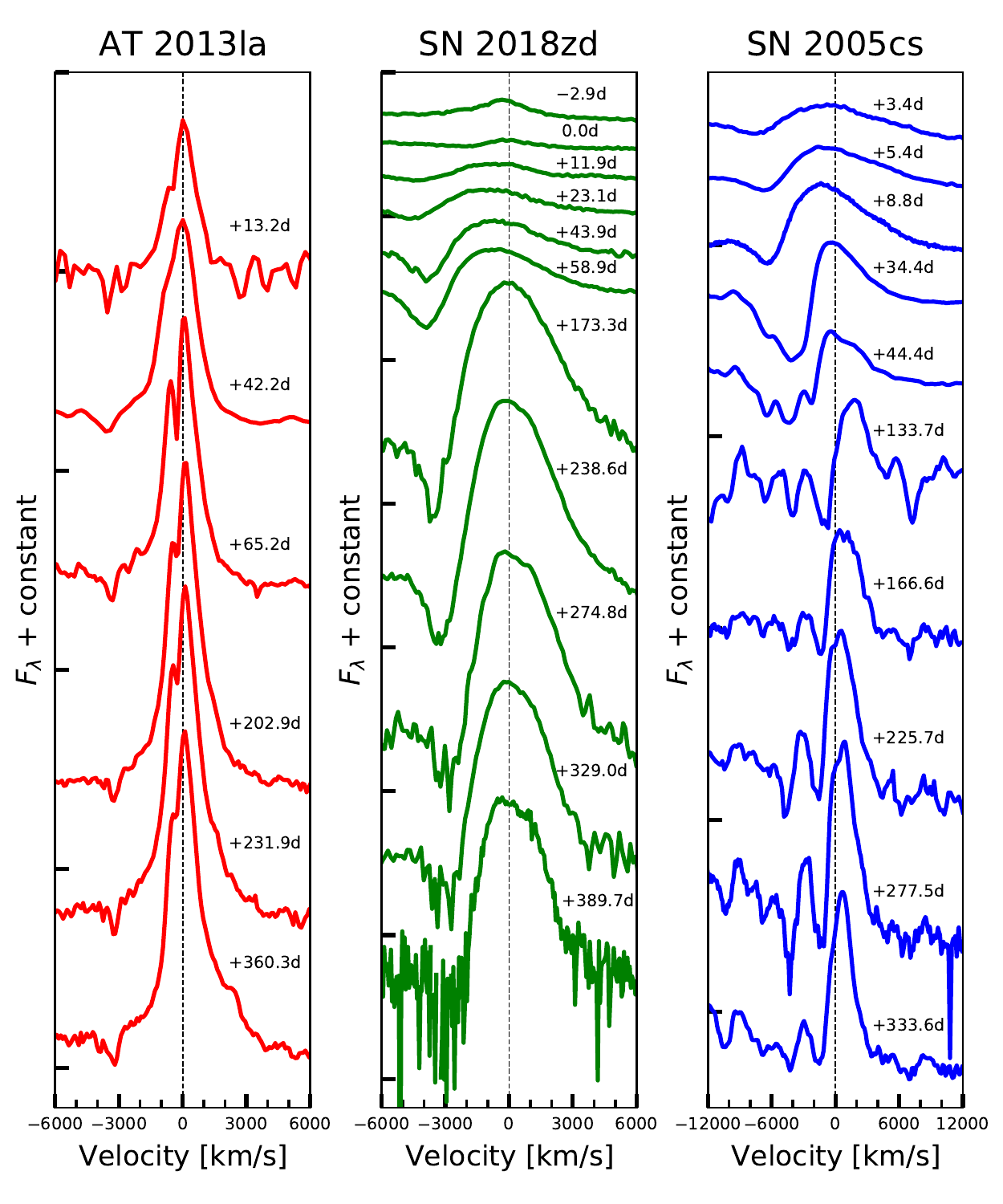}\quad
	\caption{Evolution of the H$\alpha$~profile in selected spectra of the ILRT AT~2013la (left panel), SN~2018zd \citep[][ middle panel]{Zhang2020MNRAS.498...84Z}, and the underluminous Type IIP SN~2005cs \citep[][ right panel]{Pastorello2006MNRAS.370.1752P, Pastorello2009MNRAS.394.2266P}. The H$\alpha$ lines are shown in the velocity space, with zero velocity (rest wavelength) being marked by a grey dashed line.   }
	\label{pic:Ha_profile}
\end{figure}

The [Ca {\sc ii}] doublet and Ca {\sc ii} NIR triplet lines were fitted with Lorentzian functions.  The [Ca {\sc ii}] doublet is produced by quadrupole transitions from $4s^{2}S$ to the metastable $3d^{2}D$ level, in which $\lambda_1$ at 7291 $\AA$ and $\lambda_2$ at 7324 $\AA$ are associated to $^{2}S_{1/2}$ -  $^{2}D_{5/2}$ and $^{2}S_{1/2}$ -  $^{2}D_{3/2}$ transitions, respectively \citep[e.g. ][]{Osterbrock1951ApJ...114..469O, Lambert1969SoPh....7...11L, Chevalier1994ApJ...420..268C}. The [Ca {\sc ii}]  doublet likely originates from extremely low-density gas, possibly in a slow-moving circumstellar shell. The Ca {\sc ii} NIR triplet lines with $\lambda$$_{1,2,3}$= 8498, 8542, 8662 \AA~are common features in many cool transients, and are produced by transitions from $4p ^{2}P_{1/2,3/2}$ to $3d^{2}D_{3/2,5/2}$ levels \citep[e.g. ][]{Mallik1997A&AS..124..359M, Mallik1998BASI...26..479M, Andretta2005A&A...430..669A, Bus2007A&A...466.1089B, Martin2017A&A...605A.113M}. 

At early phases (see Figure \ref{pic:spcttra_phases}, left panel), the H$\alpha$ FWHM velocities of the transients presented in this paper are 600-700 \kms~(with upper limits up to 1000 \kms~in cases of unresolved features), which are comparable to those of other ILRTs (e.g. $\sim$750 \kms~and~$\sim$600 \kms~for SN~2008S and AT~2017be, respectively). We conclude that in all ILRTs the ejected material expands with similar velocities of several hundred \kms. In most cases the FWHM of the [Ca {\sc ii}] doublet is below the resolution limits, and we do not have  reliable estimates for the outflow velocity inferred from this feature. The doublet is resolved only in the spectra of AT 2010dn at $-3.0$ d, providing an FWHM velocity of  145 \kms. The feature was also resolved in a spectrum of AT 2012jc at phase $-$6.2 d, where we measured $v_{\mathrm{FWHM}}$ $\sim$  220 \kms. For AT 2018aes, a velocity of $v_{\mathrm{FWHM}}$ $\sim$ 310 \kms~ was derived for the [Ca {\sc ii}] feature at phase -3.0 d. 
We conclude that the [Ca {\sc ii}] doublet originates in a slow-moving CSM  ($\sim$  145 - 310 \kms), while the broader H$\alpha$ component forms in fast-expanding ejected gas.  

At 3-5 weeks after maximum (Figure \ref{pic:spcttra_phases}, middle panel), the spectra become redder. We also see a strengthening of a narrow P-Cygni profile superimposed on H$\alpha$ in AT 2013la, the minimum of which is blueshifted by about 360 - 410 \kms. However, we cannot rule out that the non-detection of this feature in other objects is a mere resolution effect. The FWHM velocities of H$\alpha$ are in the range 600-850  \kms~for all objects, with M85 OT being the only outlier with  $v_{\mathrm{FWHM}}$~$\sim$ 350 \kms. A resolved velocity of 115 \kms\ was reported at this phase  for [Ca {\sc ii}] in AT 2017be \citep{Cai2018MNRAS.480.3424C}.

At late phases (+70 to +80 d; Figure \ref{pic:spcttra_phases}, right panel}), the H$\alpha$ velocity does not change significantly, still ranging from 400 (in AT 2013la) up to 850 \kms~(in the other objects). The spectra of AT 2013la also provide an averaged [Ca {\sc ii}]  $v_{\mathrm{FWHM}}$ $\sim$  365 \kms. In most cases, the low resolution or the low  S/N of our spectra prevent us from probing the Ca {\sc ii} triplet lines in detail, with an exception being the Magellan spectrum of AT 2012jc obtained on 2012 June 23 (phase +77.7 d), where the triplet emission lines show a symmetric profile, with an average $v_{\mathrm{FWHM}}$ $\simeq$  905 \kms.  The  [Ca {\sc ii}] profile appears to show very little evolution in the FWHM until $t\sim$ 70 days, and this is a further argument in favour of the hypothesis that these lines form in the circumstellar environment. We also note that none of the typical nebular lines, such as [O {\sc iii}] ($\lambda\lambda$= 4959, 5007 \AA) and [S {\sc ii}] ($\lambda\lambda$= 6716, 6731 \AA) lines \citep{Lundqvist1996ApJ...464..924L, Lundqvist2015A&A...577A..39L, Lundqvist2020MNRAS.496.1834L}, were detected in spectra. This implies that the wind density is high enough for those nebular lines to be collisionally de-excited, but that [Ca {\sc ii}] can exist at much higher density \citep[e.g. critical density: $\rho$$_{\rm{[S~II]}}$ = 1 $\times$ 10$^4$ cm$^{-3}$, $\rho_{\rm{[Ca~II]}}$ = 1 $\times$ 10$^9$ cm$^{-3}$; ][]{Li1993ApJ...405..730L}. In such a dense wind, the [Ca {\sc ii}] lines could also be broadened by Thomson scattering, albeit to a lesser extent  than the Balmer lines because these latter are formed further out in the wind/CSM.

We use the few spectra with higher S/N to highlight the evolution of the H$\alpha$ to [Ca {\sc ii}] luminosity ratio. The values for the different ILRTs are reported in Table~\ref{spectra_para} and shown in Figure~\ref{pic:spcttra_ratio}. The values for AT 2012jc and AT 2013lb show a faster decline (by a  factor of three) from maximum (when H$\alpha$ largely dominates) to one month later.
However, for AT 2010dn and AT~2018aes, the above ratio shows a much flatter evolution (ratio from about 3 to 2) over the same temporal window. Interestingly, in these two ILRTs, the [Ca {\sc ii}] doublet is very luminous with respect to H$\alpha$. In addition, the H$\alpha$ luminosity evolution is also shown in the bottom panel of Figure~\ref{pic:spectraparameter}.\\     
 
\begin{figure*}[htp]
\centering
\includegraphics[width=.9\textwidth]{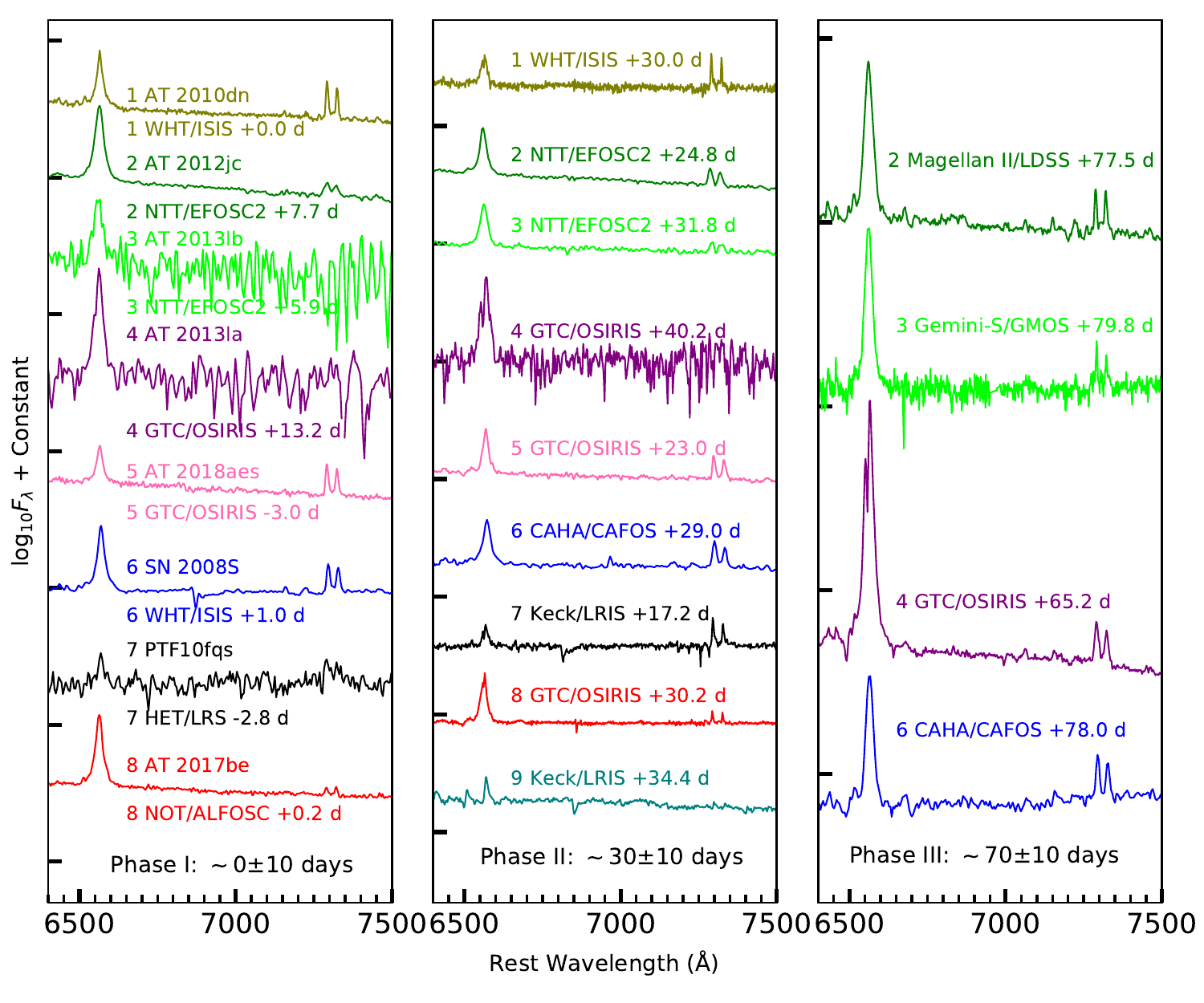}\quad
\caption{Evolution of H$\alpha$ and [Ca {\sc ii}] in ILRTs. The spectra of M85 OT, SN 2008S, PTF10fqs, and AT~2017be,  downloaded from {\textit WISeREP} archive \protect\footnotemark~\citep{Yaron2012PASP..124..668Y}, were published by \citet{Kulkarni2007Natur.447..458K}, \citet{Botticella2009MNRAS.398.1041B}, \citet{Kasliwal2011ApJ...730..134K}, and \citet{Cai2018MNRAS.480.3424C}. All  spectra were corrected for redshift and reddening.}
\label{pic:spcttra_phases}
\end{figure*}
\footnotetext{\url{https://wiserep.weizmann.ac.il/}}

\begin{figure}[htp]
\centering
\includegraphics[width=.45\textwidth]{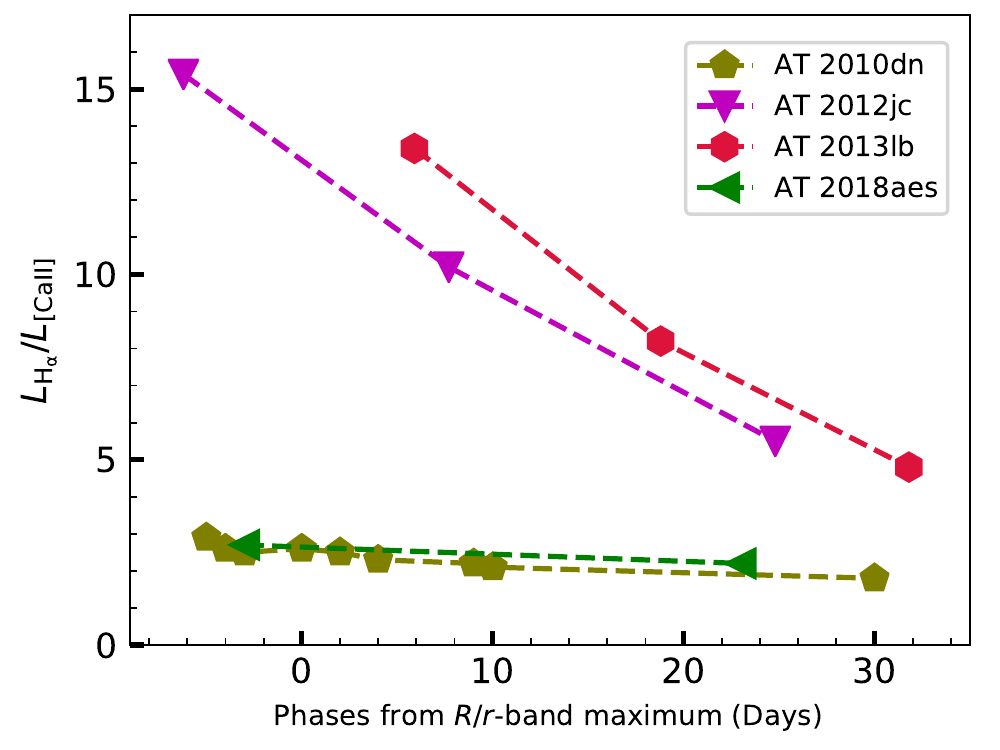}\quad
\caption{Ratio of H$\alpha$ to [Ca {\sc ii}] (total profiles at $\lambda~\approx$~7300~\AA) luminosities until phase $\sim$ 30 d.}
\label{pic:spcttra_ratio}
\end{figure}


\section{Rate estimates for ILRTs}\label{ILRTrates}

In the past 12 years (2008 - 2019), 12 objects have been confirmed to be ILRT events\footnote{We note that we only include confirmed ILRTs prior to 2020, as systematic surveys and monitoring campaigns were almost globally suspended because of the Covid-19 pandemic in 2020. In addition to the objects listed in Table \ref{table_parameters} (M85 OT was excluded in the rate estimate sample because of its controversial nature as discussed by \citet{Kulkarni2007Natur.447..458K}, \citet{Pastorello2007Natur.449E...1P}, and \citet{Rau2007ApJ...659.1536R}), there are three confirmed ILRTs to be added in 2019: AT 2019abn \citep{De2019ATel12433....1D,Nordin2019TNSTR.141....1N,Burke2019TNSCR.328....1B,Fremling2019TNSCR2789....1F}, AT 2019ahd \citep{Jha2019ATel12454....1J,Jha2019TNSCR1237....1J, Tonry2019TNSTR.161....1T}, and AT 2019udc \citep{Malesani2019TNSAN.123....1M,Siebert2019TNSCR2287....1S,Malesani2019TNSCR2881....1M,Valenti2019TNSTR2263....1V}.    }; the observed rate of discoveries is 1.0 event per year, but this does not account for the luminosity distribution of ILRTs, or the efficiency or observing strategy of the surveys. The ILRTs in our sample are distributed in a radius of 30 Mpc, corresponding to a volume of 1.13 $\times$~10$^{5}$ Mpc$^{3}$. Hence, we obtain a volumetric rate of ILRTs of 0.09 $\times$ 10$^{-4}$ Mpc$^{-3}$yr$^{-1}$.  In the same temporal window and volume, 147 CC SNe were discovered\footnote{Bright SNe archive:~\url{http://www.rochesterastroncamomy.org}; \\ Asiago SN group archive:~\url{http://graspa.oapd.inaf.it/asnc.html}. } corresponding to a rate of 1.1 $\times$ 10$^{-4}$ Mpc$^{-3}$yr$^{-1}$. Therefore, assuming that the discovery of ILRTs suffers the same selection effects as all CC SNe, the rate of ILRTs is 8\% that of CC SNe. We note that \citet{Thompson2009ApJ...705.1364T} estimated an EC SN rate of 2\%-10\% of  CC SNe in 10 years (before 2008), while \citet{Poelarends2008ApJ...675..614P} obtained an EC SN fraction of CC SNe of only about 6\%~based on a standard Salpeter initial mass function \citep[IMF; ][]{Salpeter1955ApJ...121..161S} and an EC SN mass range of 9.00 - 9.25 \msun.  
Assuming a narrow mass range $\Delta \approx$ 0.2 \msun~and a metallicity range $Z$ = 0.0001 - 0.02, \citet{Doherty2015MNRAS.446.2599D, Doherty2017PASA...34...56D} calculated that 2\%-5\% of CC SNe are EC SNe. However, theoretical estimates for EC SN rates are dependent on the uncertainties on the initial mass range, metallicity, and the details of stellar models. While theoretical calculations show some variations, they are roughly consistent with our rate estimate based on the observations. We note that we did not take into account the fact that ILRTs are on average fainter than CC SNe and surveys are incomplete up to 30 Mpc. Therefore, the true relative rate is larger than the current observed rate, and future advanced surveys could increase this estimate.   \\

\section{Discussion and conclusions}  
\label{sec:discussion}

\subsection{Observables and parameter correlations}
\label{parameter_correlations}

In this paper, we systematically analyse a sample of ILRTs. They show relatively homogeneous spectrophotometric properties that can be summarised as follows:\\

\begin{enumerate}[I]
\item They all show single-peaked light curves, resembling those of Type IIP/IIL SNe but scaled down in luminosity (see Figure \ref{pic:absLC}). The duration of the plateau is determined by the effective recombination time of the hydrogen envelope \citep[e.g. see][]{Grassberg1971Ap&SS..10...28G, Eastman1994ApJ...430..300E, de2010MNRAS.401.2081D}. \label{i} \\
\item They show a homogeneous colour (hence temperature) evolution. The temperature evolution shows an initial fast decline (usually lasting $<$30 days), followed by a much slower decline which is still ongoing at the end of the follow up. The temperature evolution of ILRTs is similar to what is observed in the hydrogen recombination of classical Type II SNe (e.g. see the Type IIP SN 2005cs in the top panel of Figure \ref{pic:spectraparameter}). We note that ILRTs are usually redder than low-luminosity SNe IIP  and other Type II events proposed to be EC SN candidates (see Figure \ref{pic:colour}).      \label{ii}\\

\item Their peak magnitudes range from $-$11~($\pm$ 0.5) to $-$15~($\pm$ 0.5)~mag. Their quasi-bolometric ($B$ to $I$) light curve peaks are in the range $\sim 5\times 10^{39}$ to $\sim 9\times 10^{40}$ erg s$^{-1}$~and the radiated energies span from 0.30 $\times$~10$^{47}$ to 2.94 $\times$~10$^{47}$~erg.  ILRTs are typically fainter than most SNe II, although they are comparable in luminosity to the most subluminous Type IIP SN~1999br (see Figure \ref{pic:absLC} and \ref{fig:PseudoBolom}).  \label{iii} \\

\item The SED of ILRTs with optical to MIR observations shows evidence of prominent IR excesses both at early and late phases. This is consistent with the expectation that the progenitors of ILRTs have dusty local environments. Furthermore, it corroborates the use of the nomenclature: `red' transients. Another common feature in the SED evolution of ILRTs is the monotonic decrease of the photospheric radius, which could be a diagnostic tool to distinguish ILRTs from LRNe \citep[see Figure 4 of][]{Cai2019A&A...632L...6C}.          \label{v}\\
\item Their spectra experience relatively slow evolution. Prominent lines include the Balmer series, along with Ca {\sc ii} (e.g. Ca H\&K,  [Ca {\sc ii}] doublet, and Ca NIR triplet), Na {\sc i} D, Fe {\sc ii}, and possibly O {\sc i}. The [Ca {\sc ii}] doublet is normally visible during the entire monitoring campaign. However, the study of AT~2019abn challenges this paradigm with a barely detectable [Ca {\sc ii}] doublet at early phases \citep{Jencson2019ApJ...880L..20J,Williams2020A&A...637A..20W}.  In addition,  this feature is never detected in SNe IIP at early phases (see Figure \ref{pic:spcttra_comparison}), while it is always visible in the nebular phases.     \label{vi}  \\
\item The ejecta velocities inferred from the H$\alpha$ FWHM lie between about 400 and 800 \kms. The measurements of the minimum of P-Cygni profiles indicate a modest wind velocity of about 360 - 410 \kms.  In addition, the [Ca {\sc ii}] feature width suggests the existence of slow-moving CSM with $v_{\mathrm{FWHM}}$ $\sim$ 170 - 300 \kms.  In ILRTs, the H$\alpha$ velocity is much lower than those inferred for SNe II  
at all phases (see Figure \ref{pic:Ha_t}).     \label{vii} \\        
\end{enumerate}

\begin{figure*}[htp]
\centering
\includegraphics[width=0.9\textwidth]{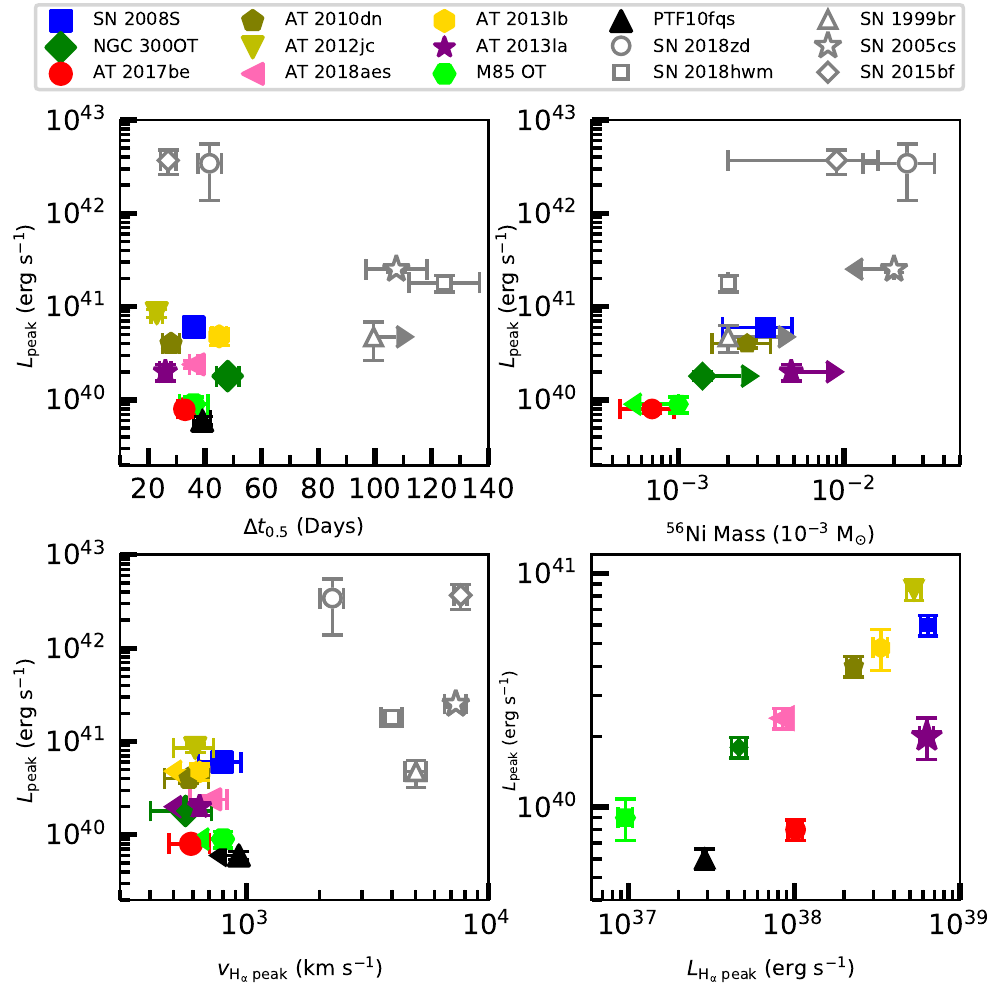}\quad
\caption{Peak luminosity ($L_{\mathrm{peak}}$) vs. decline time ($\Delta t_{0.5}$) from $L_{\mathrm{peak}}$ to 0.5$L_{\mathrm{peak}}$  (top-left); $L_{\mathrm{peak}}$ vs. $^{56}$Ni mass (top-right); $L_{\mathrm{peak}}$ vs.  $v_{\mathrm{H\alpha}}$ at peak (bottom-left); and  $L_{\mathrm{peak}}$ vs. \Ha peak luminosity (bottom-right). }
\label{pic:Para_relation}
\end{figure*}

In order to better characterise ILRTs, we looked for possible correlations among their physical parameters. In Figure~\ref{pic:Para_relation}, we plot peak luminosity ($L_{\rm{peak}}$; reported in Table \ref{table:peak}) against different physical parameters of ILRTs, namely the time ($\Delta \mathrm{t}_{0.5}$) for the luminosity to decline by a factor of two from maximum, the $^{56}$Ni mass, the \Ha~ velocity, and the \Ha~luminosity at maximum. Unfortunately, several factors (e.g. poorly sampled light curve peaks; lack of late-time observations; low spectral resolution or modest S/N; incomplete wavelength coverage) limit the information available for the sample. More specifically, there seems to be no correlation between $L_{\rm{peak}}$ and decline time $\Delta \mathrm{t}_{0.5}$, which is in the range of  20-50 days (Figure~\ref{pic:Para_relation}, top-left panel). The lack of a correlation may point to variation in the progenitor mass loss or the strength of CSM-ejecta interaction. A trend may exist linking the peak luminosity to the ejected $^{56}$Ni mass (Figure~\ref{pic:Para_relation} top-right panel), although the lower limit for AT 2013la is somewhat discrepant with those of other ILRTs. We cannot rule out that the ejecta-CSM interaction significantly affects the late-time luminosity, therefore biasing the $^{56}$Ni mass estimates of the sampled objects. The limited range of $^{56}$Ni masses ($\sim$ 1-5 $\times 10^{-3}$ \msun) may be attributed to their progenitor masses lying within a narrow mass range ($\sim$ 8-10 \msun). As a comparison, both Type II and Type Ib/c SNe also show a similar correlation, but they have higher energies and wider $^{56}$Ni mass ranges than ILRTs \citep[][]{Hamuy2001PhDT.......173H, Hamuy2003ApJ...582..905H, Pastorello2005coex.conf..195P}.  SN~2018zd \footnote{\citet{Hiramatsu2021NatAs.tmp..107H} estimated a much lower $^{56}$Ni mass ($\sim 8.6 \times 10^{-3}$ \msun) for SN~2018zd, adopting a very low distance obtained through the standard candle method (SCM). We note however that the SCM distance is discrepant with the significantly larger kinematic estimates \citep[e.g. ][ Callis et al., in preparation]{Zhang2020MNRAS.498...84Z}.}, which was also proposed as a possible EC SN, has a significantly higher $^{56}$Ni mass \citep[0.013 - 0.035~\msun; ][]{Zhang2020MNRAS.498...84Z}, exceeding that of ILRTs  ($\sim$ 1-5 $\times 10^{-3}$ \msun) by nearly one order of magnitude. In the bottom-left panel, the \Ha~velocity seems to be uncorrelated with $L_{\rm{peak}}$. The fact that $v$ (H$\alpha$) $=$ 600 $\pm$ 200 \kms~for all objects suggests a common explosive or eruptive mechanism for all ILRTs, and again a relatively small range of progenitor masses. From this, the LBV giant eruption scenario appears less likely. The diagram showing bolometric versus  H$\alpha$ luminosity at peak reveals a trend, with ILRTs clustered along two different lines. In particular, we note that ILRTs along the lower line of the diagram (PTF10fqs, AT~2017be, and AT 2013la) show a clear plateau in their light curves.  A wider sample of well-monitored ILRTs is necessary to confirm or rule out the putative trends mentioned above.  Figure~\ref{pic:Para_relation} shows how the different classes of objects discussed in this paper distribute in the $L_{\rm{peak}}$ versus $\Delta \mathrm{t}_{0.5}$~and~\Ha~velocity diagrams. The clear separation of ILRTs from the comparison SNe II suggests that they are a distinct class of transients.  \\
        
\begin{figure}[htp]
\centering
\includegraphics[width=0.45\textwidth]{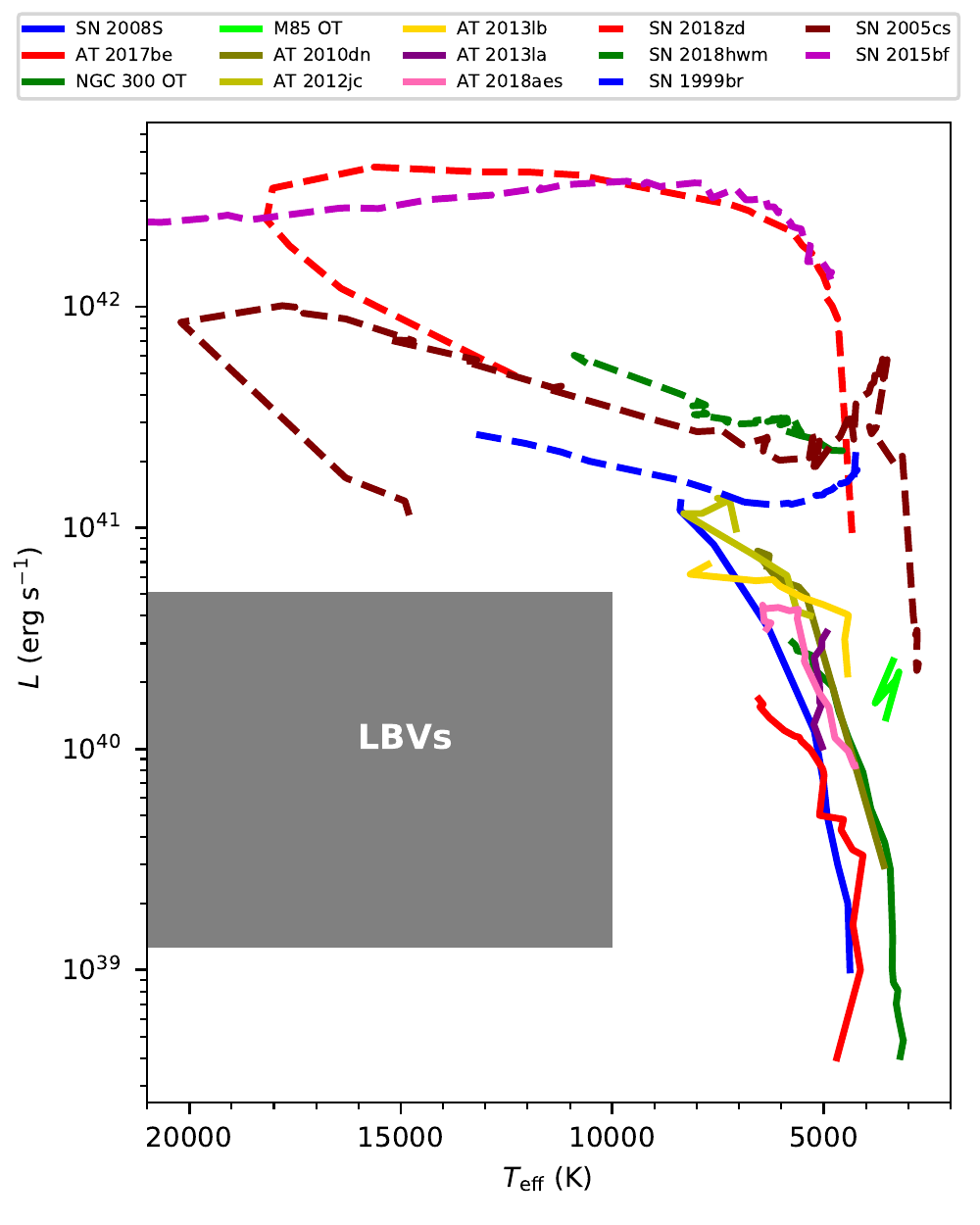}\quad
\caption{Luminosity ($L$) vs. effective temperature ($T_{\mathrm{eff}}$). ILRTs are marked as solid lines with different colours, while the area where LBVs lie is marked in grey. Comparison SNe II lie in the upper part of the diagram and are labelled with dashed lines. }
\label{pic:Para_relationLT}
\end{figure}

In Figure \ref{pic:Para_relationLT}, we also investigated the evolution of luminosity ($L$) against effective temperature ($T_{\mathrm{eff}}$) of ILRTs, and compared with giant eruptions of  LBVs, claimed EC SNe, and some representative SNe II. The ILRTs tend to show a homogeneous evolution, and lie in a narrow strip in which the luminosity declines monotonically with temperature. SNe II lie in a region that is distinct from that where ILRTs and LBVs are found, and span a wide range of luminosities and temperatures. The physical homogeneity in this L-T diagram is suggestive of a common mechanism triggering the ILRTs and seems to exclude the LBV origin for ILRTs and enable us to distinguish ILRTs from traditional Type II SNe.     \\

\subsection{Plausible scenarios for ILRTs and conclusions}
\label{ILRTsenario}

Although several plausible scenarios have been proposed to explain ILRTs, their nature is still debated. Possible scenarios include LBV-like eruptions, LRN-like events (stellar mergers), and EC-induced SN explosions.  \\

Luminous blue variables originate from massive star ($>$ 20 \msun) eruptions, and show irregular light curves and SN IIn-like spectra\footnote{Although the [Ca {\sc ii}] doublet is sometimes visible in LBV spectra, no persistent signature of this feature has been observed in these events.}. A luminous progenitor is expected to be detected both in pre-outburst and post-outburst optical and NIR images. No $^{56}$Ni is expected to be produced. In addition, LBVs and ILRTs are distributed in two separate regions in the $L$-$T_{\rm{eff}}$ diagram (see Figure \ref{pic:Para_relationLT}).  Our observational arguments disfavour a LBV-like eruption scenario for ILRTs. However, we note that \citet{Andrews2020arXiv200913541A} proposed that a recent gap transient, AT~2019krl, was an unobscured ILRT eruption from a relatively massive star.     \\

Luminous red novae are another subclass of gap transients, and show double-peaked (or triple-peaked) light curves and a remarkable spectral evolution. Specifically, the major drop in continuum temperature, the appearance of narrow metal lines in absorption, the dramatic change in the H$\alpha$ strength and profile, and the emergence of molecular bands are typical spectral features of LRNe. In addition, the typical H$\alpha$ FWHM velocities of LRNe are lower than 500 \kms~\citep[][]{Pastorello2019A&A...630A..75P}. Their colours become progressively redder; for example, the intrinsic $B-V$ colour of an LRN can even reach 1.8 mag at late phases \citep[see Figure 4 of][]{Pastorello2019A&A...630A..75P}.  The evolution of the SED and the radius of LRNe is markedly different from that of ILRTs \citep{Cai2019A&A...632L...6C,Stritzinger2020A&A...639A.104S}. LRNe are likely produced by the coalescence of two stars, however we disfavour such a merging scenario for ILRTs.  \\

Theory predicts that some stars with initial mass 8-10 \msun~form a S-AGB star with a strongly degenerate oxygen–neon–magnesium (O-Ne-Mg) core surrounded by thin He and H envelopes \citep[e.g. ][]{Nomoto1984ApJ...277..791N, Nomoto1987ApJ...322..206N,  Pumo2009ApJ...705L.138P, Takahashi2013ApJ...771...28T, Moriya2014A&A...569A..57M, Doherty2017PASA...34...56D, Nomoto2017hsn..book..483N, Leung2019PASA...36....6L, Leung2020ApJ...889...34L}. The final O-Ne-Mg core mass is determined by the competition between core increase from the He and H shell burning and mass loss through thermal pulses \citep{Siess2007A&A...476..893S, Langer2012ARA&A..50..107L,  Leung2020ApJ...889...34L}. Once the core reaches the Chandrasekhar mass \citep[1.37 \msun; ][]{Nomoto1984ApJ...277..791N}, electron capture reactions on $^{20}$Ne and $^{24}$Mg take place, and eventually ignite an O-Ne deflagration that propagates outward. This type of CC explosion is referred to as an EC SN \citep{Nomoto1984ApJ...277..791N, Zha2019ApJ...886...22Z, Leung2020ApJ...889...34L}.  According to models, EC SNe are expected to explode with low energy, low ejected $^{56}$Ni mass, and typically in a dense and dusty CSM \citep[e.g. ][]{Poelarends2008ApJ...675..614P, Woosley2015ApJ...810...34W, Moriya2016MNRAS.461.2155M}. Although Fe-core progenitors at the low-mass end of the CC SNe also explode with low energy ($\sim~0.5 - 1.0 \times$ 10$^{50}$ erg) and small $^{56}$Ni mass (4-6 $\times$ 10$^{-3}$ \msun) \citep{Stockinger2020MNRAS.496.2039S}, some of them \cite[e.g. low-lominosity Type IIP SNe 2003gd, 2005cs, and 2008bk; ][]{ Eldridge2007MNRAS.376L..52E, Maund2014MNRAS.438.1577M, Maund2014MNRAS.438..938M} can be distinguished from EC SNe through the explosion mechanism, chemical composition, and nucleosynthesis \citep{Jerkstrand2018MNRAS.475..277J}. The possible correlations among physical parameters, the homogenous observational properties and the L-T evolution suggest that all ILRTs are regulated by the same mechanism, and the EC SN explosion is a plausible scenario. In particular, this scenario for ILRTs is supported by the following arguments: \\

\begin{itemize}
	\item The late-time light curves of SN 2008S and NGC 300 OT2008-1 decline approximately following the $^{56}$Co decay rate (i.e. $\sim$ 0.98 mag/100d). The $^{56}$Ni masses inferred from the bolometric light curve are of the order of 10$^{-4}$ to 10$^{-3}$~\msun, consistent with the expected EC SN yields. \label{iv}\\
	
	\item Investigation of archival optical and IR images of the transient locations obtained years before the explosion suggests that the ILRT progenitors are moderately massive, in the range between 8 and 15~\msun.  Their progenitors appear to be embedded in dusty cocoons \citep[i.e. SN~2008S; ][]{Botticella2009MNRAS.398.1041B, Prieto2009ApJ...705.1425P}.   \label{viii} \\

	\item Inspection of the transient sites a few years after the explosion suggests that the objects are at least 15 times fainter than the quiescent progenitors \citep[i.e. SN~2008S and NGC~300~OT; ][]{Adams2016MNRAS.460.1645A}. This supports the terminal SN explosion nature.   \label{viiii}\\  
	
	\item The estimated ILRT event rate is in fair agreement with EC SN theoretical predictions.
\end{itemize}

The observational properties described in this paper agree well with theoretical expectations, and we favour terminal EC SN explosions as the possible origin of ILRTs. Future facilities, such as the Vera C. Rubin Observatory\footnote{\url{https://www.lsst.org/}}~and the Nancy Grace Roman Space Telescope\footnote{\url{https://www.nasa.gov/content/goddard/nancy-grace-roman-space-telescope}}, will be essential for expanding the sample of ILRTs, and will be crucial in fine-tuning existing theoretical models.    \\


\section*{Acknowledgments}
We thank the reviewer for his/her insightful comments that have improved the paper. 
Y.-Z. Cai thanks AP/AR/EC/MT/NER/SB for their training in the photometric and spectral reduction techniques.
We are grateful to Han Lin (THU-THCA) and JuJia Zhang (CAS-YNAO) for providing the data of SN~2015bf and SN~2018zd, respectively.
We thank F. Taddia and D. E. Wright for their few images taken with NOT/ALFOSC. We also thank M. Sullivan for his contribution in a spectrum observed with WHT/GB-GR.
This work is funded by China Postdoctoral Science Foundation (grant no. 2021M691821).
This work is supported by National Natural Science Foundation of China (NSFC grants 12033003, 11633002, 11325313, and 11761141001), National Program on Key Research and Development Project (grant no. 2016YFA0400803).  
DAH is supported by NSF grant AST-1911225. 
F. H. is supported by NSFC grant 11803021 and the Startup Fund for Youngman Research at SJTU.
KM and SJP are supported by H2020 ERC grant no.~758638.
MF is supported by a Royal Society - Science Foundation Ireland University Research Fellowship. 
SB, PO, and MT are partially supported by PRIN-INAF 2017 of Toward the SKA and CTA era: discovery, localization, and physics of transient sources. EC, NER, LT, FO acknowledges support from MIUR, PRIN 2017 (grant 20179ZF5KS) "\textit{The new frontier of the Multi-Messenger Astrophysics: follow-up of electromagnetic transient counterparts of gravitational wave sources}". 
LBo acknowledges the funding support from Italian Space Agency (ASI) regulated by ``Accordo ASI-INAF n. 2013-016-R.0 del 9 luglio 2013 e integrazione del 9 luglio 2015 CHEOPS Fasi A/B/C''. 
L.G. acknowledges financial support from the Spanish Ministry of Science, Innovation and Universities (MICIU) under the 2019 Ram\'on y Cajal program RYC2019-027683 and from the Spanish MICIU project PID2020-115253GA-I00.
T.R. acknowledges the financial support of the Jenny and Antti Wihuri and the Vilho, Yrj{\"o} and Kalle V{\"a}is{\"a}l{\"a} foundations. 
AR acknowledges support from ANID BECAS/DOCTORADO NACIONAL 21202412.
MG is supported by the Polish NCN MAESTRO grant 2014/14/A/ST9/00121.
MS is supported by generous grants from Villum FONDEN (13261,28021) and by a project grant (8021-00170B) from the Independent Research Fund Denmark.
L.W. is sponsored (in part) by the Chinese Academy of Sciences (CAS)  through a grant to the CAS South America Center for Astronomy (CASSACA) in Santiago, Chile. 
AMG acknowledges financial support from the 2014-2020 ERDF Operational Programme and by the Department of Economy, Knowledge, Business and University of the Regional Government of Andalusia through the FEDER-UCA18-107404 grant.
E.Co. acknowledges support from ANID project Basal AFB-170002.
S.Mo. acknowledges the financial support of the Magnus Ehrnrooth Foundation.
Research by S.V. is supported by NSF grants AST–1813176  and AST-2008108.

This paper is partially based on observations obtained under  the European supernova collaboration involved in ESO-NTT large programme 184.D-1140 led by Stefano Benetti and the ESO-NTT Large Program 188.D-3003 (the Public ESO Spectroscopic Survey for Transient Objects - PESSTO). 
Based on observations made with the Nordic Optical Telescope, operated by the Nordic Optical Telescope Scientific Association at the Observatorio del Roque de los Muchachos, La Palma, Spain, of the Instituto de Astrofisica de Canarias. Observations from the NOT were obtained through the NUTS and NUTS2 collaboration which are supported in part by the Instrument Centre for Danish Astrophysics (IDA). The data presented here were obtained in part with ALFOSC, which is provided by the Instituto de Astrofisica de Andalucia (IAA) under a joint agreement with the University of Copenhagen and NOTSA. The Liverpool Telescope is operated on the island of La Palma by Liverpool John Moores University in the Spanish Observatorio del Roque de los Muchachos of the Instituto de Astrofisica de Canarias with financial support from the UK Science and Technology Facilities Council.
The Italian Telescopio Nazionale Galileo (TNG) operated on
the island of La Palma by the Fundación Galileo Galilei of the INAF (Istituto Nazionale di Astrofisica) at the Spanish Observatorio del Roque de los Muchachos of the Instituto de Astrofísica de Canarias.
Based on observations obtained with the Gran  Telescopio Canarias (GTC), installed in the Spanish Observatorio del Roque de los Muchachos of the Instituto de Astrofisica de Canarias, in the island of La Palma.
Based on observations collected at Copernico and Schmidt telescope (Asiago, Italy) of the INAF - Osservatorio Astronomico di Padova; and the Galilei Telescope of the Padova University in Asiago.
Based on observations obtained at the international Gemini Observatory, a program of NSF’s NOIRLab, which is managed by the Association of Universities for Research in Astronomy (AURA) under a cooperative agreement with the National Science Foundation. on behalf of the Gemini Observatory partnership: the National Science Foundation (United States), National Research Council (Canada), Agencia Nacional de Investigaci\'{o}n y Desarrollo (Chile), Ministerio de Ciencia, Tecnolog\'{i}a e Innovaci\'{o}n (Argentina), Minist\'{e}rio da Ci\^{e}ncia, Tecnologia, Inova\c{c}\~{o}es e Comunica\c{c}\~{o}es (Brazil), and Korea Astronomy and Space Science Institute (Republic of Korea).
This work makes use of data from the Las Cumbres Observatory Network and the Global Supernova Project.  
This work is based on observations collected at the William Herschel  Telescope  (WHT),  operated  on  the  island  of  La  Palma by the Isaac Newton Group of Telescope. 
This paper includes data gathered with the 6.5 meter Magellan Telescopes located at Las Campanas Observatory, Chile.
Data for this study were gathered with the du Pont telescope at Las  Campanas Observatory, Chile.  
This publication makes use of data products from the Two Micron All Sky Survey, which is a joint project of the University of Massachusetts and the Infrared Processing and Analysis Center/California Institute of Technology, funded by the National Aeronautics and Space Administration and the National Science Foundation.
This work has made use of data from the Asteroid Terrestrial-impact Last Alert System (ATLAS) project. The Asteroid Terrestrial-impact Last Alert System (ATLAS) project is primarily funded to search for near earth asteroids through NASA grants NN12AR55G, 80NSSC18K0284, and 80NSSC18K1575; byproducts of the NEO search include images and catalogs from the survey area. This work was partially funded by Kepler/K2 grant J1944/80NSSC19K0112 and HST GO-15889, and STFC grants ST/T000198/1 and ST/S006109/1. The ATLAS science products have been made possible through the contributions of the University of Hawaii Institute for Astronomy, the Queen’s University Belfast, the Space Telescope Science Institute, the South African Astronomical Observatory, and The Millennium Institute of Astrophysics (MAS), Chile.
The Pan-STARRS1 Surveys (PS1) and the PS1 public science archive have been made possible through contributions by the Institute for Astronomy, the University of Hawaii, the Pan-STARRS Project Office, the Max-Planck Society and its participating institutes, the Max Planck Institute for Astronomy, Heidelberg and the Max Planck Institute for Extraterrestrial Physics, Garching, the Johns Hopkins University, Durham University, the University of Edinburgh, the Queen’s University Belfast, the Harvard-Smithsonian Center for Astrophysics, the Las Cumbres Observatory Global Telescope Network Incorporated, the National Central University of Taiwan, the Space Telescope Science Institute, the National Aeronautics and Space Administration under Grant No. NNX08AR22G issued through the Planetary Science Division of the NASA Science Mission Directorate, the National Science Foundation Grant No. AST-1238877, the University of Maryland, Eotvos Lorand University (ELTE), the Los Alamos National Laboratory, and the Gordon and Betty Moore Foundation.
The operation of Xingming Observatory was made possible by the generous support from the Xinjiang Astronomical Observatory of the Chinese Academy of Sciences.\\

{\sc iraf} was distributed by the National Optical Astronomy Observatory, which was managed by the Association of Universities for Research in Astronomy (AURA) under a cooperative agreement with the National Science Foundation. This research has made use of the NASA/IPAC Extragalactic Database (NED), which is operated by the Jet Propulsion Laboratory, California Institute of Technology, under contract with the National Aeronautics and Space Administration. 



\bibliographystyle{aa} 
\bibliography{ILRTs}

\begin{appendix}
\onecolumn

\section[]{Basic information for observational facilities used}
\label{sec:facilities}
We report the basic information for observational facilities in Table \ref{table_setup}, which were used for the five ILRTs.

\begin{table*}[h]
\centering
\caption{Information on the instrumental setups.}
\label{table_setup}
\scalebox{0.8}{
\begin{tabular}{@{}lllll@{}}
\hline \hline
Code & Diameter&Telescope & Instrument & Site \\
                & $\mathrm{m}$&            &                   & \\
\hline
QHY9           & 0.35 & Celestron C14 35-cm reflector   &      QHY-9                   & Xingming Observatory,  Xinjiang, China \\
EM01$^*$       & 0.40 & LCO (FTN site)              & EM01       & LCO node at Haleakala Observatory, Hawaii, USA \\
PROMPT3/5      & 0.41 &PROMPT Telescope & Apogee Alta & Cerro Tololo Inter-American Observatory, Cerro Tololo, Chile   \\
TRAPPIST       & 0.60 &TRAPPIST-S Telescope & FLI ProLine&       ESO La Silla Observatory, La Silla, Chile\\
Moravian  & 0.67/0.92 & Schmidt Telescope        &Moravian &  Osservatorio Astronomico di Asiago, Asiago, Italy\\
SBIG      & 0.67/0.92 & Schmidt Telescope        &  SBIG   &  Osservatorio Astronomico di Asiago, Asiago, Italy\\
fl03$^*$       & 1.00 &  LCO (LSC site) &   Sinistro    &LCO node at Cerro Tololo Inter-American Observatory, Cerro Tololo, Chile   \\
fl05$^*$       & 1.00 &  LCO (ELP site)        &   Sinistro  & LCO node at McDonald Observatory, Texas, USA  \\
fl06$^*$       & 1.00 &  LCO (CPT site)   &   Sinistro  & LCO node at South African Astronomical Observatory, Cape Town, South Africa \\
fl12$^*$       & 1.00 &   LCO (COJ site)   &   Sinistro  & LCO node at Siding Spring Observatory, New South Wales, Australia  \\
B$\&$C         & 1.22 & Galileo Telescope           & B$\&$C    &  Osservatorio Astronomico di Asiago, Asiago, Italy\\
ANDICAM-CCD    & 1.30 & SMARTS Telescope & ANDICAM &Cerro Tololo Inter-American Observatory, Cerro Tololo, Chile   \\
ANDICAM-IR     & 1.30 & SMARTS Telescope & ANDICAM &Cerro Tololo Inter-American Observatory, Cerro Tololo, Chile   \\
AFOSC          & 1.82 & Copernico Telescope    & AFOSC     &  Osservatorio Astronomico di Asiago, Asiago, Italy    \\
IO:O           & 2.00 & Liverpool Telescope          & IO:O     &  Observatorio Roque de Los Muchachos, La Palma, Spain\\
RATCam         & 2.00   & Liverpool Telescope            & RATCam & Observatorio Roque de Los Muchachos, La Palma, Spain \\
SupIRCam       & 2.00   & Liverpool Telescope            & SupIRCam & Observatorio Roque de Los Muchachos, La Palma, Spain \\ 
WFCCD          & 2.50 & Ir$\rm{\acute{e}}$n$\rm{\acute{e}}$e du Pont Telescope & WFCCD & Las Campanas Observatory, Atacama Region, Chile \\
ALFOSC         & 2.56 & Nordic Optical Telescope & ALFOSC  &   Observatorio Roque de Los Muchachos, La Palma, Spain\\
NOTCam         & 2.56 & Nordic Optical Telescope & NOTCam &  Observatorio Roque de Los Muchachos, La Palma, Spain\\
EFOSC2         & 3.58 & New Technology Telescope& EFOSC2 & ESO La Silla Observatory, La Silla, Chile\\
SOFI           & 3.58 & New Technology Telescope& SOFI  & ESO La Silla Observatory, La Silla, Chile\\   
LRS            & 3.58 & Telescopio Nazionale Galileo & LRS  &  Observatorio Roque de Los Muchachos, La Palma, Spain\\
NICS           & 3.58 & Telescopio Nazionale Galileo & NICS   &  Observatorio Roque de Los Muchachos, La Palma, Spain\\
ISIS           & 4.20 & William Hershel Telescope    & ISIS    &  Observatorio Roque de Los Muchachos, La Palma, Spain\\
LDSS           & 6.50 & Magellan II - Clay Telescope &LDSS & Las Campanas Observatory, Atacama Region, Chile \\
GMOS           & 8.10 & Gemini South Telescope       & GMOS-S & Gemini Observatory, Cerro Pachon, Chile\\
OSIRIS         & 10.40& Gran Telescopio CANARIAS     & OSIRIS  & Observatorio Roque de Los Muchachos, La Palma, Spain\\
\hline \hline
\end{tabular}
}
\medskip
\\ 
\begin{flushleft}
$*$ They are distributed globally at different sites and form part of the LCO global telescope network \citep{Brown2013PASP..125.1031B}.      
\end{flushleft}
\end{table*}

\section{Reddenings and distances adopted for SNe II}\label{appendix:Dist_Reden}
In this section, the reddenings and distances adopted for the comparison SNe II are reported in Table  \ref{Dist_Reden_II}, along with their respective references.

\begin{table*}[h]
	\caption{Reddenings and distances of SN 2018zd, SN~2018hwm, SN~2015bf, SN~1999br, and SN~2005cs.  }
	\label{Dist_Reden_II}
	\scalebox{0.7}{
		\begin{tabular}{@{}cccccc@{}}
			\hline
			
			\hline
			Object & Distance & Distance Modulus  &   $E(B-V)_{\mathrm{Gal}}$ &$E(B-V)_{\mathrm{Host}}$   & Sources \\ 
		       	   & ($\mathrm{Mpc}$)    &  ($\mathrm{mag}$)       & ($\mathrm{mag}$)            &($\mathrm{mag}$) & Codes\\ 
			\hline
			
			\hline
			SN 2018zd  & 18.4 (4.5) & 31.32 (0.48)  &   0.085    & 0.085  &1    \\
			SN~2018hwm & 52 (5)     & 33.58 (0.19) &   0.003     & 0      &2  \\
			SN~2015bf  & 60.1 (1.4) & 33.89 (0.05) &   0.059     & 0.089  &3   \\
			SN~1999br$^{a}$ & 15.63       & 30.97        &  0.021      &  0     &4 \\
			SN~2005cs &  8.4 (0.5)       & 29.62 (0.15) &  0.031     &  0.032  &5     \\
			\hline
			
			\hline
			
		\end{tabular}
	}
	\medskip
	
	$Notes$: References for each object are labelled by numbers in the last column. The uncertainties are reported in the parentheses.\\
	1 = \citet{Zhang2020MNRAS.498...84Z}; 2 = \citet{Reguitti2021MNRAS.501.1059R}; 3 = \citet{Lin2021MNRAS.505.4890L}; 4 = \citet{Pastorello2004MNRAS.347...74P}; 5 = \citet{Pastorello2006MNRAS.370.1752P}.  \\
	$^{a}$ The distance has been rescaled to $H_{0}=73$ \kms $\rm{Mpc}^{-1}$. The reddening has been updated to the value in \citet{Schlafly2011ApJ...737..103S}.   
\end{table*}

\newpage
\section[]{Photometric data of ILRTs}
\label{sec:spectphot}
\begin{figure*}[htp]
	\centering
	\includegraphics[width=.80\textwidth]{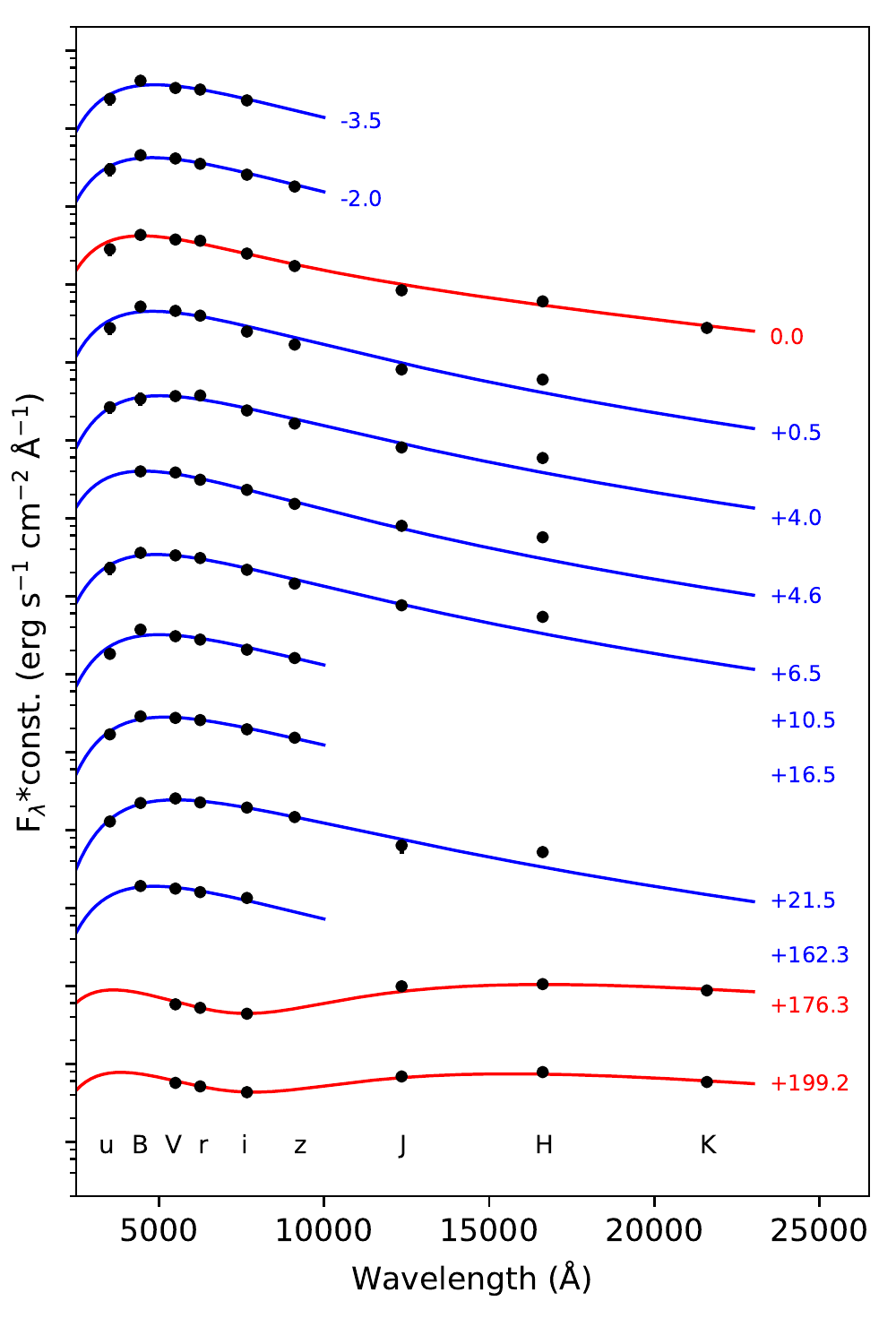}\quad
	\caption{SED evolution of AT 2010dn. The lines are the best-fit BBs, which are overplotted on each SED. Blue lines are the fits of the single BB, while red lines are the fits of the double BB.    The epochs marked to the right of each SED are relative to the $r$-band maximum. Each SED has been sifted by an arbitrary constant for clarity.}
	\label{figSED2010dn}
\end{figure*}

The photometric data of five ILRTs and associated errors are reported in this section. In addition, black body parameters and SED evolution of individual ILRT AT~2010dn are presented in Table \ref{AT2010dnSed}~and Figure \ref{figSED2010dn}, respectively. Our observations will be made public via the Weizmann Interactive Supernova Data Repository  \citep[WISeREP; ][]{Yaron2012PASP..124..668Y}.

\begin{table*}[h]
\begin{minipage}{175mm}                                  
\caption{Optical ($BVuriz$) photometric measurements of AT~2010dn.}
\label{2010dn_opt_LC}  
\scalebox{0.9}{
\begin{tabular}{@{}ccccccccccl@{}}
\hline
Date         &     MJD       &Phase$^{a}$&  $B$(err)  &  $V$(err)   &  $u$(err)    &  $r$(err)$^{bc}$  &  $i$(err)$^{b}$   & $z$(err)   & Instrument \\
\hline
20010302& 51970.50&-3384.4 & --         & $>$24.1        & --           &  --           &  --        & --            & 1      \\
20090330& 54921.11&-433.8 & --& --& --                   &   $>$19.0      & --          & --                     & 2        \\
20100521& 55338.00&-16.9   & --& --& --                     &   $>$18.0      & --         & --        & 2 \\    
20100531& 55348.04& -6.9   & --& --& --                    & 17.546(0.123) & --         & --      &   2 \\ 
20100601& 55348.98& -5.9   & --& --& --                    & 17.407(0.211)  & --         & --      &   2 \\ 
20100601& 55349.45&-5.4     & --& --& --                    &  17.593(0.172)  & --         & --      & ALFOSC\\
20100602& 55350.36&-4.5   & --& --& --                     &  17.41                 & --         & --       & 3     \\
20100603& 55350.65&-4.2   & -- &  17.4  &   --           & --                    & --         & --          & 4  \\ 
20100603& 55350.99& -3.9   & --& --& --                    & 17.412(0.266)    & --         & --      &   2 \\ 
20100603& 55351.40&-3.5   &  18.036(0.028)&  17.649(0.051) & 18.780(0.105) &   17.465(0.040)    & 17.379(0.043)         & --      &LRS	   \\
20100605& 55352.85&-2.0   & 17.923(0.036)&  17.411(0.044) &18.700(0.206) &    17.346(0.040) & 17.260(0.036) & 17.252(0.065) &  EM01   \\
20100607& 55355.42&+0.5 &  17.781(0.046)&  17.298(0.047) &   18.677(0.080) &  17.215(0.060)    &   17.292(0.040)         & --      &LRS	   \\
20100609& 55357.40&+2.5 &  -- &  --  &  --    &    17.264(0.167)  &  --           &  --           &  ALFOSC \\
20100611& 55359.40&+4.5 &  18.102(0.034)&  17.584(0.017) &  --   &    17.477(0.037) & 17.370(0.017) &  --           &  RATCam \\
20100611& 55359.43&+4.5 &  18.066(0.047)&  17.481(0.092) & --   &  17.424(0.086)   &  17.347(0.052)         & --      &LRS	   \\
20100612& 55360.41&+5.5 &  18.074(0.016)&  17.607(0.029) &  --  &    17.498(0.018) & 17.386(0.038) &  --           &  RATCam \\
20100613& 55361.45&+6.6  &  18.177(0.021)&  17.638(0.025) & 18.990(0.207)   &    17.490(0.032) & 17.430(0.039) & 17.486(0.073) &  RATCam \\
20100616& 55363.84&+8.9 &  18.141(0.053)&  17.663(0.031) & 19.158(0.358)   &    17.552(0.034) & 17.467(0.038) & 17.366(0.048) &  EM01   \\
20100616& 55364.39&+9.5 &  --  &  -- &  --  &    17.669(0.248) &   --          &   --          &  LRS    \\
20100617& 55365.40&+10.5 &  18.139(0.028)&  17.730(0.025) & 19.237(0.114)   &    17.605(0.057) & 17.495(0.040) & 17.376(0.060) &  RATCam \\
20100618& 55366.38&+11.5 &  18.294(0.029)&  17.790(0.028) &    --          &    17.607(0.031) & 17.486(0.039) &    --                 &  RATCam \\
20100620& 55367.82&+12.9 &  18.261(0.041)&  17.701(0.033) & 19.233(0.148)   &    17.700(0.040) & 17.442(0.037) & 17.378(0.068) &  EM01   \\
20100621& 55368.79&+13.9 &  18.325(0.051)&  17.752(0.025) & 19.186(0.109)   &    17.650(0.049) & 17.550(0.046) & 17.432(0.072) &  EM01   \\
20100623& 55370.80&+15.9 &  18.367(0.096)&  17.797(0.090) &     --          &    17.665(0.048) & 17.419(0.072) & 17.313(0.088) &  EM01   \\
20100623& 55371.41&+16.5 &  18.422(0.022)&  17.850(0.033) & 19.322(0.137)   &    17.687(0.049) & 17.550(0.039) & 17.428(0.056) &  RATCam \\
20100625& 55373.41&+18.5 &  18.449(0.039)&  17.907(0.036) & 19.405(0.160)   &    17.619(0.058) & 17.504(0.039) & 17.445(0.058) &  RATCam \\
20100626& 55373.80&+18.9 &  18.468(0.089)&  17.810(0.049) &  --             &    17.667(0.029) & 17.574(0.046) & 17.395(0.074) &  EM01   \\
20100627& 55374.80&+19.9 &  18.489(0.054)&  17.930(0.044) &  --             &    17.854(0.047) & 17.637(0.043) & 17.374(0.073) &  EM01   \\
20100628& 55375.79&+19.9 &  -- &  -- &  --  &   --  &    --    & 17.465(0.174) &  EM01   \\
20100628& 55376.39&+21.5 &  18.699(0.052)&  17.934(0.019) & 19.612(0.123)   &    17.828(0.012) & 17.558(0.044) & 17.473(0.062) &  RATCam \\
20100629& 55377.42&+22.5 &  18.658(0.030)&  17.922(0.027) & 19.559(0.130)   &    17.799(0.010) & 17.553(0.027) & 17.466(0.115) &  RATCam \\	
20101117& 55517.18&+162.3 &  22.555(0.101)&  22.019(0.040) & -- &  21.461(0.066)  & 20.618(0.053)  & --                    &LRS	   \\	
20101213& 55543.08&+188.2  &  --   &  22.057(0.116) & --  &  21.485(0.132)  &   20.679(0.075)     & --      &LRS	   \\	
20101231& 55561.20&+206.3 &  --  &  22.053(0.149)  & --   &  21.494(0.101)  &   20.721(0.125)   & --      &LRS	   \\	
20110105& 55566.14&+211.3 &  --  &  --   & --   &  21.542(0.036)    &   --    & --      &LRS	   \\	
20111228& 55923.11&+568.2  &  $>$23.0   &  $>$22.9    & --   &  $>$22.8    &  --    & --   & AFOSC 	\\
\hline
\end{tabular}   
}
\medskip
\\
$^a$ Phases are relative to the $r$-band maximum light: MJD=55354.9.\\ 
$^b$ Johnson-Cousins $R$ and $I$ filter data were converted to Sloan $r$ and $i$ band magnitudes respectively, following the relations of \citet{Jordi2006A&A...460..339J} \\ 
$^c$ Unfiltered data obtained by amateur astronomers, calibrated to Sloan $r$-band magnitudes. \\ 
$1$ This approximation is based on the similar throughputs between wide filters F555W/F814W and $V$/$I$ . Original data published by \citet{Berger2010ATel.2655....1B}, and reported to Johnson-Cousins $V$ and $I$ ($I>$24.9~$\mathrm {mag}$) magnitudes.  \\
$2$ Amateur data (K. Itagaki) scaled to $r$ band  . \\
$3$ Bright Supernova website (http://www.rochesterastronomy.org/sn2010/index.html\#2010dn; Observation: J. Nicolas). \\
$4$ Bright Supernova website (http://www.rochesterastronomy.org/sn2010/index.html\#2010dn; Observation: J. Brimacombe). \\

\end{minipage}
\end{table*}

\begin{table*}[h]
\begin{minipage}{175mm}
\caption{NIR ($JHK$) photometric measurements of AT~2010dn. }
\label{2010dn_nir_LC}
\begin{tabular}{@{}ccccccl@{}}
\hline
Date & MJD & Phase$^a$ & $J$(err) & $H$(err) & $K$(err) & Instrument \\
\hline
20100606& 55353.89&-1.0 & 16.611(0.067)& 15.690(0.065) & --            &  SupIRCam\\
20100607& 55354.89&+0.0 & 16.412(0.133)& 15.704(0.053) & 15.278(0.068) &  NICS	\\
20100607& 55354.96&+0.1 & 16.490(0.091)& 15.703(0.119) & --            &  SupIRCam\\
20100611& 55358.89&+4.0 & 16.491(0.064)& 15.753(0.066) & --            &  SupIRCam\\
20100614& 55361.90&+7.0 & 16.564(0.095)& 15.830(0.091) & --            &  SupIRCam\\
20100627& 55374.89&+20.0 & 16.744(0.340)& 15.838(0.082) & --            &  SupIRCam\\
20100629& 55376.90&+22.0 & 16.758(0.192)& 15.871(0.112) & --            &  SupIRCam\\
20101201& 55531.22&+176.3 & 18.974(0.167)& 17.774(0.152) &  16.868(0.128) & NICS	\\
20101224& 55554.14&+199.2 & 19.341(0.127)& 18.053(0.141) & 17.465(0.101)  &  NICS \\
\hline
\end{tabular}

\medskip
$^a$ Phases are relative to the $r$-band maximum light: MJD=55354.9.\\  
\end{minipage} 
\end{table*}

\begin{table*}
\begin{minipage}{175mm}
	\caption{{\it Spitzer} IRAC photometry of AT~2010dn (AB mag). Uncertainties are given in parentheses. }
	\label{table:SpitzerMag}
	{
		\scalebox{1.0}{
			\begin{tabular}{@{}lcccc@{}}
				\hline
				
				\hline
				Date &MJD  &Phase$^a$  & CH1 (3.6 $\mu$m) &  CH2 (4.5 $\mu$m) \\
				&        &  (d)        & (mag)   & (mag) \\
				
				\hline
				20100708 & 55385.5 & $+30.6 $  & 16.88 (0.01) &16.76 (0.01)\\    
				20101230 & 55560.6 & $+205.7$  & 18.24 (0.03) &17.94 (0.02) \\
				20110203 & 55595.7 & $+240.8$  & 18.04 (0.03) &17.78 (0.01) \\
				20120114 & 55940.8 & $+585.9$  & 18.33 (0.03) &18.04 (0.02)\\
				20120616 & 55094.9 & $+740.0$  & 18.83 (0.05) &18.30 (0.02)\\
				20140130 & 56687.4 & $+1332.5$ & 21.39 (0.45) &20.09 (0.12)\\
				20140303 & 56719.1 & $+1364.2$ & 20.99 (0.33) &20.00 (0.11)   \\
				20140703 & 56841.0 & $+1486.1$ & 22.12 (0.76) &20.25 (0.13)\\
				20140706 & 56844.5 & $+1489.6$ & 21.76 (0.59) &20.24 (0.13)  \\
				20150211 & 57064.9 & $+1710.0$ & 22.44 (0.93) &21.12 (0.28)  \\
				\hline		
				
				\hline
			\end{tabular}
		}
	}
	\medskip
	
	\begin{flushleft} 
		$^a$ Phases are relative to the $R/r$-band maximum (MJD=55354.9).\\ 
	\end{flushleft}

\end{minipage}	
\end{table*}




 
\begin{table*}
	\caption{Parameters of blackbody fit to the uBVrizJHK bands of AT~2010dn. Uncertainties are given in parentheses.}
	\label{AT2010dnSed}
	\begin{tabular}{@{}cccccc@{}}
		\hline
		Date & MJD & Phase$^a$  &   Luminosity (hot)   &  Temperature (hot)    &    Radius (hot)    \\ 
		&          & (d)       &  (10$^{39}$erg~s$^{-1}$) &  (K)               &    (10$^{13}$~cm)    \\ 
		\hline
		20100603 & 55351.40 & $-3.5$   &  66.6 (28.0)     &  6350 (320)   &   24.0 (5.1)        \\
		20100605 & 55352.85 & $-2.0$  &   75.5 (10.1)    & 6405 (100)    &   25.1 (1.7)    \\
		20100607$^b$ & 55354.89 & $0.0$  &   75.2 (37.5)    & 6490 (1200)    &   24.4 (12.2)    \\	
		20100607 & 55355.42 & $+0.5$  &   78.8 (34.5)    & 6460 (360)    &   25.2 (5.5)    \\
		20100611 & 55358.89 & $+4.0$  &   72.4 (26.9)    & 6290 (295)    &   25.5 (4.7)    \\
		20100611 & 55359.40 & $+4.5$  &   69.7 (32.4)    & 6035 (350)    &   27.1 (6.3)    \\
		20100613 & 55361.45 & $+6.6$  &   63.7 (17.8)    & 6075 (200)    &   25.6 (3.6)    \\	
		20100617 & 55365.40 & $+10.5$  &   59.6 (26.2)    & 6090 (310)    &   24.6 (5.4)    \\
		20100623 & 55371.41 & $+16.5$  &   54.3 (9.5)    & 5795 (110)    &   26.0 (2.3)    \\		
		20100628 & 55376.39 & $+21.5$  &   48.8 (10.7)    & 5390 (135)    &   28.5 (3.1)    \\
		20101117 & 55517.18 & $+162.3$  &   3.5 (5.1)    & 3380 (420)    &   19.3 (14.2)    \\		
		20101201$^b$ & 55531.22 & $176.3$  &   3.2 (1.6)    & 3420 (170)    &   18.2 (9.1)    \\
		20101224$^b$ & 55554.14 & $199.2$  &   2.8 (1.4)    & 3540 (280)    &   15.9 (8.0)    \\
		\hline
	\end{tabular}
	\medskip
	
	$^a$ Phases are relative to the $r$-band maximum light curve (MJD=55354.9).\\ 
	$^b$ Data are obtained by fitting a double black body function, but only the first hot-component parameters are reported.  The second warm component has large uncertainties because its data are incompletely sampled.\\

\end{table*}

\clearpage
\begin{table*}
\begin{minipage}{175mm}                                  
\caption{Optical ($UBVRI$) photometric measurements of AT 2012jc.}
\label{NGC5775OT_opt_LC}  
\scalebox{0.95}{
\begin{tabular}{@{}cccccccccl@{}}
\hline
Date       &     MJD   &Phase$^{a}$& $U$(err) &  $B$(err) &  $V$(err)   & $R$(err)   & $I$(err)            & Instrument \\
\hline
20120313& 55999.27&-24.2  & --         &   --          &   --                & $>$19.9          &  --            &  PROMPT3/5$^{b}$\\
20120317& 56003.39&-20.1  & --            &   --         & $>$19.5        &     --               &  --            &  1\\
20120327& 56013.49&-10.0  & --            &   --         &  18.7             &    --                 &  --            &  1\\
20120328& 56014.07&-9.4  & --            &   --          &  18.5            &   --                   &  --            &  1\\
20120329& 56015.06&-8.4  & 20.308(0.114) & 18.823(0.022) & 18.206 (0.027) & 17.855 (0.031) & 17.460 (0.058) &  RATCam    \\
20120329& 56015.14&-8.4  & --            & 18.744(0.053) & 18.259 (0.044) & 17.924 (0.034) & 17.521 (0.129) &  AFOSC     \\
20120329& 56015.35&-8.2  & --            &   --          &  18.0 (0.1)       &     --        &  --            &  2\\
20120331& 56017.26&-6.2  & --            & 18.390(0.113) & 17.888 (0.051) & 17.503 (0.064) & 17.129 (0.056) &  PROMPT3   \\
20120412& 56029.24&+5.7  & --            &  --           & 17.776 (0.069) & 17.403 (0.057) & 17.102 (0.045) &  PROMPT3/5  \\
20120414& 56031.37&+7.9  & --            & 18.422(0.020) & 17.901 (0.020) & 17.564 (0.013) & 17.164 (0.029) &  EFOSC2     \\
20120415& 56032.41&+8.9  & --            &  --           &  --            & 17.614 (0.062) &  --            &  PROMPT3/5  \\
20120416& 56033.24&+9.7  & --            &  --           & 17.904 (0.055) & 17.561 (0.059) & 17.239 (0.049) &  PROMPT3/5  \\
20120416& 56033.97&+10.5  &19.391(0.093)  & 18.494(0.028) & 17.902 (0.029) & 17.552 (0.032) & 17.285 (0.112) &  RATCam    \\
20120417& 56034.05&+10.6  & --            & 18.443(0.034) & 17.947 (0.030) & 17.591 (0.017) & 17.287 (0.055) &  AFOSC     \\
20120417& 56034.19&+10.7  & --            &  --           & 17.900 (0.069) & 17.634 (0.056) & 17.201 (0.067) &  PROMPT3/5  \\
20120421& 56038.25&+14.8  & --            & 18.694(0.062) & 18.064 (0.049) & 17.825 (0.084) & 17.445 (0.057) &  PROMPT3/5  \\
20120422& 56039.01&+15.5  &19.917(0.109)  & 18.598(0.024) & 18.117 (0.025) & 17.856 (0.031) & 17.511 (0.053) &  RATCam    \\
20120425& 56042.11&+18.6  & --            &  --           & 18.478 (0.109) & 18.094 (0.168) & 17.710 (0.126) &  SBIG      \\
20120501& 56048.17&+24.7  & --            & 19.011(0.084) & 18.436 (0.080) & 18.083 (0.055) & 17.649 (0.071) &  PROMPT3/5  \\
20120501& 56048.20&+24.7  & --            & 19.155(0.013) & 18.479 (0.018) & --             &  --            &  ALFOSC\\
20120501& 56048.38&+24.9  & --            & 19.064(0.031) & 18.487 (0.032) & 18.112 (0.029) & 17.696 (0.074) &  EFOSC2     \\
20120502& 56049.21&+25.7  & --            & 19.134(0.113) & 18.485 (0.088) & 18.073 (0.058) & 17.647 (0.094) &  PROMPT3/5  \\
20120509& 56056.10&+32.6 & --            & 19.386(0.033) & 18.595 (0.035) & 18.154 (0.076) & 17.700 (0.088) &  LRS	      \\
20120515& 56062.11&+38.6   &21.101(0.212)  & 19.560(0.039) & 18.743 (0.044) & 18.292 (0.050) & 17.701 (0.226) &  RATCam    \\
20120515& 56062.24&+38.7  & --            &  --           & 18.707 (0.068) & 18.226 (0.072) & 17.647 (0.042) &  PROMPT3/5  \\
20120520& 56067.18&+43.7  & --            &  --           &   --           &     --         & 17.788 (0.052) &  PROMPT3/5  \\
20120521& 56068.17&+44.7  & --            & 19.710(0.132) & 18.807 (0.068) & 18.325 (0.078) & 17.819 (0.035) &  PROMPT3/5  \\
20120601& 56079.97&+56.5  & $>$21.3         & 20.142(0.111) & 19.217 (0.051) & 18.699 (0.058) & 18.089 (0.057) &  RATCam    \\
20120603& 56081.11&+57.6  & --            &   --          &  --            & 18.804 (0.154) &  --            &  PROMPT3/5$^{b}$  \\
20120603& 56081.96&+58.5  & $>$21.3         & 20.196(0.148) & 19.263 (0.093) & 18.741 (0.040) & 18.115 (0.079) &  RATCam    \\
20120620& 56098.24&+74.7  & --            & 20.501(0.148) & 19.512 (0.101) & $>$19.7          &   --           &  TRAPPIST  \\
\hline
\end{tabular}   
}
\medskip
$^a$ Phases are relative to the $R$-band maximum light: MJD=56023.5.  \\ 
$^b$ These unfiltered data were reported to Johnson-Cousins $R$ magnitudes, as suggested by the quantum efficiency of the CCD.  \\
$1$ Observations from The Astronomer's Telegram \citep[ATel, No. 4004; see][]{Howerton2012ATel.4004....1H}.\\
$2$ Bright Supernova website (http://www.rochesterastronomy.org/sn2012/index.html\#PSNJ14535395+0334049; Observation: J. Brimacombe) 
\end{minipage}
\end{table*}


\clearpage
\begin{table*}
\begin{minipage}{175mm}                                  
\caption{Optical ($UBVRI$) photometric measurements of AT 2013lb.}
\label{NGC5917OT_opt_LC}  
\scalebox{0.9}{
\begin{tabular}{@{}cccccccccl@{}}
\hline
Date         &     MJD       &Phase$^{a}$& $U$(err) &  $B$(err)         &  $V$(err)          & $R$(err)            & $I$(err)            &  Instrument \\
\hline
20120624& 56102.05& -222.4& --          &  --            &  --           & $>$19.2     &  --          &   PROMPT35$^b$   \\
20130127& 56319.34&-5.2 & --          &  --            &  --           & 18.897(0.138) &  --          &   PROMPT35$^b$	  \\
20130131& 56323.34&-1.2 & --          &  --            &  --           & 18.728(0.099) &  --          &   PROMPT35$^b$	  \\
20130204& 56327.36&+2.9 & --          &  --            &  --           & 18.601(0.143) &  --          &   PROMPT35$^b$	  \\
20130204& 56327.41&+2.9 & --          &  --            &  --           & $>$18.3          &  --          &   1	  \\
20130207& 56330.20&+5.7 & --          &  --            & 19.132(0.144) & 18.833(0.078) & 18.422(0.090)&   AFOSC	  \\
20130207& 56330.39&+5.9 & --          &  --            & 19.143(0.035) &  --           &  --          &   EFOSC2	  \\
20130208& 56331.23&+6.7 & --          &  --            &  --           & 18.859(0.029) & 18.450(0.038)&   RATCam	  \\
20130209& 56332.28&+7.8 & --          &  --            & 19.237(0.035) & --            &    --        &   EFOSC2	  \\
20130211& 56334.23&+9.7 & --          &  --            &  --           & 18.998(0.030) & 18.466(0.035)&   RATCam	  \\
20130215& 56338.28&+13.8 & --          &  19.878(0.046) & 19.368(0.029) & 18.882(0.043) & 18.455(0.064)&   ANDICAM-CCD\\
20130218& 56341.23&+16.7&19.541(0.160)&  19.997(0.060) &  --           & 19.005(0.037) & 18.415(0.038)&   RATCam	  \\
20130218& 56341.32&+16.8&  --         &  19.963(0.047) & 19.444(0.034) & 18.888(0.062) & 18.346(0.068)&   ANDICAM-CCD\\
20130220& 56343.30&+18.8 &  --         &  19.937(0.042) & 19.350(0.031) & 19.050(0.031) & 18.474(0.042)&   ANDICAM-CCD\\
20130221& 56344.22&+19.7 & $>$19.9   &  19.998(0.046) & 19.307(0.034) & 18.987(0.050) & 18.455(0.040)&   RATCam	  \\
20130223& 56346.20&+21.7 & $>$19.9   &  20.084(0.110) & 19.413(0.046) & 18.937(0.078) & 18.458(0.064)&   RATCam	  \\
20130225& 56348.14&+23.6 & $>$19.9   &  20.188(0.178) & 19.385(0.082) & 19.001(0.088) & 18.550(0.075)&   RATCam	  \\
20130227& 56350.30&+25.8 &        --   &  20.226(0.124) & 19.510(0.093) &  --           & --           &   ANDICAM-CCD\\
20130301& 56352.32&+27.8 &        --   &   --           & --            & 19.111(0.071) & 18.695(0.087)&   ANDICAM-CCD\\
20130305& 56356.25&+31.8 &        --   &   --           & --            & 19.113(0.054) & --           &   EFOSC2$^b$ 	  \\
20130305& 56356.26&+31.8 &        --   &  20.367(0.077) & 19.616(0.056) & 19.167(0.045) & 18.558(0.065)&   ANDICAM-CCD\\
20130310& 56361.22&+36.7 &        --   &  $>$20.4          & $>$19.6     & 19.115(0.226) & 18.496(0.293)&   RATCam	  \\
20130311& 56362.26&+37.8 &        --   &  20.509(0.052) & 19.618(0.052) & 19.076(0.050) & 18.594(0.063)&   ANDICAM-CCD\\
20130312& 56363.09&+38.6 & $>$19.9   &  20.590(0.080) & 19.688(0.051) & 19.143(0.054) & 18.625(0.051)&   RATCam	  \\
20130315& 56366.28&+41.8 &   --        &  20.689(0.051) & 19.857(0.057) & 19.082(0.038) & 18.755(0.043)&   ANDICAM-CCD\\
20130318& 56369.07&+44.6 &$>$20.0    &  20.850(0.138) & 19.802(0.071) & 19.284(0.075) & 18.675(0.090)&   RATCam	  \\
20130320& 56371.33&+46.8 &  --         &  20.949(0.060) & 19.984(0.063) & 19.257(0.030) & 18.735(0.084)&   ANDICAM-CCD\\
20130321& 56372.11&+47.6 &$>$20.1    &  20.921(0.088) & 19.991(0.049) & 19.295(0.067) & 18.721(0.057)&   RATCam	  \\
20130324& 56375.29&+50.8 &        --   &  21.018(0.120) & 20.111(0.094) & 19.484(0.056) & 18.634(0.071)&   ANDICAM-CCD\\
20130328& 56379.26&+54.8 &        --   &  21.277(0.358) & 19.913(0.131) & 19.469(0.159) & 18.610(0.125)&   ANDICAM-CCD\\
20130401& 56383.12&+58.6 &        --   &  $>$21.2     & 20.089(0.195) & 19.386(0.202) & 18.656(0.130)&   RATCam	  \\
20130407& 56389.07&+64.6 &        --   &  21.584(0.195) & 20.280(0.083) & 19.682(0.085) & 19.049(0.059)&   RATCam	  \\
20130409& 56392.00&+67.5 &        --   &  21.746(0.225) & 20.476(0.097) & 19.863(0.104) & 18.963(0.102)&   RATCam	  \\
20130414& 56396.30&+71.8 &        --   &  21.839(0.028) & 20.794(0.046) & 20.054(0.037) & 19.391(0.063)&   EFOSC2    \\
20150214& 57067.31& +742.8&        --   &   --           & --            & $>$19.8     & --           &   PROMPT35$^b$  \\

\hline
\end{tabular}   
}

\medskip
$^a$ Phases are relative to the $R$-band maximum light: MJD=56324.5.\\ 
$^b$ Unfiltered data  reported to broad-band Johnson-Cousins $R$ magnitudes. \\    
$1$ This limit was obtained by L. Elenin (Lyubertsy, Russia) and I. Molotov (Moscow, Russia) through a 0.45-m f/2.8 telescope plus KAF09000 CCD, which is also reported in the Central Bureau for Astronomical Telegrams (CBAT, see http://www.cbat.eps.harvard.edu/unconf/followups/J15213475-0722183.html). \\
\end{minipage}
\end{table*} 

\begin{table*}
\begin{minipage}{175mm}
\caption{NIR ($JHK$) photometric measurements of AT 2013lb. }
\label{NGC5917OT_nir_LC}
\begin{tabular}{@{}ccccccl@{}}
\hline
Date & MJD & Phase$^a$ & $J$(err) & $H$(err) & $K$(err) & Instrument \\
\hline
20130208& 56331.29& +6.8 &17.700(0.038) & 17.239(0.039) & 16.761(0.047) &  SOFI \\
20130218& 56341.30& +16.8 &17.967(0.165) &  --                    &  --                    &  ANDICAM-IR  \\
20130220& 56343.29& +18.8 &17.988(0.184) &  --                    &  --                    &  ANDICAM-IR  \\
20130222& 56345.36& +20.9 &17.977(0.096) & 17.339(0.039) & 16.975(0.052) &  SOFI \\
20130227& 56350.29& +25.8 &17.947(0.161) & --                     & --                     &  ANDICAM-IR \\
20130301& 56352.30& +27.8 &17.936(0.179) & --                     & --                    &  ANDICAM-IR \\
20130305& 56356.25& +31.8 &18.036(0.145) &--                      &--                     &  ANDICAM-IR \\
20130306& 56357.31& +32.8 &18.002(0.096) & 17.505(0.101) & 16.981(0.060) &  SOFI \\
20130311& 56362.25& +37.8 &18.137(0.205) &--                      &--                     &  ANDICAM-IR  \\
20130315& 56366.27& +41.8 &18.070(0.135) &--                      &--                    &  ANDICAM-IR  \\
20130320& 56371.32& +46.8 &18.004(0.153) &--                     &--                     &  ANDICAM-IR  \\
20130324& 56375.27& +50.8 &18.068(0.136) &--                     &--                     &  ANDICAM-IR \\
20130328& 56379.25& +54.8&18.182(0.215) &--                      &--                    &  ANDICAM-IR \\
20130412& 56394.33& +69.8&18.292(0.081) & 17.705(0.067) & 17.362(0.065) &  SOFI \\
\hline
\end{tabular}

\medskip
$^a$ Phases are relative to the $R$-band maximum light: MJD=56324.5.\\ 
\end{minipage}
\end{table*}

\clearpage

\clearpage

\begin{table*}
\begin{minipage}{175mm}                                  
\caption{Optical ($BVugriz$) photometric measurements of AT 2013la.}
\label{UGC8246OT_opt_LC}  
\scalebox{0.9}{
\begin{tabular}{@{}ccccccccccl@{}}
\hline
Date         &     MJD       &Phase$^{a}$ &  $B$(err) &  $V$(err)         & $u$(err) &  $g$(err)         &  $r$(err)$^b$        &  $i$(err)$^b$          & Instrument \\
20130331 &56382.71&-269.3 & --   &   --  & --   &   --  &  $>$18.8         &  --      &  QHY9$^c$ \\
20131220 &56646.93&-5.4    &  --   &   --  &--   &  --    & 18.246(0.251)  & --      & QHY9$^c$ \\
20131226 &56652.96& +1.0  & --   &   --  &--    &  --     &17.992(0.230)   &--      & QHY9$^c$  \\
20131227 &56653.93 &+1.9  & --   &   --  &--    &  --    & 17.964(0.180)   &--      & QHY9$^c$ \\
20131229 &56655.87 &+3.9  &--   &   --  &--     &  --    & 18.062(0.165)   & --     & QHY9$^c$ \\
20131230 &56656.87&  +4.9 & --   &   --  &--    &  --     & $>$17.8           & --     &  QHY9$^c$ \\
20140101& 56658.21&+6.2 &    19.393(0.035) & 18.634(0.013) &   20.840(0.186) &  --           & 18.314(0.011) & 18.101(0.010)  &   IO:O    \\
20140102& 56659.16&+7.2 &  19.386(0.045) & 18.691(0.030) & 20.519(0.218) &  --           & 18.351(0.015) & 18.133(0.016)  &   IO:O    \\
20140104& 56661.26&+9.3 &  19.373(0.045) & 18.702(0.014) & 20.227(0.136) &  --           & 18.360(0.013) & 18.202(0.012)  &   IO:O    \\
20140107& 56664.26&+12.3 &  19.522(0.034) & 18.763(0.051) & 20.160(0.108) &  --          & 18.490(0.039) & 18.293(0.121)  &   IO:O   \\
20140107& 56664.84& +12.8 &  --                  &   --                     & --                   &  --             &  18.467(0.396) & --                   & QHY9$^c$   \\
20140108& 56665.16&+13.2 &   19.432(0.112) & 18.774(0.055) &  $>$19.6       &  --             & 18.511(0.069) & 18.208(0.044)  &   IO:O   \\
20140108& 56665.23&+13.2  &  --                    & 18.811(0.180)  &    --              &  --             & 18.417(0.218)  &  --                  & AFOSC	 \\
20140109& 56666.96& +15.0 &  --               &  --             &--               &  --             &18.501(0.224)    &  --                     & QHY9$^c$\\
20140111& 56668.88 & +16.9  &  --               &  --             & --               &  --             & 18.529(0.213)  &  --                   & QHY9$^c$\\
20140115& 56672.15&+21.1 &  19.559(0.066) & 18.888(0.073) & 20.093(0.211) &  --           & 18.555(0.028) & 18.404(0.022)  &   IO:O   \\
20140122& 56679.14&+27.1  &   19.650(0.085) & 19.083(0.058) &  --                & --             &18.636(0.256) & 18.672(0.058)  &  SBIG	 \\
20140127& 56684.27&+32.3 &  19.957(0.042) & 19.154(0.026) &  --                &  --              & 18.803(0.022) & 18.751(0.028)  &   IO:O   \\
20140204& 56692.23& +40.2 &   --                     & --                    &   --                &  --             &   19.111(0.008) & --                     & LRS \\
20140207& 56695.24&+43.2 &  20.223(0.037) & 19.357(0.022) & 20.386(0.186) &  --           & 19.030(0.019) & 18.985(0.023)  &   IO:O   \\
20140209& 56697.61&+45.6 &--            & --            &  --            &   20.06       & --            & --             &   1      \\
20140209& 56697.62&+45.6 & --            & --            & --            &   20.13       & --            & --             &   1$^{d}$  \\
20140210& 56698.24&+46.2 &  20.187(0.038) & 19.480(0.032) &  20.485(0.150) &  --           & 19.037(0.018) & 19.002(0.027)  &   IO:O   \\
20140213& 56701.10&+49.1 &  20.213(0.093) & 19.556(0.050) & --            & --            & 19.054(0.037) & 19.086(0.033)  &   IO:O   \\
20140221& 56709.60&+57.6 &  --            & --            & --            & --            & 19.14         & --             &   1      \\
20140221& 56709.61&+57.6 &  --            & --            & --            & --            & 19.22         & --             &   1      \\
20140223& 56711.22&+59.2 &  20.350(0.051) & 19.607(0.029) &  --            & --            & 19.205(0.022) & 19.244(0.027)  &   IO:O   \\
20140226& 56714.28&+62.3 & --            & --            &  --            & --            & 19.263(0.017) & 19.258(0.038)  &   OSIRIS \\
20140228& 56716.19&+64.2 &  20.550(0.054) & 19.653(0.033) &  $>$20.6 &  --         & 19.277(0.022) & 19.306(0.039)  &   IO:O   \\
20140309& 56725.17&+73.2 &  20.603(0.238) & 19.923(0.104) &   --            & --            & 19.535(0.104)  & 19.389(0.049)  &   AFOSC  \\
20140323& 56739.10&+87.1 &  21.051(0.038) & 20.250(0.061) & --            & --            & 19.644(0.008) & 19.781(0.010)  &   LRS    \\
20140401& 56748.08&+96.1 &  21.267(0.158) & 20.291(0.219) & --            & --            & --                 & 19.934(0.032)  &   AFOSC  \\
20140425& 56772.99&+121.0 &    --                  & 20.697(0.023) &   --        & --            &19.838(0.012) & 20.107(0.019)&  ALFOSC\\
20140428& 56775.94&+123.9 &--             & --            &  --            & --            & 19.849(0.080) & 20.173(0.085)  &   OSIRIS\\
20140514& 56791.93&+139.9 &  --                    & 20.985(0.086) &--            & --            & 19.927(0.039) & 20.357(0.064)  &   LRS   \\
20140615& 56823.94&+171.9 &  21.713(0.048) & 21.340(0.033) &  --            & --            & 20.151(0.015) & 20.545(0.017)  &   LRS   \\
20140723& 56861.92&+209.9 &  21.949(0.054) & 21.454(0.035) & --            & --            & 20.310(0.013) & 20.619(0.023)  &   LRS   \\
20140808& 56877.88&+225.9 &  22.257(0.070) & 21.693(0.047) & --            & --            & 20.447(0.020) & 20.816(0.025)  &   LRS   \\
20140814& 56883.88&+231.9 &--            & --            &  --            & --            & 20.540(0.165) & 20.925(0.046)  &   OSIRIS\\
20141217& 57008.20&+356.2 & --            &  --           &  --            &  --           & 21.531(0.106) & --                    & ALFOSC\\
20141220& 57011.19&+359.2 & --            & 22.302(0.539) & --            & --            & 21.406(0.106) & 21.540(0.196)  &   AFOSC \\
20141223& 57014.25&+362.2 &--            & --            &  --            & --            & 21.450(0.168) & 21.612(0.124)  &   OSIRIS\\
20150125& 57047.27&+395.3 &    --            & $>$22.5     &   --            & --            & 21.815(0.034) & 21.967(0.041)  &   ALFOSC\\
20150311& 57092.27&+440.3 &   --             &   $>$22.6   &  --          &  --           & 22.436(0.070)   & --                    & ALFOSC	\\
20150428& 57140.77&+488.8 & --            & --            & --            & --            & $>$23.0           & --             &   ALFOSC\\ 
\hline

\hline
\end{tabular}   
}
\medskip

$^a$ Phases are relative to the $r$-band maximum light: MJD=56652.0.\\ 
$^b$ Johnson-Cousins $R$ and $I$ filter data were transformed to Sloan $r$ and $i$ band magnitudes respectively, following the relations of \citet{Jordi2006A&A...460..339J} \\ 
$^c$ Measurements on unfiltered images, calibrated to the Sloan $r$-band magnitudes. \\ 
$^d$ This epoch also has a $z$-band detection at 18.69 $\mathrm{mag}$. In addition, there is a very early (2013-03-07; MJD=56476.29) $z$-band limit of $>$ 20.8~mag. \\
$1$ These data were taken from the Pan-STARRS Survey for Transients (see website: \\ https://star.pst.qub.ac.uk/ps1threepi/psdb/candidate/1131007351341051000/). \\
\end{minipage}
\end{table*} 

\clearpage

\clearpage

\begin{table*}
\begin{minipage}{175mm}                                  
\caption{Optical ($BVgriz$) photometric measurements of AT~2018aes.}
\label{AT2018aes_opt_LC}  
\scalebox{0.9}{
\begin{tabular}{@{}ccccccccccl@{}}
\hline
Date  & MJD &Phase$^{a}$&  $B$(err)  &  $V$(err) &  $g$(err) & $r$(err) & $i$(err) & $z$(err)  & Instrument \\
\hline
20180215 & 58164.65 &-46.5 & --           & --              & --             & $>$20.1    & --            & --        & ATLAS$^*$ \\
20180218 & 58167.57 &-43.5 & --           & --              & --             & $>$18.7    & --            & --        & 1 \\
20180307 & 58184.61 &-26.5 & --           & --              & --             & $>$20.1    & --            & --        & ATLAS \\
20180311 & 58188.54 &-22.6 & --           & --              & --             & 18.51(0.3) & --            & --        & 1 \\
20180311 & 58188.55 &-22.6 & --           & --              & --             & --         & 19.103(0.029) & --        & 2 \\
20180311 & 58188.61 &-22.5 & --           & --              & --             & 19.21(0.60) & --            & --        & ATLAS \\
20180314 & 58191.73 &-19.4 &20.004(0.089) & 19.321(0.084) & 19.613(0.054)   & 19.051(0.061) & 18.881(0.103) & --        & fl12 \\
20180317 & 58194.59 &-16.5 & --           & --              & 19.52(0.44)   & --            & --            & --        & ATLAS \\
20180318 & 58195.09 & -16.0&19.794(0.074) & 19.185(0.084) & 19.558(0.021) & 18.997(0.052) & 18.727(0.080) & --        & fl06 \\
20180319 & 58196.59 & -14.5& --           & --              & --          & 18.78(0.34) & --            & --        & ATLAS \\
20180322 & 58199.12 & -12.0&19.841(0.027) & 19.140(0.020) & 19.509(0.011) & 18.949(0.023) & 18.896(0.023) & --        & ALFOSC \\
20180322 & 58199.38 & -11.7&19.868(0.131) & 19.082(0.078) & 19.402(0.047) & 18.919(0.061) & 18.775(0.080) & --        & fl05 \\
20180326 & 58203.23 & -7.9&$>$19.6   & 18.904(0.064) & 19.314(0.045) & 18.830(0.048) & 18.663(0.072) & --             & fl03 \\
20180329 & 58206.50 & -4.6& --           & --           & --          & 18.84(0.31) & --            & --        & ATLAS \\
20180330 & 58207.11 & -4.0&19.714(0.068) & 19.089(0.065) & 19.383(0.064) & 18.827(0.045) & 18.716(0.041) & 18.634(0.075) & ALFOSC \\
20180405 & 58213.09 & +2.0&19.830(0.020) & 19.019(0.010) & 19.386(0.010) & 18.858(0.010) & 18.546(0.049) & 18.569(0.027) & ALFOSC \\
20180407 & 58215.19 & +4.1&--            & 19.069(0.026) & 19.465(0.031) & 18.750(0.050) & 18.575(0.027) & 18.479(0.045) & ALFOSC \\
20180408 & 58216.16 & +5.1&19.923(0.029) & 19.068(0.019) & 19.486(0.010) & 18.871(0.015) & 18.652(0.025) & 18.605(0.041) & ALFOSC \\
20180411 & 58219.14 & +8.0&19.934(0.025) & 19.011(0.011) & 19.559(0.010) & 18.876(0.019) & 18.632(0.030) & 18.596(0.035) & ALFOSC\\
20180412 & 58220.53 & +9.4& --           & --           & --          & 18.77(0.10) & --            & --        & ATLAS \\
20180417 & 58225.03 & +13.9&20.104(0.042) & 19.220(0.024) & 19.590(0.025) & 18.952(0.024) & 18.725(0.018) & 18.574(0.037) & ALFOSC \\
20180418 & 58226.10 & +15.0&20.184(0.075) & 19.224(0.029) & 19.653(0.018) & 19.021(0.022) & 18.773(0.040) & 18.758(0.051) & IO:O \\
20180418 & 58226.98 & +15.9&--            & 19.138(0.104) & 19.648(0.094) & 19.039(0.078) & 19.005(0.135) & --             & Moravian \\
20180419 & 58227.98 & +16.9&20.121(0.175) & 19.205(0.088) & 19.736(0.090) & 18.969(0.099) & 18.811(0.219) & --             & Moravian \\
20180426 & 58234.13 & +23.0&20.389(0.033) & 19.563(0.017) & 19.885(0.025) & 19.172(0.021) & 18.899(0.038) & 18.785(0.043) & ALFOSC \\
20180501 & 58239.04 & +27.9&20.584(0.049) & 19.570(0.022) & 20.099(0.019) & 19.390(0.043) & 19.089(0.032) & 19.033(0.046) & ALFOSC\\
20180504 & 58242.98 & +31.9&20.771(0.134) & 19.654(0.050) & 20.124(0.054) & 19.476(0.031) & 19.200(0.050) & 18.999(0.078) & IO:O \\
20180507 & 58245.98 & +34.9&20.820(0.104) & 19.781(0.028) & 20.478(0.033) & 19.538(0.013) & 19.313(0.061) & 19.021(0.090) & IO:O \\
20180510 & 58248.48 & +37.4& --           & --            & --            & 19.66(0.62)   & --            & --          & ATLAS \\
20180513 & 58251.02 & +39.9&21.111(0.146) & 20.175(0.043) & 20.566(0.065) & 19.798(0.041) & 19.568(0.082) & 19.145(0.068) & IO:O \\
20180515 & 58254.00 & +42.9&21.182(0.086) & 20.189(0.041) & 20.693(0.041) & 19.938(0.036) & 19.473(0.037) & 19.181(0.044) & ALFOSC \\
20180517 & 58255.95 & +44.8&21.239(0.210) & 20.276(0.080) & 20.972(0.051) & 20.008(0.039) & 19.651(0.121) & 19.313(0.070) & IO:O \\
20180522 & 58260.92 & +49.8&$>$21.2       & 20.468(0.120) & 21.261(0.373) & 20.140(0.087) & 19.985(0.116) & 19.498(0.118) & IO:O \\
20180525 & 58263.05 & +52.0&21.714(0.111) & 20.559(0.105) & 21.343(0.046) & 20.356(0.071) & 20.092(0.082) & 19.728(0.071) & ALFOSC\\
20180531 & 58269.02 & +57.9&22.048(0.164) & 20.986(0.060) & 21.813(0.062) & 20.608(0.024) & 20.257(0.121) & 19.910(0.093) & ALFOSC \\
20180531 & 58269.94 & +58.8&$>$21.8       & $>$20.9          & $>$21.6       & $>$20.7       & 20.332(0.275) & 19.900(0.267) & IO:O \\
20180605 & 58274.05 & +62.9&22.189(0.128) & 21.106(0.070) & 21.898(0.073) & 20.907(0.032) & 20.567(0.114) & 19.917(0.060) & ALFOSC \\
20180605 & 58274.91 & +63.8&$>$21.8       & 21.280(0.075) & 21.956(0.120) & 20.831(0.044) & 20.708(0.435) & 19.911(0.161) & IO:O \\
20180610 & 58279.99 & +68.9&$>$21.8       & 21.454(0.193) & 22.124(0.147) & 21.304(0.123) & 20.712(0.258) & 20.258(0.235) & IO:O \\
20180626 & 58295.97 & +84.9&--            & --            &  $>$22.5      & 22.335(0.104) & $>$21.8       & 20.996(0.226) & ALFOSC \\
20180704 & 58303.97 & +92.9&--            & --            &    --         & 22.865(0.221) & 21.866(0.379) & 21.414(0.211) & ALFOSC \\
\hline
\end{tabular}   
}
\medskip

$^a$ Phases are relative to the $r$-band maximum light: MJD= 58211.1.\\ 
$^*$ ATLAS \citep[Asteroid Terrestrial-impact Last
Alert System; ][]{Heinze2018AJ....156..241H, Tonry2018PASP..130f4505T, Smith2020PASP..132h5002S} observations in {\sl cyan}- and {\sl orange}-band magnitudes are reported in the sloan g and r columns respectively.\\
$1$ Unfiltered KAIT data  reported to the sloan $r$ magnitudes (AB system). Data from the Transient Name Server (TNS; https://wis-tns.weizmann.ac.il/object/2018aes).  \\ 
$2$ The $i$-band magnitude is the weighted mean value obtained from the Pan-STARRS Survey for Transients (see website: https://star.pst.qub.ac.uk/sne/ps13pi/psdb/lightcurve/1134817771035644300/).\\ 
\end{minipage}
\end{table*} 

\begin{table*}
\begin{minipage}{175mm}
\caption{NIR ($JHK$) photometric measurements of AT~2018aes. }
\label{AT2018aes_nir_LC}
\begin{tabular}{@{}ccccccl@{}}
\hline
Date & MJD & Phase$^a$ & $J$(err) & $H$(err) & $K$(err) & Instrument \\
\hline
20180331& 58209.00& -2.1& 17.781(0.040)&  17.387(0.075)&  16.235(0.088) &NOTCam\\
20180511& 58249.92& +38.8& 18.257(0.065)&  17.994(0.063)&  17.148(0.076) & NOTCam\\
20180627& 58296.95& +85.9& 19.752(0.209)&  19.363(0.299)&  17.957(0.160) & NOTCam\\
20180731& 58330.88& +119.8& $>$ 20.5       &  19.662(0.392)&    --                  & NOTCam\\
20180819& 58349.86& +138.8&  --                 &         --             &  18.442(0.186) & NOTCam\\
\hline
\end{tabular}

\medskip
$^a$ Phases are relative to the $r$-band maximum light: MJD= 58211.1.\\ 
\end{minipage}
\end{table*}

\clearpage


\newpage
\end{appendix}
%
%



\end{document}